\documentclass[12pt,manuscript]{aastex}
\usepackage{natbib}
\usepackage{comment}

\begin{document}
\title{Chromosphere to 1 AU Simulation of the 2011 March 7th Event: A Comprehensive Study of Coronal Mass Ejection Propagation}
\author{M. Jin\altaffilmark{1,2}, W. B. Manchester\altaffilmark{3}, B. van der Holst\altaffilmark{3}, I. Sokolov\altaffilmark{3},  G. T\'{o}th\altaffilmark{3}, A. Vourlidas\altaffilmark{4}, C. A. de Koning\altaffilmark{5}, and T. I. Gombosi\altaffilmark{3}}

\altaffiltext{1}{Lockheed Martin Solar and Astrophysics Lab, Palo Alto, CA 94304, USA; jinmeng@lmsal.com}
\altaffiltext{2}{NASA/UCAR LWS Jack Eddy Fellow}
\altaffiltext{3}{Climate and Space Sciences and Engineering, University of Michigan, Ann Arbor, MI 48109, USA; chipm@umich.edu}
\altaffiltext{4}{The Johns Hopkins University Applied Physics Laboratory, Laurel, MD 20723, USA; angelos.vourlidas@jhuapl.edu}
\altaffiltext{5}{Cooperative Institute for Research in Environmental Sciences, University of Colorado, Boulder, CO 80309, USA; curt.a.dekoning@noaa.gov
}

\begin{abstract}
We perform and analyze results of a global magnetohydrodyanmic (MHD) simulation of the fast coronal mass ejection (CME) that occurred on 2011 March 7. The simulation is made using the newly developed Alfv\'{e}n Wave Solar Model (AWSoM), which describes the background solar wind starting from the upper chromosphere and extends to 24 R$_{\odot}$. Coupling AWSoM to an inner heliosphere (IH) model with the Space Weather Modeling Framework (SWMF) extends the total domain beyond the orbit of Earth.  Physical processes included in the model are multi-species thermodynamics, electron heat conduction (both collisional and collisionless formulations), optically thin radiative cooling, and Alfv\'{e}n-wave turbulence that accelerates and heats the solar wind. The Alfv\'{e}n-wave description is physically self-consistent, including non-Wentzel-Kramers-Brillouin (WKB) reflection and physics-based apportioning of turbulent dissipative heating to both electrons and protons. Within this model, we initiate the CME by using the Gibson-Low (GL) analytical flux rope model and follow its evolution for days, in which time it propagates beyond STEREO A.  A detailed comparison study is performed using remote as well as \textit{in situ} observations. Although the flux rope structure is not compared directly due to lack of relevant ejecta observation at 1 AU in this event, our results show that the new model can reproduce many of the observed features near the Sun (e.g., CME-driven extreme ultraviolet (EUV) waves, deflection of the flux rope from the coronal hole, ``double-front" in the white light images) and in the heliosphere (e.g., shock propagation direction, shock properties at STEREO A).
\end{abstract}

\keywords{interplanetary medium -- magnetohydrodynamics (MHD) -- methods: numerical -- solar wind -- Sun: corona -- Sun: coronal mass ejections (CMEs)}

\section{introduction}
Coronal mass ejections (CMEs) are a major source of potentially destructive space weather conditions, in which 10$^{15}$--10$^{16}$ g of plasma are ejected from the Sun with a kinetic energy of order 10$^{31}$--10$^{32}$ erg. The interplanetary CMEs (ICMEs) that pass Earth can disturb the Earth's magnetosphere and trigger geomagnetic storms \citep{gosling93}. Also, fast CMEs can drive shocks in the heliosphere (e.g., \citealt{sime87, vour03}) that are believed to be responsible for gradual solar energetic particle (SEP) events \citep{reames99} through the diffusive shock acceleration (DSA) mechanism. The SEPs can pose major hazards for spacecraft and human life in outer space. Due to the limited observations of CMEs/ICMEs, numerical models play a vital role for interpreting observations, testing theories, and providing forecasts. In particular, the ability to realistically simulate events with global MHD models is critical for the development of more accurate space weather forecast models.

The first attempts to predict CME evolution were achieved with empirical and kinematic models. These kinds of models utilize the remote observations near the Sun to predict the arrival time of CMEs at 1 AU. By using Solar and Heliospheric Observatory (SOHO) coronagraph measurements of CMEs, \citet{gopalswamy01} established an empirical model to estimate the arrival time of the CMEs at 1 AU with an average uncertainty of $\sim$10.7 hours. Another successful example is the kinematic 3-D Hakamada-Akasofu-Fry version 2 (HAFv.2) model \citep{hakamada82, fry01, dryer04}, in which type II radio burst, soft X-ray, and solar image data are used to derive shock speed and direction. The prediction error of the HAFv.2 model is also around 10 hours. In the past, the most frequently used predictive kinematic model was the cone model, which fits CME observations with three free parameters: angular width, speed, and central CME position \citep{zhao02}. Additional improvements were made to the cone model by allowing for non-uniform density and velocity \citep{hayashi06}. The cone model has been widely used by the research community to predict the CME/CME-driven shock velocity (e.g., \citealt{xie04, michalek07, luhmann10, vrsnak14}). With STEREO observations, the cone model has been significantly improved for application in an operational setting through the use of multi-view fitting, resulting in the CME Analysis Tool (CAT) \citep{millward13}.

In order to provide more accurate forecasts, the kinematic models are routinely combined with 3D MHD models. Typically, the kinematic models provide the inner boundary conditions (e.g., velocity, pressure, and density) to the MHD models. Then the CME disturbance in the MHD model can propagate to 1 AU and provide the forecast. Successful examples include combining the ENLIL heliosphere model with the CME cone model (e.g., \citealt{xie04, ods05}) or coupling the 3D MHD model by \citet{han88} with the HAFv.2 model \citep{wu07a, wu07b}. Both model combinations give density, temperature and velocity predictions at 1 AU with an arrival time error in the order of 10 hours. The average error in the CAT-Wang-Sheeley-Arge (WSA)-ENLIL operational model is 7.5 hours \citep{pizzo11, millward13}. While very useful, this type of model does not include the magnetic field of the CME since the heliosphere MHD model always starts outside of the magneto-sonic point, at which radial distance there are no magnetic observations available for use as inner boundary conditions.

In order to improve the capability of forecasting models, especially the ability to forecast geomagnetic storms, realistic 3D coronal models are needed to take into account the magnetic structure of CMEs. Therefore, the most sophisticated research models to date have inner boundaries lower in the solar corona and incorporate magnetically driven models of CME initiation. Several solar wind models with coronal inner boundaries have been developed in the past decade (e.g., \citealt{mikic99, groth00, rou03, cohen07, feng11, evans12}). By applying data-driven boundary conditions from synoptic magnetograms, these solar wind models can reproduce realistically the steady state solar wind. Some data-driven models can also couple with a surface flux transport model to capture the global coronal evolution \citep{feng12}. 

Here, we use a new state-of-the-art chromosphere-coronal model, the Alfv\'{e}n wave solar model (AWSoM), and couple it with the Space Weather Modeling Framework (SWMF) to self-consistently simulate the space environment beyond 1 AU \citep{bart14}. AWSoM is developed from previous works \citep{bart10, jin12, sokolov13, oran13}, which began with a two-temperature Alfv\'{e}n wave-driven solar wind model \citep{bart10}. In order to mimic the turbulent heating of electrons, \citet{jin12} partitioned 40\% of the dissipation energy to electrons in the model and did a validation study with multiple observations. \citet{sokolov13} further developed the model by incorporating the balanced turbulence at the top of the closed field lines and by extending it down to the chromosphere and including radiative cooling. A detailed model-data comparison was done for this model by \citet{oran13}. By separating the electron and proton thermodynamics, the CME and CME-driven shocks can be correctly simulated with this model \citep{chip12, jin13}.  The newly developed AWSoM further incorporated physically-based wave reflection, energy partition, and collisionless heat conduction. More details of the model will be shown in \S 2.1.

There are three major types of CME initiation models in the SWMF: the analytical flux rope model, the breakout model, and the flux-emergence model. For the first type (e.g., \citealt{titov99, gibson98}), the flux ropes are implemented into the background solar wind solution and will erupt due to force-imbalance. Recently, \citet{titov14} developed a modified Titov-D\'{e}moulin (TD) flux rope model that can reach a numerically exact equilibrium in a subsequent MHD relaxation therefore represents a more self-consistent modeling of pre-eruptive configuration. For the second type \citep{anti99}, photospheric shear flows are applied around the polarity inversion line (PIL) until a current sheet forms and reconnection drives the eruption. The advantage of the breakout model lies on the realistic CME acceleration process during the initiation. For the flux-emergence model, CMEs are triggered by Lorentz-force-driven shearing motions that transport axial flux and energy to the expanding field (e.g., \citealt{chip04c}). All three initiation models have been successfully used in CME simulations (e.g., \citealt{chip04a, chip04b, rou04, chip08, mac04, bart09, karpen12, chip14, jin16}).

Here, we use a flux rope model, which has the advantages of being both data-driven (more details in \S 2.2) and computationally efficient (the system starts from a state of force imbalance and does not require a long and costly energy build-up phase). By initiating a TD flux rope, \citet{chip08} simulated the Halloween CME event from the corona to the Earth and did the first quantitative comparison between the synthetic coronagraph images and LASCO observations, in which the strong CME-driven shock was simulated and validated. In a description of the same simulation by \citet{toth07}, the arrival time of the simulated CME is within $\sim$1.8 hours comparing with the observed arrival time. Due to the realistic CME and shock structures, this type of model has also been used to investigate shock-driven SEP acceleration \citep{rou04, chip05, kozarev13} and CME-CME interaction \citep{lugaz05, lugaz07, lugaz13}. For a recent review of the numerical modeling of ICMEs, one can refer to \citet{lugaz11a}.

In this paper, we describe a realistic CME simulation of an event that occurred on 2011 March 7 from active region (AR) 11164. The simulation covers the CME propagation from the Sun to 1 AU by initiating the CME in the AWSoM with the Gibson-Low (GL) flux rope model \citep{gibson98}. Good observational coverage of this event from SDO, SOHO, and STEREO A/B (STA/B), provides an excellent opportunity to validate our CME simulation from the Sun to 1 AU. Detailed analysis of the simulation and observational data will help us get a better understanding of the important physical processes at play during the CME propagation in the heliosphere. The 2011 March 7 CME event is fast, with a speed over 2000 km s$^{-1}$, it drives a shock and produces a strong SEP event. The major part of the CME-driven shock is toward STEREO A (STA). The shock structure passes STA at $\sim$6:50 UT on March 9 without the flux rope structure behind the shock. The CME-driven shock in the slow speed stream did hit the Earth at 7:44 UT on March 10, with a lengthy period of negative Bz and triggered a geomagnetic storm of Kp = 6. However, the ICME at Earth may result from the interaction between the ICME of our chosen event with that from an earlier, slower event \citep{wood12}, which is not included in our model. Therefore, we do not show the \textit{in situ} comparison at Earth in this paper but rather at STA, where an isolated shock structure was observed.

The paper is organized as follows: In Section 2, we describe AWSoM for the background solar wind as well as the GL flux rope model for the CME initiation. The 2011 March 7 CME event simulation results and the comparison by observations are shown in Section 3, followed by the summary and conclusion in Section 4.

\section{Models}
\subsection{Background Solar Wind Model}
The global simulation of the CME to 1 AU is performed with two individual models that comprise the solar corona (SC) and inner heliosphere (IH), each of which is based on the MHD model Block-Adaptive-Tree-Solarwind-Roe-Upwind-Scheme (BATS-R-US; \citealt{powell99}). The eruptive event generator (EE) is a suite of CME models specified as both initial and boundary conditions. These models are just three of almost a dozen components that can be run under the SWMF that was developed at the Center for Space Environment Modeling (CSEM; \citealt{toth05, toth12}). The SWMF allows for different physical domains of the space environment to be simultaneously simulated and coupled to form a more complete description than could be attained by any single numerical model. In this case, the coupled models extend from the solar upper chormosphere to interplanetary space extending beyond 1 AU.

The SC model used in this study is the newly developed AWSoM \citep{bart14}, which is a data-driven model extending from the upper chromosphere to the corona and solar wind. The steady state solar wind solution is obtained with the local time stepping and second-order shock-capturing scheme \citep{toth12}. The inner boundary condition of the magnetic field is specified by GONG synoptic magnetograms, while the initial magnetic field configuration is calculated by the Potential Field Source Surface (PFSS) model using a finite difference method \citep{toth11b}. The model starts from the upper chromosphere with fixed temperature $T =$ 50,000 K and density $n =$2$\times$10$^{17}$ m$^{-3}$. At the base of the atmosphere, the temperature is fixed at 50,000 K while the density falls off exponentially until it reaches a level where the radiative losses are sufficiently low that the temperature increases monotonically with height. Above this height, the temperature increases rapidly forming the transition region. This procedure allows chromospheric evaporation to self-consistently populate the corona with an appropriately high plasma density. The inner boundary density and temperature do not otherwise have a significant influence on the global solution \citep{lio09}. The Alfv\'{e}n wave turbulence is launched at the inner boundary with the Poynting flux scaling with the surface magnetic field. The solar wind is heated by Alfv\'{e}n wave dissipation and accelerated by thermal and Alfv\'{e}n wave pressure. Electron heat conduction and radiative cooling are also included in the model, which self-consistently create the solar transition region. In order to produce physically correct solar wind and CME structures, such as shocks, the electron and proton temperatures are separated.  Thus, while the electrons and protons are assumed to have the same bulk velocity, heat conduction is applied only to the electrons, owing to their much higher thermal velocity. Note that AWSoM also works for three temperatures to include the ion pressure anisotropy \citep{bart14, meng15}.

The SC model uses a 3D spherical block-adaptive grid from 1 R$_{\odot}$ to 24 R$_{\odot}$. The grid blocks consist of 6$\times$4$\times$4 mesh cells. The smallest radial cell size is $\sim$10$^{-3}$ R$_{\odot}$ near the Sun to resolve the steep density and temperature gradients in the upper chromosphere. The largest radial cell size in SC is $\sim$1 R$_{\odot}$. Inside $r=1.7$ R$_{\odot}$, the angular resolution is $\sim$1.4$^{\circ}$. Outside that region, the grid is coarsened by one level to $\sim$2.8$^{\circ}$. The IH model uses a block-adaptive Cartesian grid to reach 250 R$_{\odot}$ with grid blocks consisting of 4$\times$4$\times$4 mesh cells. The smallest cell size in IH is $\sim$0.1 R$_{\odot}$ and the largest cell size is $\sim$8 R$_{\odot}$. For both the SC and IH, adaptive mesh refinement (AMR) is performed to resolve the heliospheric current sheet (HCS). The number of total cells is $\sim$3$\times$10$^{6}$ in SC, and $\sim$1$\times$10$^{6}$ in IH. In steady state, both the SC and IH domains are in heliographic rotating coordinates (i.e., Carrington coordinates).

There are three major improvements to the model that should be mentioned compared with our previous paper \citep{jin13}: First, the Alfv\'{e}n wave turbulence dissipation rate is revised to incorporate physically consistent wave reflection and dissipation. The new dissipation rate can be expressed as:
\begin{equation}
\Gamma_{\pm}=\max\left(\mathcal{R}_{imb}, \frac{2\sqrt{|{\bf B}|}}{(L_{\perp}\cdot\sqrt{|{\bf B}|})}\sqrt{\frac{w_\mp}{\rho}}\right)
\end{equation}

\begin{equation}
\mathcal{R}_{imb}=\sqrt{[({\bf V_{A}}\cdot\nabla) \log V_{A}]^2+[{\bf b}\cdot(\nabla\times{\bf u})]^2}
\end{equation}
where $w_\mp$ are the wave energy densities. The $+$ sign is for waves propagating in the direction parallel to magnetic field $\bf B$, while the $-$ sign is for waves propagating antiparallel to $\bf B$. $\bf V_{A}=\bf B/\sqrt{\mu_{0}\rho}$ is the Alfv\'{e}n speed, $\bf b=\bf B/|B|$, and $\mu_{0}$ is the permeability of vacuum. $\rho$ is the mass density, and $\bf u$ is the velocity. $L_{\perp}$ represents the transverse correlation length of turbulence. $\mathcal{R}_{imb}$ represents the wave reflection rate, which is due to Alfv\'{e}n speed gradient and vorticity along the field lines. Second, instead of using a constant value for the heat partitioning between the electrons and protons, the results of linear wave theory and stochastic heating are used \citep{chandran11}. With this specification, the majority of wave heating goes to the electrons near the Sun and around the HCS, while ion heating dominates away from the Sun and HCS due to the stochastic heating mechanism. For the detailed calculation of the heat partitioning, please refer to the Appendix B of \citet{bart14}. 

\subsection{CME Initiation Model}
Within the steady state solar wind obtained in \S 2.1, we initiate the CME using the analytical GL flux rope model implemented in the EE of SWMF. We apply the analytical flux rope to the active region along the PIL in a state of force imbalance (due to the insufficient background plasma pressure to offset the magnetic pressure of the flux rope), such that it will erupt immediately. The analytical solution of the GL flux rope is obtained by finding a solution to $(\nabla\times{\bf B})\times{\bf B}-\nabla p-\rho {\bf g}=0$ and $\nabla\cdot{\bf B}=0$, by applying a mathematical stretching transformation to an axisymmetric spherical ball of twisted magnetic flux in the pressure equilibrium. During this process, the flux rope will acquire a geometrically complex configuration. At the same time,  Lorentz forces will be introduced, which support dense filament plasma in the solar gravitational field. There are several advantages of choosing the GL flux rope: First, it can be implemented easily into any magnetic configuration so that application in an operational space weather forecast is easier than the breakout model that requires a special field configuration. Second, compared with the TD flux rope, the magnetic structure of the GL flux rope is less diffusive \citep{chip04b} and leads to a better \textit{in situ} comparison at 1 AU. Third, the most important feature of the GL flux rope is that it captures the typical 3-part density structure of the CME \citep{llling85}.

For this simulation, the GL flux rope parameters are specified as follows: the stretching parameter $a=0.6$; the radius of the flux rope torus $r_{0}=0.8 R_{\odot}$; the distance of torus center from the center of the Sun $r_{1}=1.8 R_{\odot}$; the flux rope field strength parameter $\alpha=2.25$. The flux rope is placed at 27 degree latitude and 155 degree longitude into AR 11164. The flux rope is rotated 90 degrees to match the PIL and the position of the pre-existing filament observed before the eruption in $H_{\alpha}$ \citep{gal02}. The radius of the flux rope is constrained by the size of the active region. The field strength parameter is constrained by the observed CME speed near the Sun. In order to get a proper field strength parameter, successive runs were made to give the best overall propagation time. In this study, the field strength parameter is set so that the simulated CME speed is slightly larger than the observed CME speed near the Sun in order to offset the higher simulated solar wind density in the heliosphere as shown in Table 1 (see \S 3.2 for more details). Furthermore, the results based on these successive runs provide an empirical relationship that allow the flux rope parameters to be prescribed based on observations. The model can then be used to predict the longer term evolution of the CME in interplanetary space. The details of this parameter study are presented in the companion paper by \citet{jin16b}.

After the GL flux rope is inserted into the active region, the simulation is switched to time-accurate mode to capture the CME eruption and the MHD equations are solved in conservative form to guarantee the energy conservation across the CME-driven shock. Two more levels of refinement along the CME path are performed to resolve the CME-driven shock, which doubles the number of total cells in SC to $\sim$6$\times$10$^{6}$. The SC runs 1 hour alone to let the CME propagate to $\sim$18 R$_{\odot}$ when the SC-IH coupling begins. In the time-accurate mode, the IH runs in heliographic inertial coordinates (i.e., heliocentric inertial coordinates). In order to capture the shock structure, especially the shock structure during the satellite-passing, both the grids along the CME path and around the satellite points are refined, which triples the number of total cells in IH to $\sim$3$\times$10$^{6}$. The coupling between the SC and IH runs to $\sim$8 hours when all the CME structures have passed through the SC into the IH domain. Then the SC is turned off and the IH runs alone till the CME arrives at 1 AU.

\section{Results}
\subsection{Background Solar Wind \& CME Initiation}
\begin{figure}[h]
\begin{center}$
\begin{array}{c}
\includegraphics[scale=0.65]{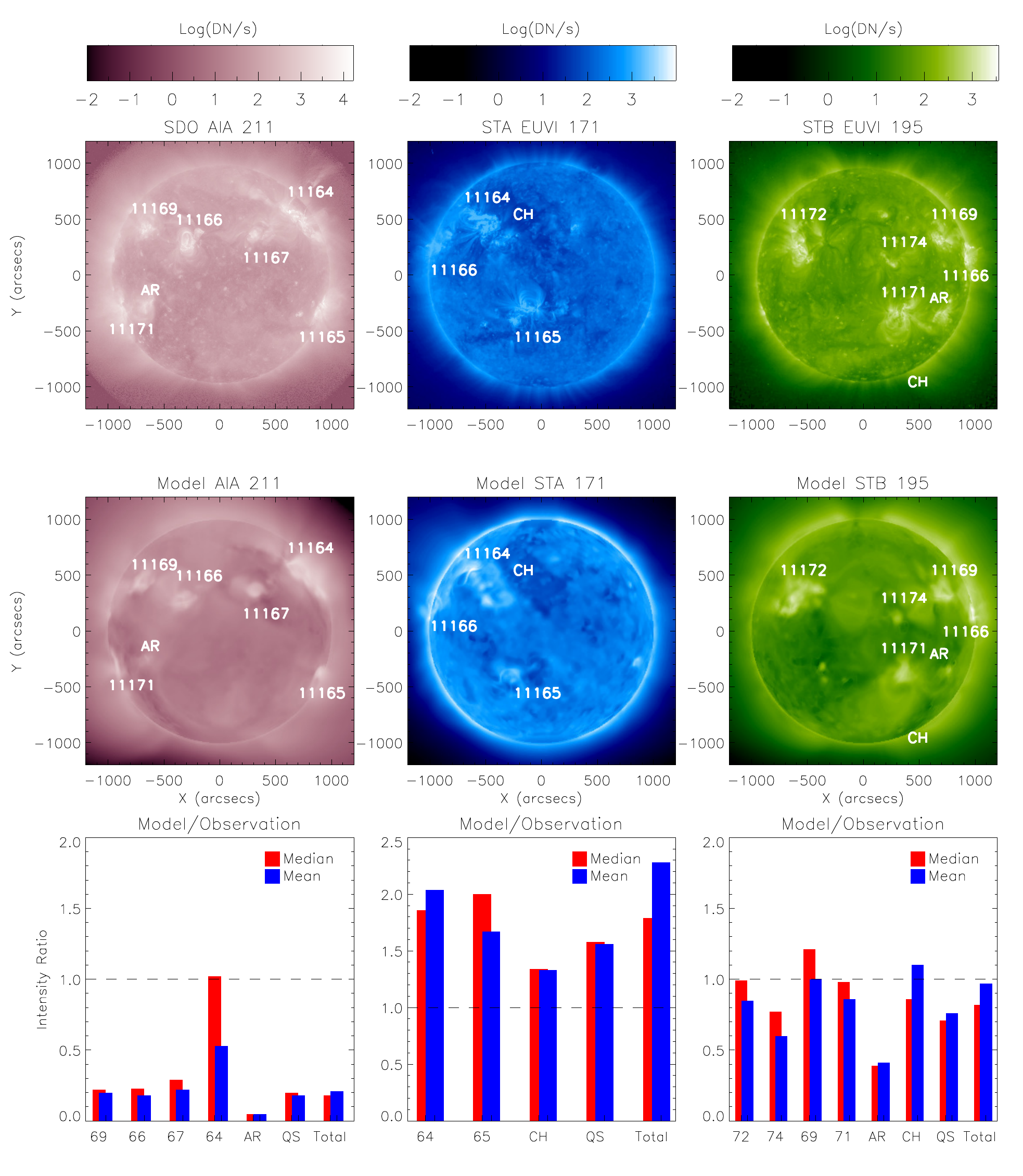}
\end{array}$
\end{center}
\caption{\label{fig:euv}The comparison between observations and synthesized EUV images of the steady state solar wind model. Top panels: Observational images from SDO AIA 211 \AA, STEREO A EUVI 171 \AA, and STEREO B EUVI 195 \AA. The observation time is 2011 March 7 $\sim$20:00 UT. Middle panels: synthesized EUV images of the model. The active regions and coronal holes are marked both in the observational and synthesized images. Bottom panels: Quantitative comparison between the model and observation for different structures of the Sun. The intensity ratio is Model/Observation. The active region numbers (without the initial ``111") are marked. AR, QS, and CH stand for active region, quiet Sun, and coronal hole respectively.}
\end{figure}

In order to validate the steady state solution of our model, we compare our model results with the available observations. Near the Sun, the model density and temperature are used to produce synthesized extreme ultraviolet (EUV) images, which are then compared with the EUV observations from SDO/AIA \citep{lemen12} and STEREO/Extreme UltraViolet Imager (EUVI; \citealt{howard08}). The comparison results are shown in Figure \ref{fig:euv}. Three EUV spectral bands (SDO AIA 211 \AA, STA EUVI 171 \AA, and STB EUVI 195 \AA) are selected that cover the temperature range from 1 MK to 2 MK. The observation time is at $\sim$20:00 UT on 2011 March 7, at which time, STA was $\sim$88$^{\circ}$ ahead of Earth and STB was $\sim$95$^{\circ}$ behind Earth. From these three view points, most of the Sun can be viewed. For both the observed and synthesized images, we use the identical log scale with unit DN s$^{-1}$. We can see clearly that the model reproduces all the major active regions and the on-disk/polar coronal holes. Compared with our previous model \citep{sokolov13}, the intensity of the active region is enhanced, which leads to a better comparison with the observations. The enhanced intensity is due to the increase of the wave reflection around the active regions, which results in greater wave dissipation and higher electron temperatures. Note that in order to resolve the active regions, the 6$\times$6$\times$6 grid block and spatially fifth-order MP5 limiter \citep{suresh97, chen16} are used. 

In order to compare the EUV emission in a more quantitative way, we further obtain the median/mean intensity ratio between model and observation for different structures on the Sun, including active regions, coronal holes, quiet Sun, and total emission. The comparison result is shown at the bottom panel of Figure \ref{fig:euv}. In general, the synthesized STB EUVI 195 \AA band has the best agreement with the observation, while the synthesized AIA 211 \AA band underestimates the emission by a factor of $\sim$5 and the synthesized STA EUVI 171 \AA band overestimates the emission by a factor of $\sim$2. Since the peak emission temperatures (log T) for 211 \AA, 195 \AA, and 171 \AA are 6.3, 6.2, and 5.8, respectively, the different model/observation ratios among the bands suggest a lower average coronal temperature therefore a smaller scale height in the model. The larger emission in the synthesized 171 \AA image could also be related to the optical thickness of that band, which is not taken into account when calculating the synthesized emission. In the same EUV band, different structures also show varying performance. For example, the AR 11164 in the AIA 211 \AA band has a better comparison than the other structures. Since the structures in the simulation highly depend on the input magnetogram, using magnetograms with more instantaneous magnetic field in the model could improve the EUV comparison and should be done in the future.

\begin{figure}[h]
\begin{center}$
\begin{array}{cc}
\includegraphics[scale=0.42]{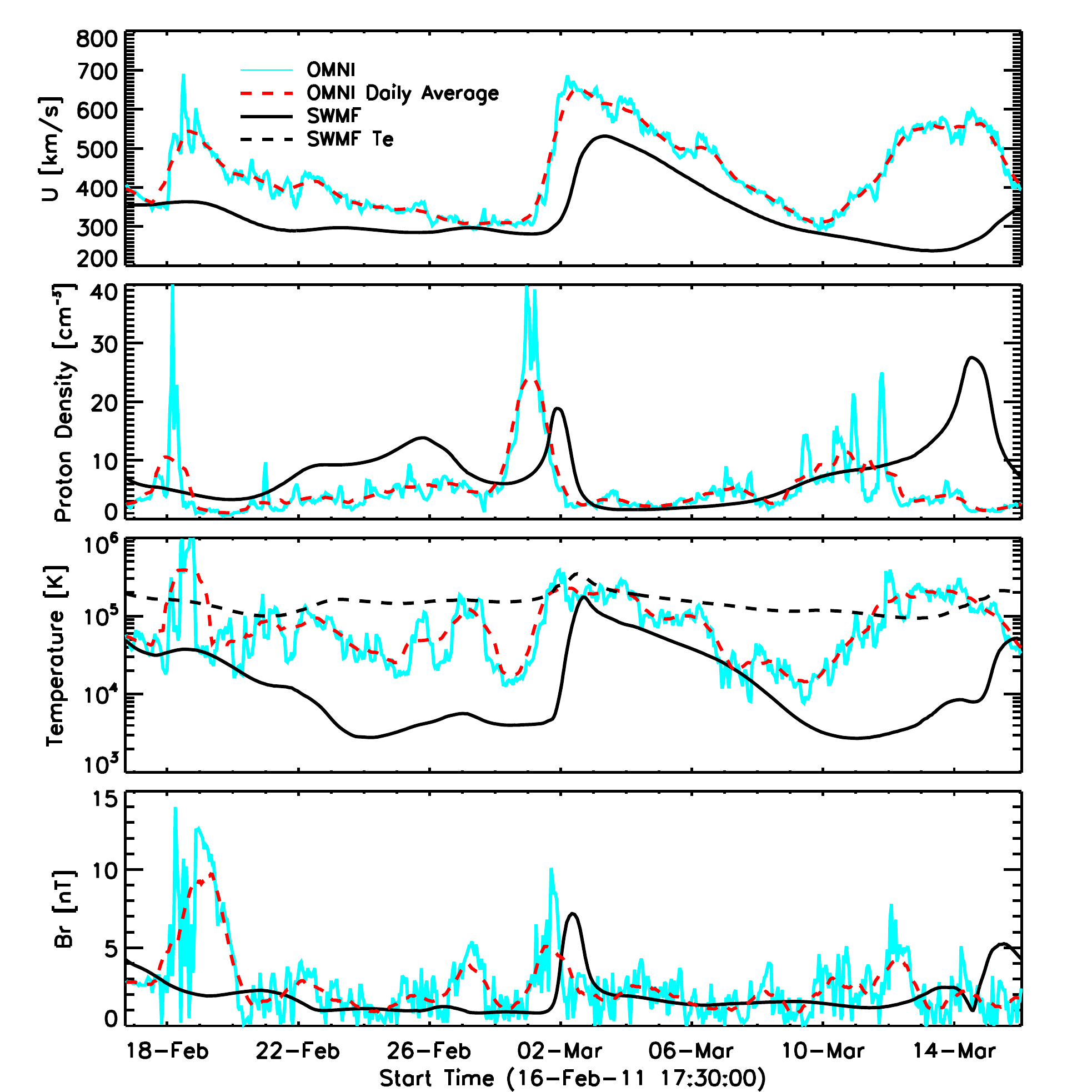} &
\includegraphics[scale=0.42]{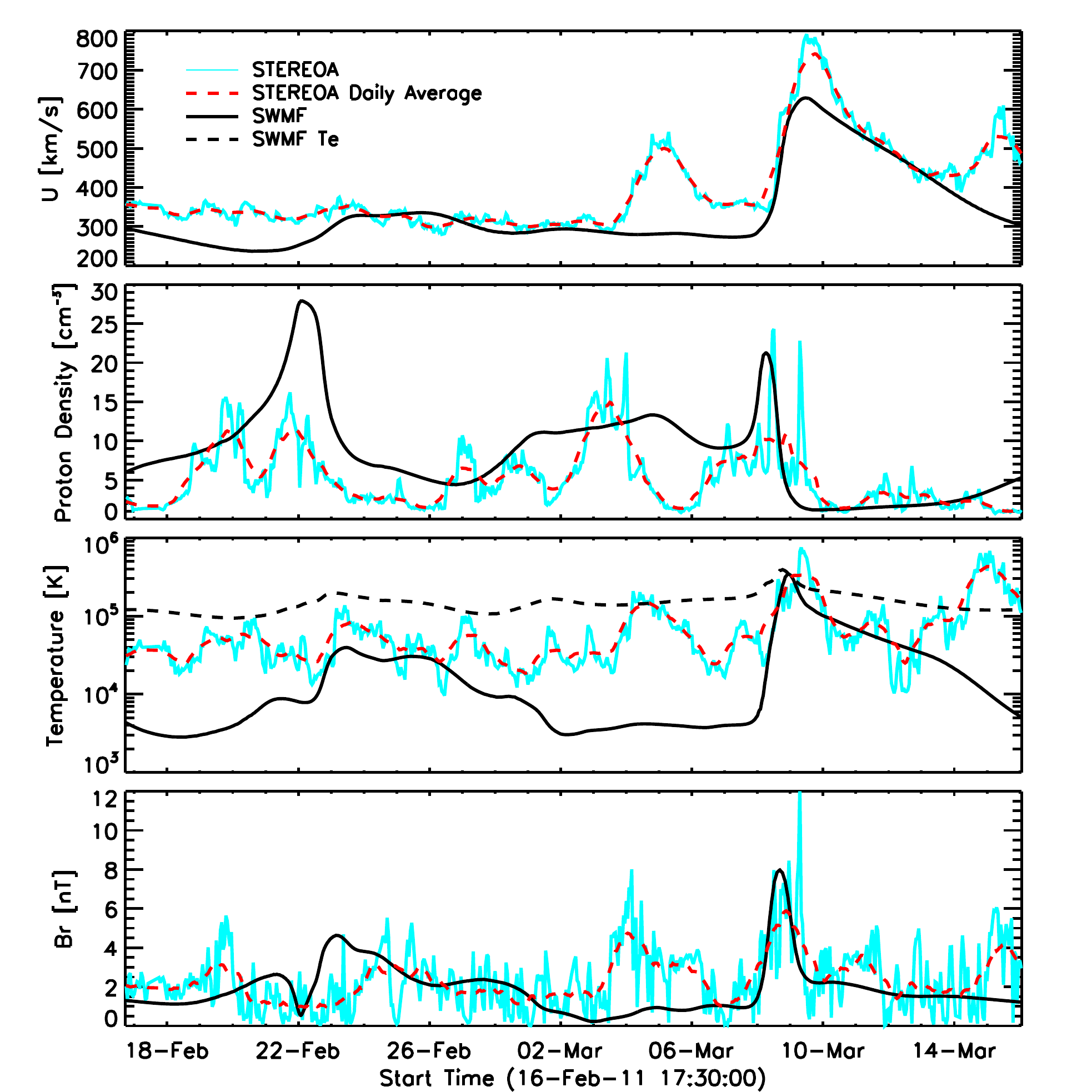}
\end{array}$
\end{center}
\caption{\label{fig:1au}Comparison of OMNI and STEREO A observed solar wind speed, proton density, proton/electron temperatures, and magnetic field with the steady state model output for CR2107.}
\end{figure}

In Figure \ref{fig:1au}, the \textit{in situ} OMNI and STA solar wind velocity, proton density, proton/electron temperatures, and magnetic field are shown with the steady state model results for comparison. The OMNI data (obtained from the National Space Science Data Center (NSSDC)) provides selected data from the Advanced Composition Explorer (ACE), Wind, Geotail, and IMP8 spacecraft (IMP8 ceased operation after October 7, 2006). The STA data comes from two instruments on board: the proton parameters are provided by the Plasma and Supra-Thermal Ion Composition Investigation (PLASTIC; \citealt{galvin08}); the magnetic field data is provided by the \textit{in situ} Measurements of Particles and CME Transients (IMPACT; \citealt{luhmann08}). We can see that the model reproduces the solar wind conditions at 1 AU. Both the location and plasma parameters of the co-rotating interaction region (CIR) are captured in the model. Note that for the 2011 March 7 CME event, the CIR and the CME-driven shock structures are very close in location and may interact with each other. Therefore, getting CIR structure correct is very important for successful CME event simulation. With the implementation of the collisionless heat conduction, the electron temperature reaches 0.1 MK at 1 AU, which is suggested by previous observations (e.g., \citealt{burlaga71}). In Table 1, we show statistics of simulated/observed solar wind parameters so that the comparison can be viewed in a more quantitative way. The mean square error (MSE) between the observed and simulated parameters are calculated: $MSE=\frac{1}{n}\sum_{t=1}^{n} (X_{t}-X_{t}^{\prime})^2$, where $X$ and $X^{\prime}$ represent observed and simulated plasma parameters. These values can be compared directly with the study by \citet{jian2015} (Figure 6 in their paper), in which they compared different solar wind models at the Community Coordinated Modeling Center (CCMC). By comparing the MSEs, it shows that the simulated solar wind speed, density, and magnetic field in this study outperform most of the models at CCMC in this regard. However, the simulated proton temperature is too low in the slow solar wind therefore underperforms the models at CCMC. Note that the study of CCMC models are based on 7 Carrington rotations. Therefore, with the single rotation result in this study, it is hard to judge the performance. The ratios between the simulated and observed median/mean/maximum/minimum are also shown in Table 1, in which we can see that the simulated solar wind is relatively slower and denser, with smaller magnetic field and lower temperature.

\begin{figure}[h]
\begin{center}$
\begin{array}{cc}
\includegraphics[scale=0.23]{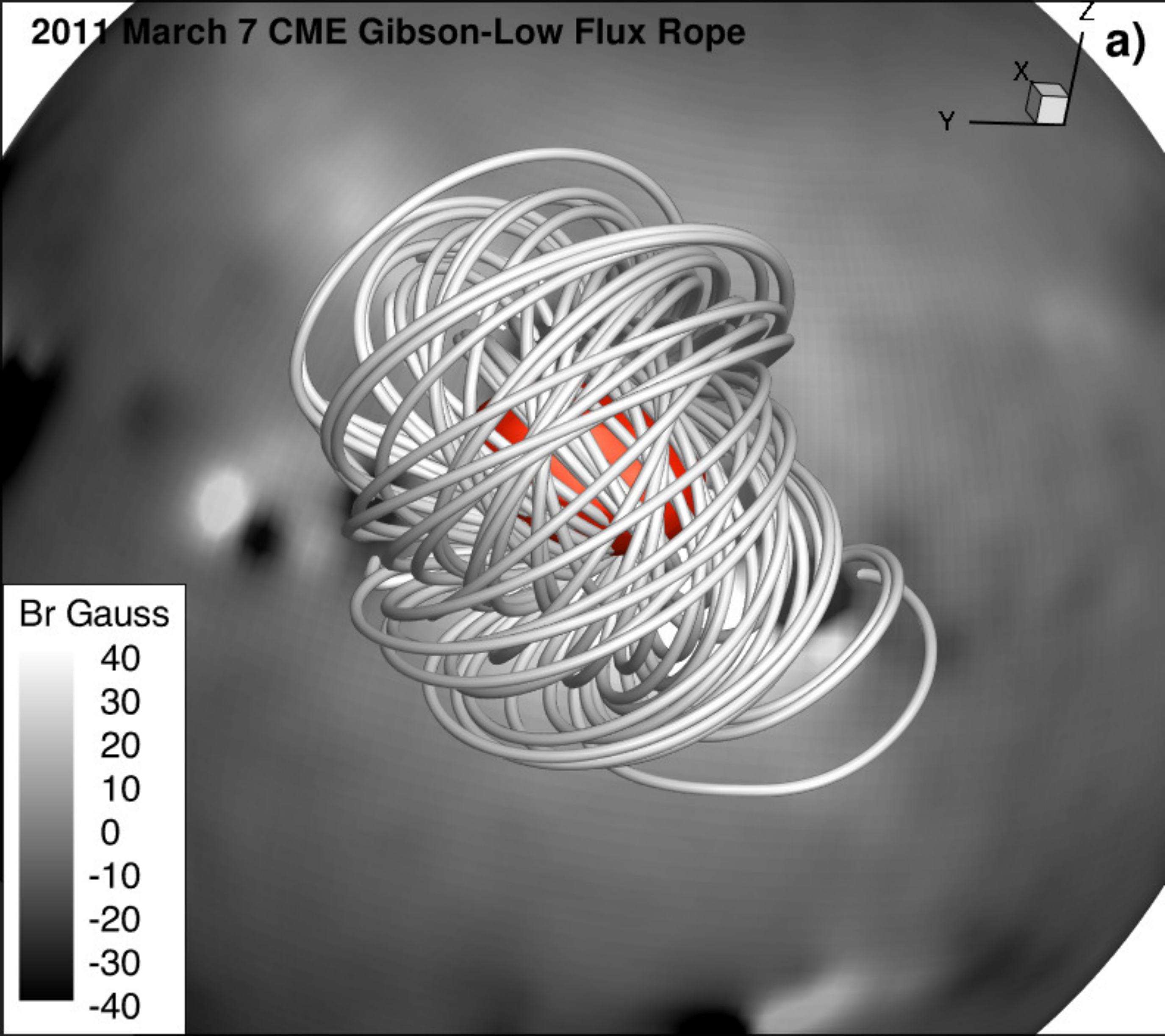} &
\includegraphics[scale=0.23]{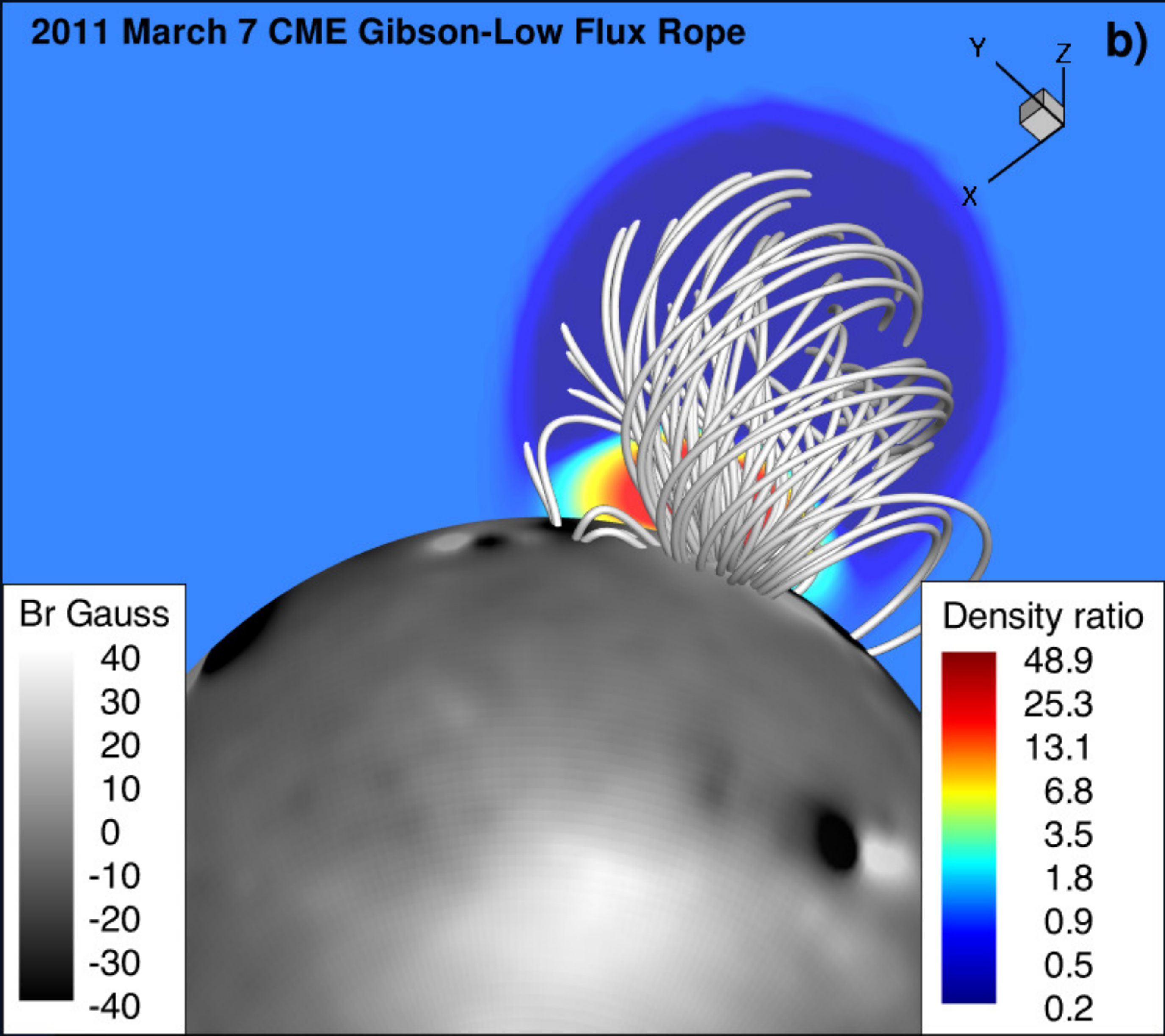} \\
\includegraphics[scale=0.23]{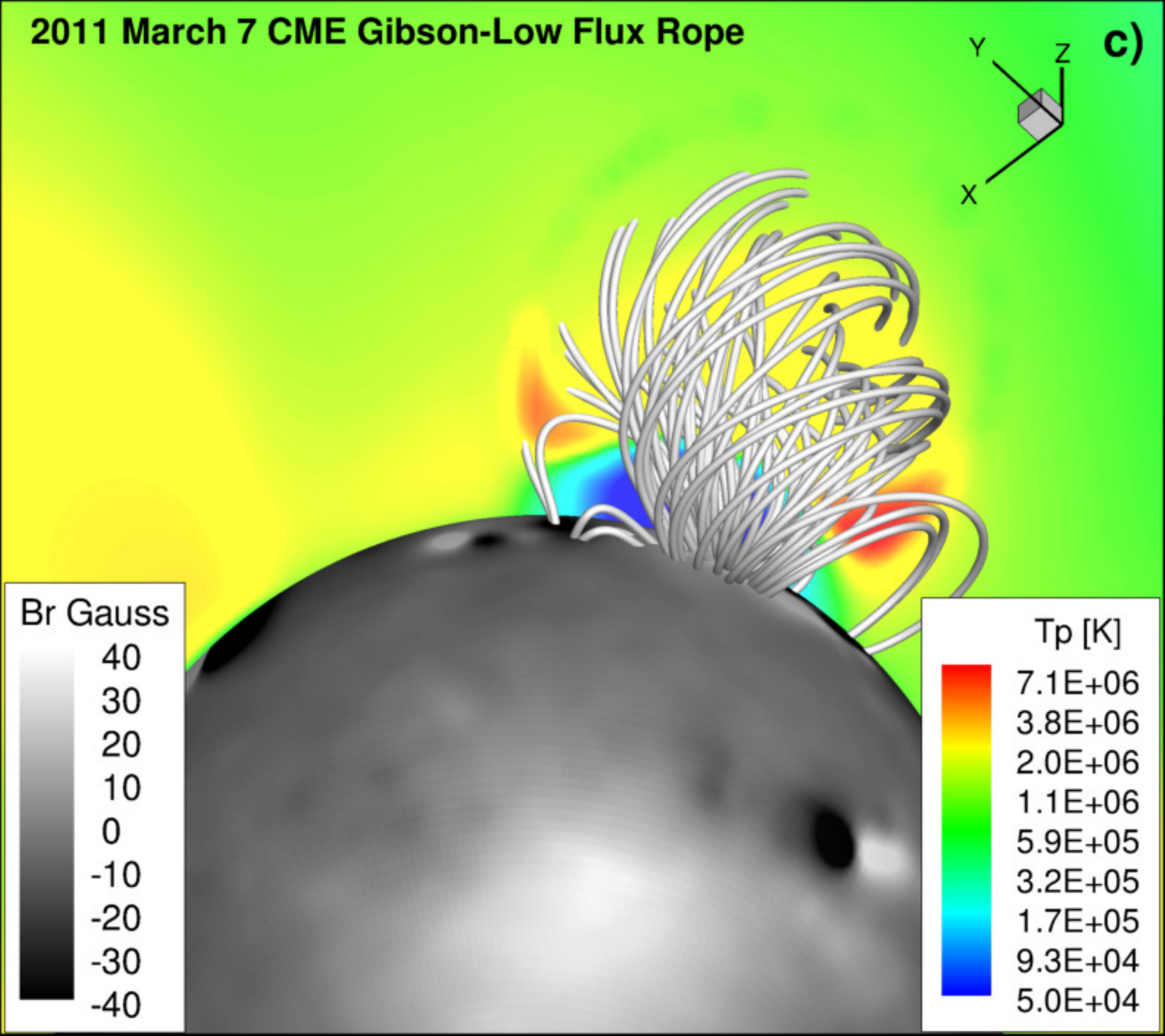} &
\includegraphics[scale=0.23]{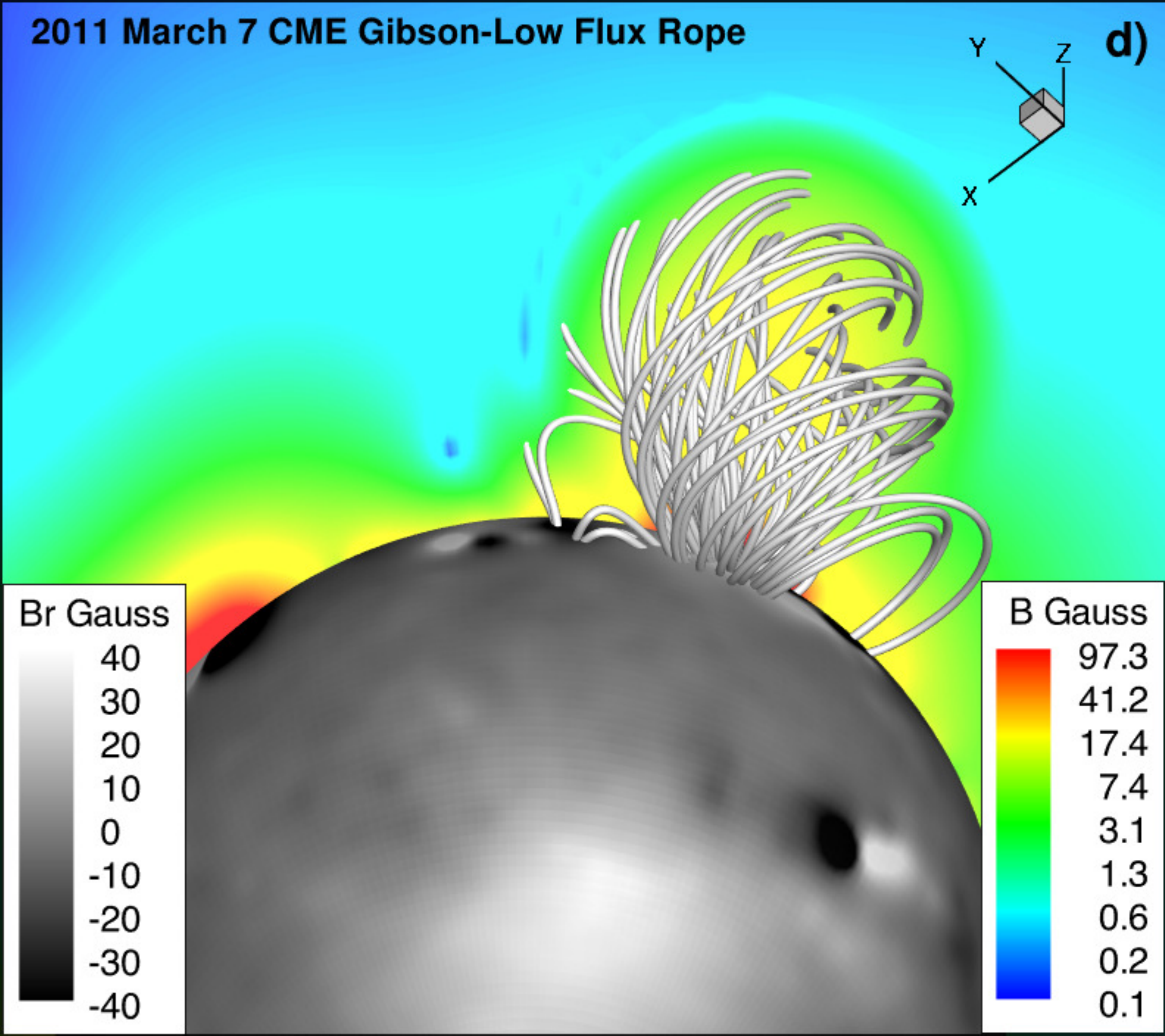} \\
\includegraphics[scale=0.23]{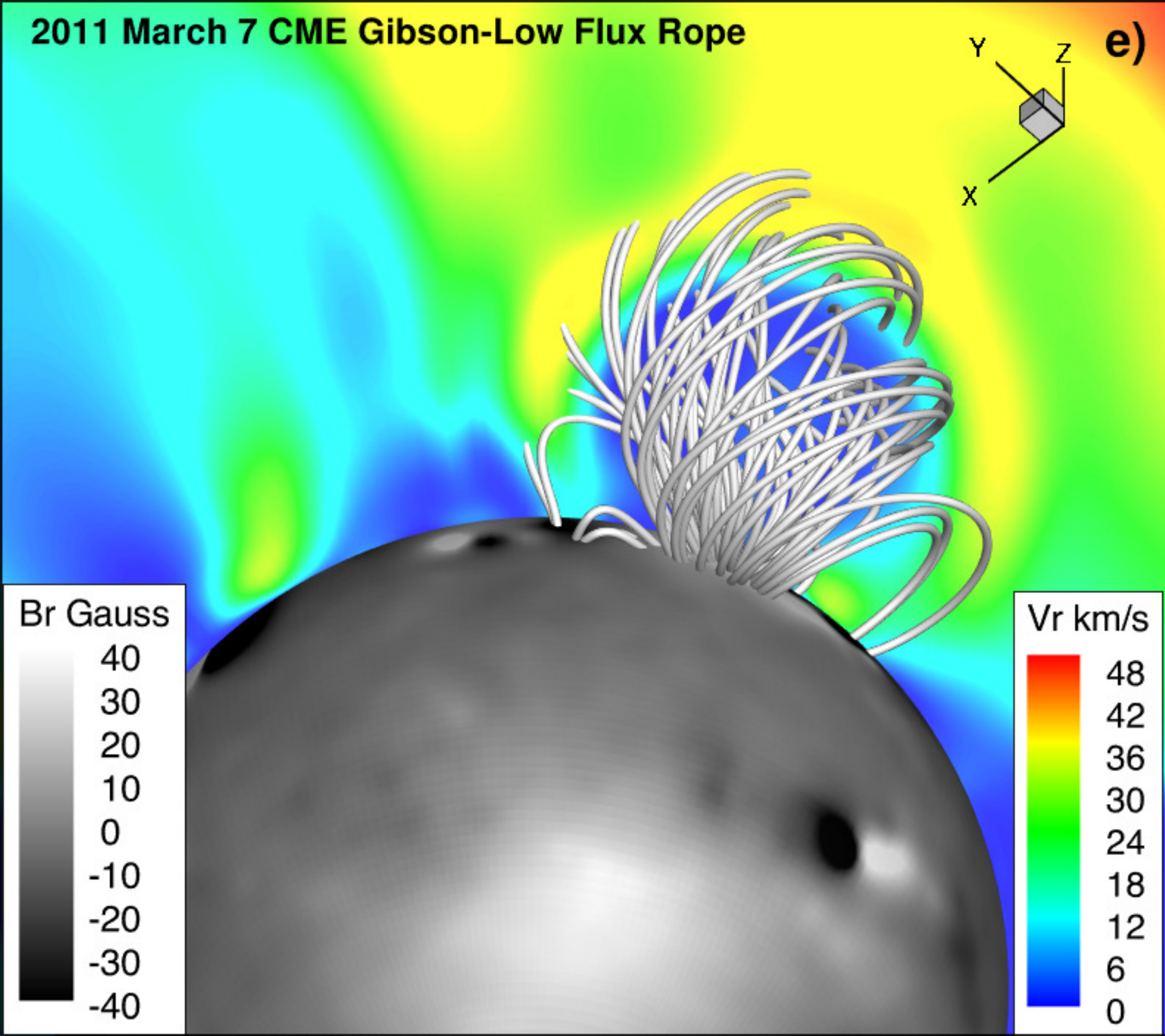} &
\includegraphics[scale=0.23]{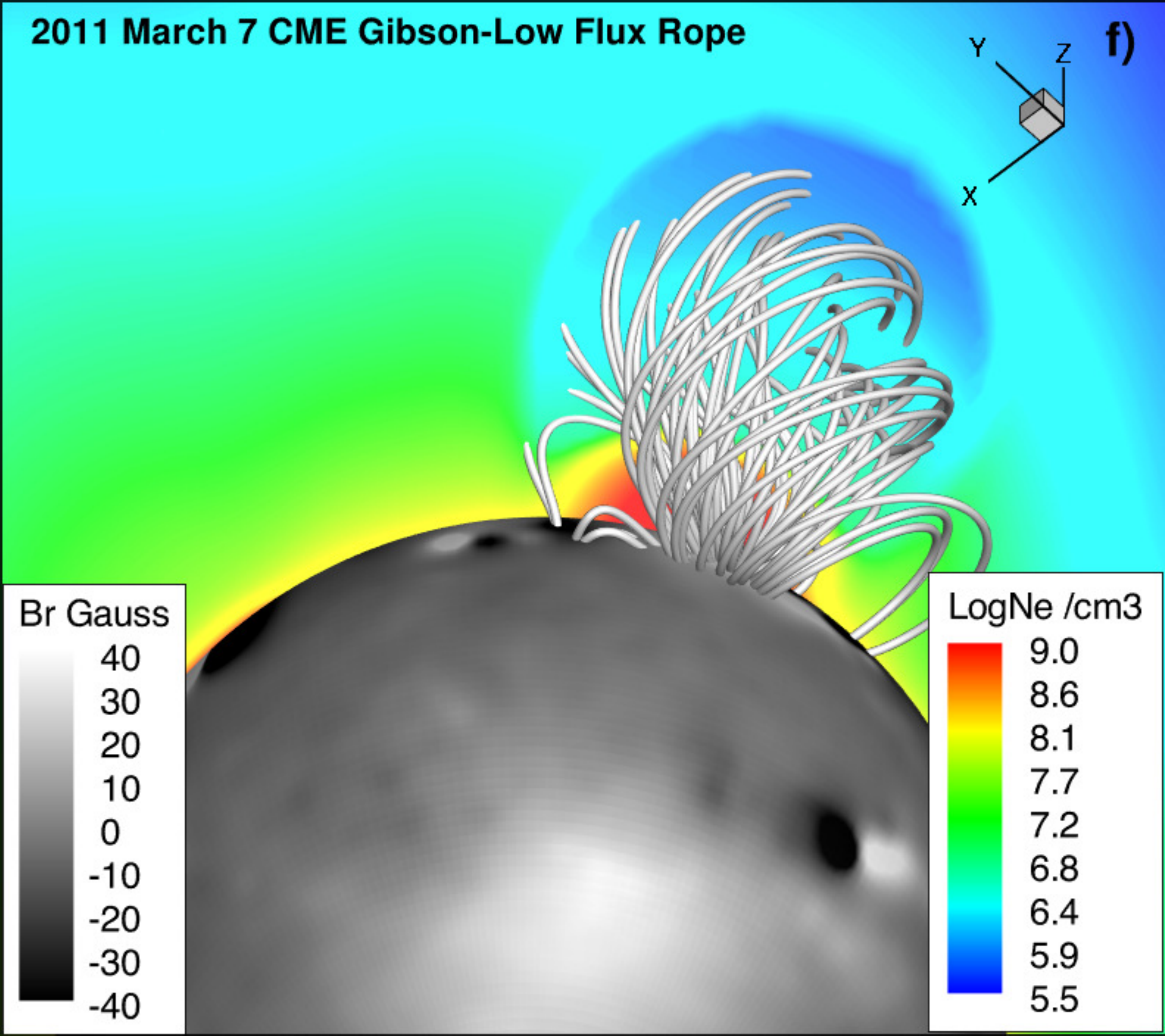} 
\end{array}$
\end{center}
\caption{\label{fig:GL}The initial GL flux rope configuration for 2011 March 7 CME. (a) 3D GL flux rope configuration viewed from the top of the active region. (b)--(f): central plane of the GL flux rope with density ratio, proton temperature, total magnetic field, radial velocity, and plasma density.}
\end{figure}

In Figure \ref{fig:GL}, we show the initial GL flux rope configuration inserted in the steady state solar wind solution. Figure \ref{fig:GL}a shows the 3D GL flux rope structure viewed from above the active region AR 11164. In order to mimic the observed filament configuration, the GL flux rope is modified so that both the filament polarity and chirality are matched with the observation \citep{martin98}. We can see both the toroidal and poloidal fields from the selected field lines. Also, the filament material is included at the bottom of the GL flux rope above the PIL. In Figure \ref{fig:GL}b-f, the density ratio, proton temperature, total magnetic field, radial velocity, and plasma density are shown on the central planes of the GL flux rope. The core of the GL flux rope has a higher density and lower temperature than the background, while the cavity of the GL flux rope has low density and higher temperature along with a higher magnetic field strength. All these features of the density match the 3-part CME structure observed in Thomson scattered white light observations (the 3-part structure in the synthetic white light images will be shown in \S 3.2).

\begin{figure}[h]
\begin{center}$
\begin{array}{cc}
\includegraphics[scale=0.28]{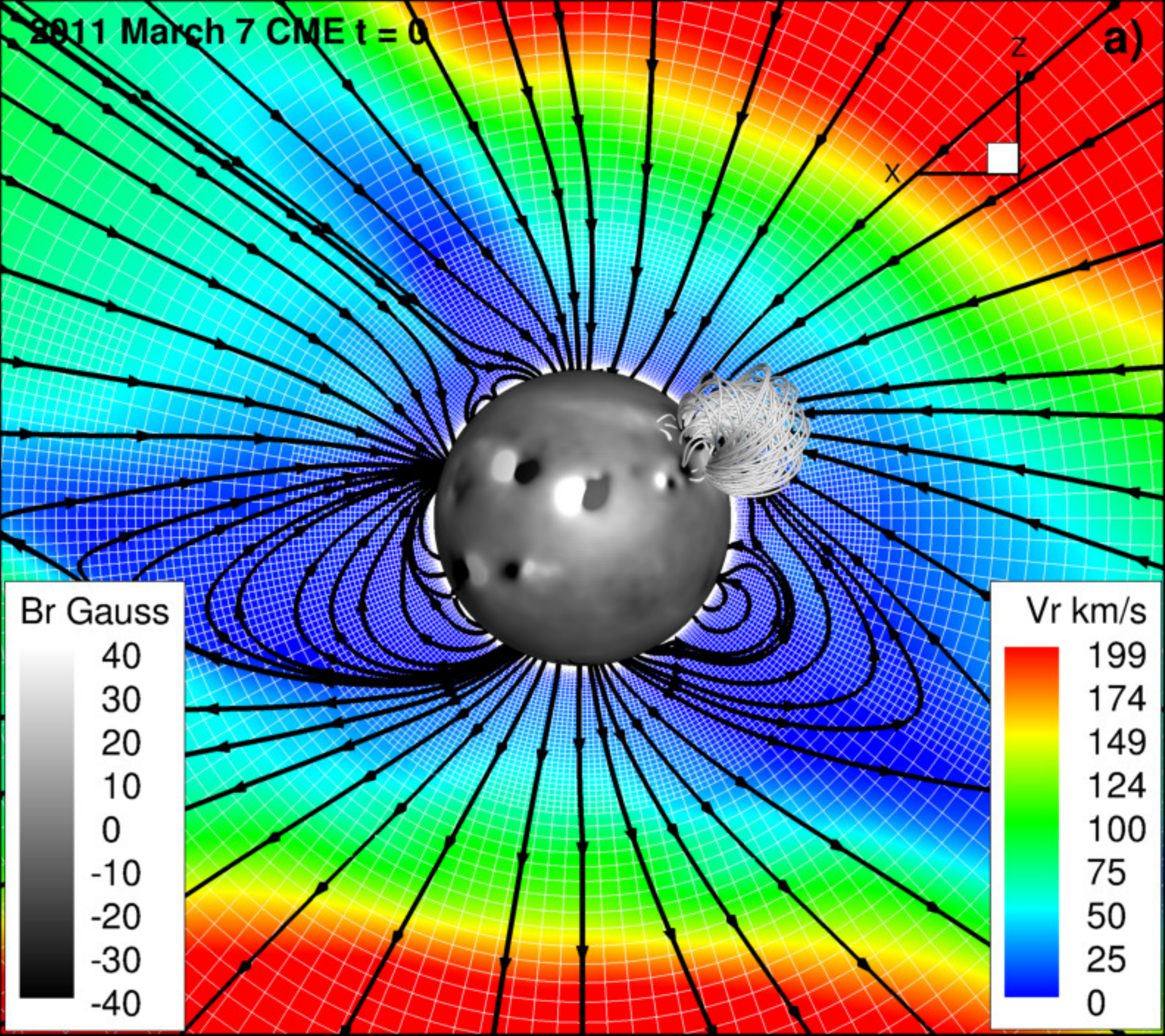} &
\includegraphics[scale=0.28]{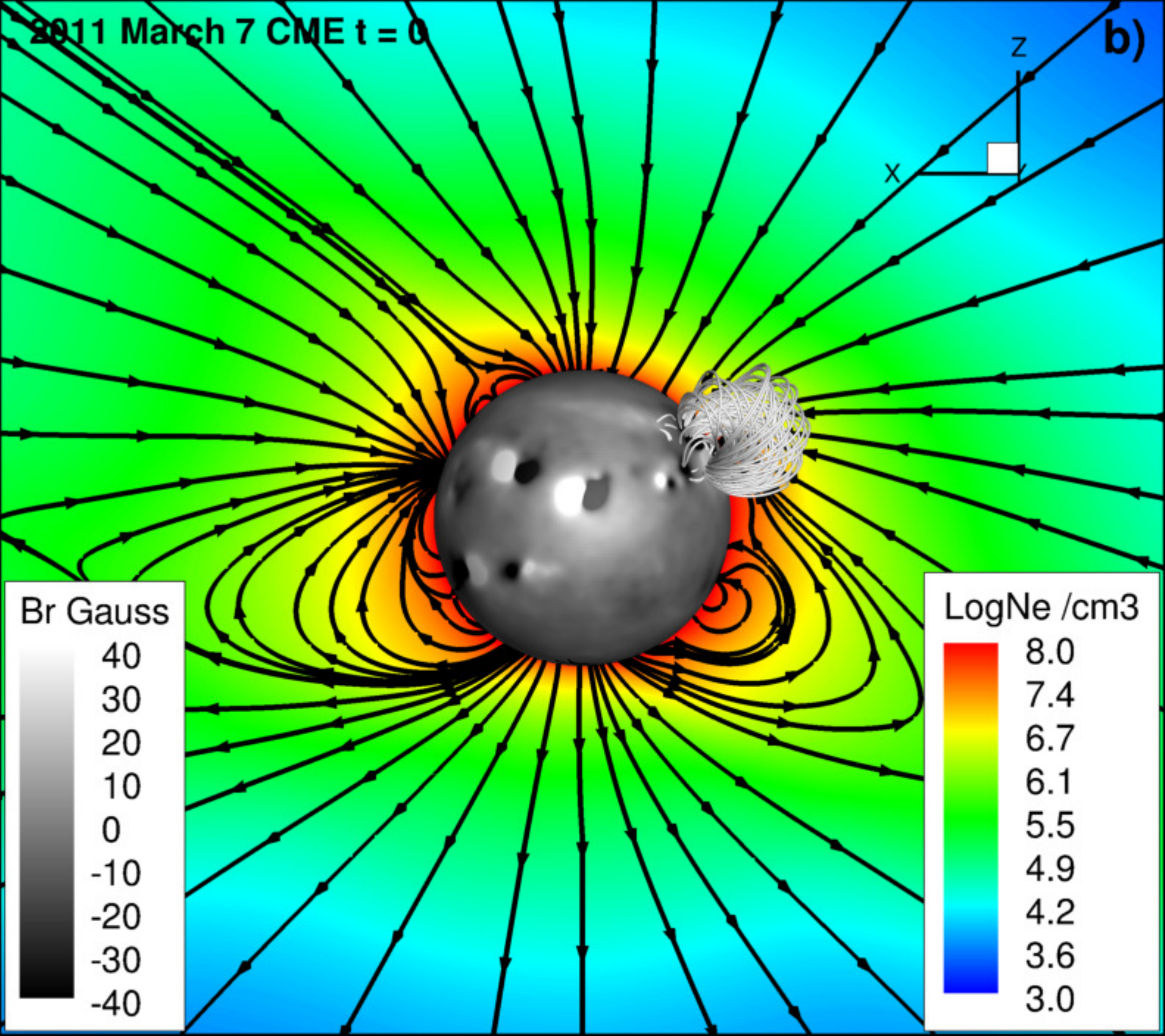} \\
\includegraphics[scale=0.28]{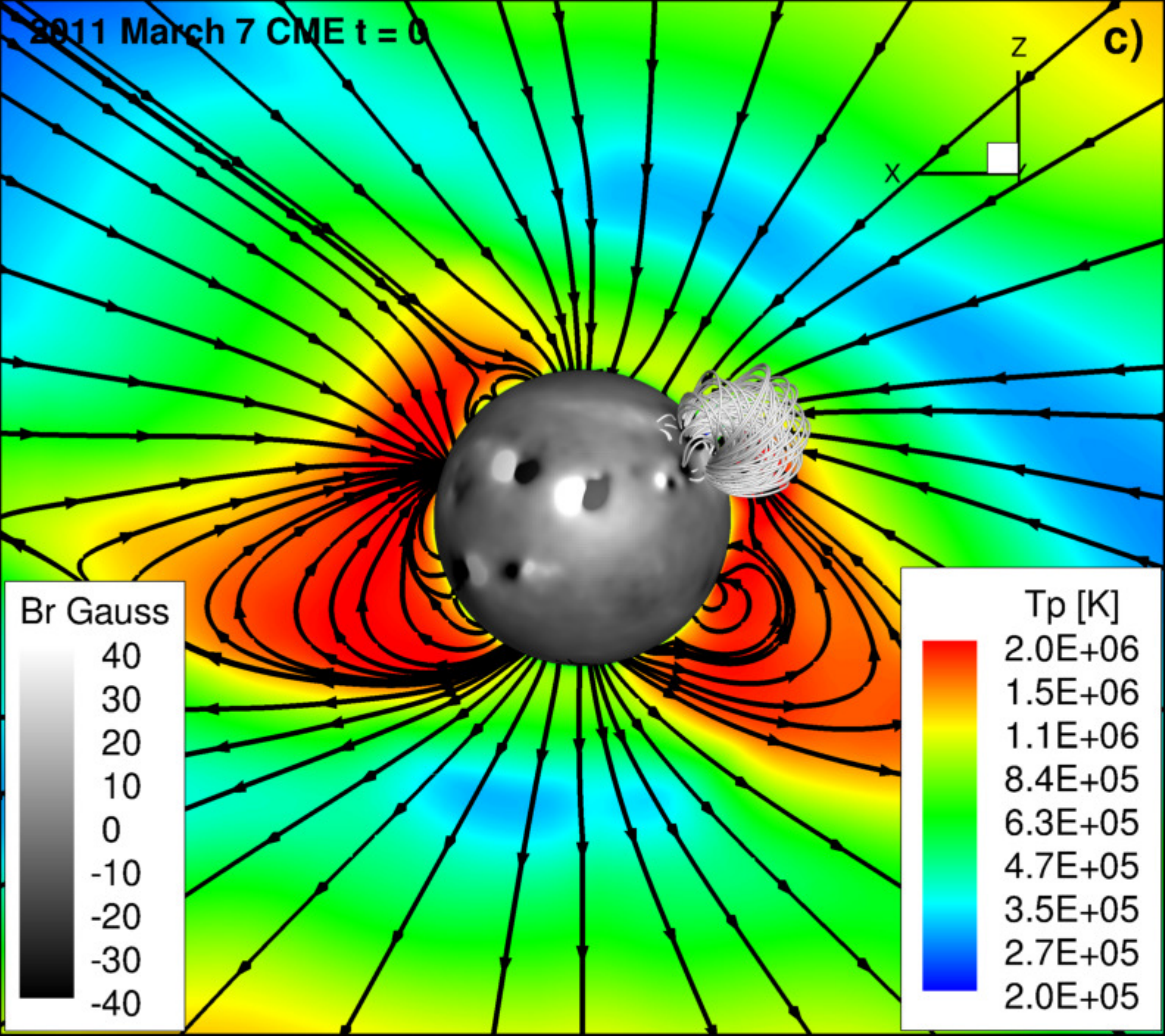} &
\includegraphics[scale=0.28]{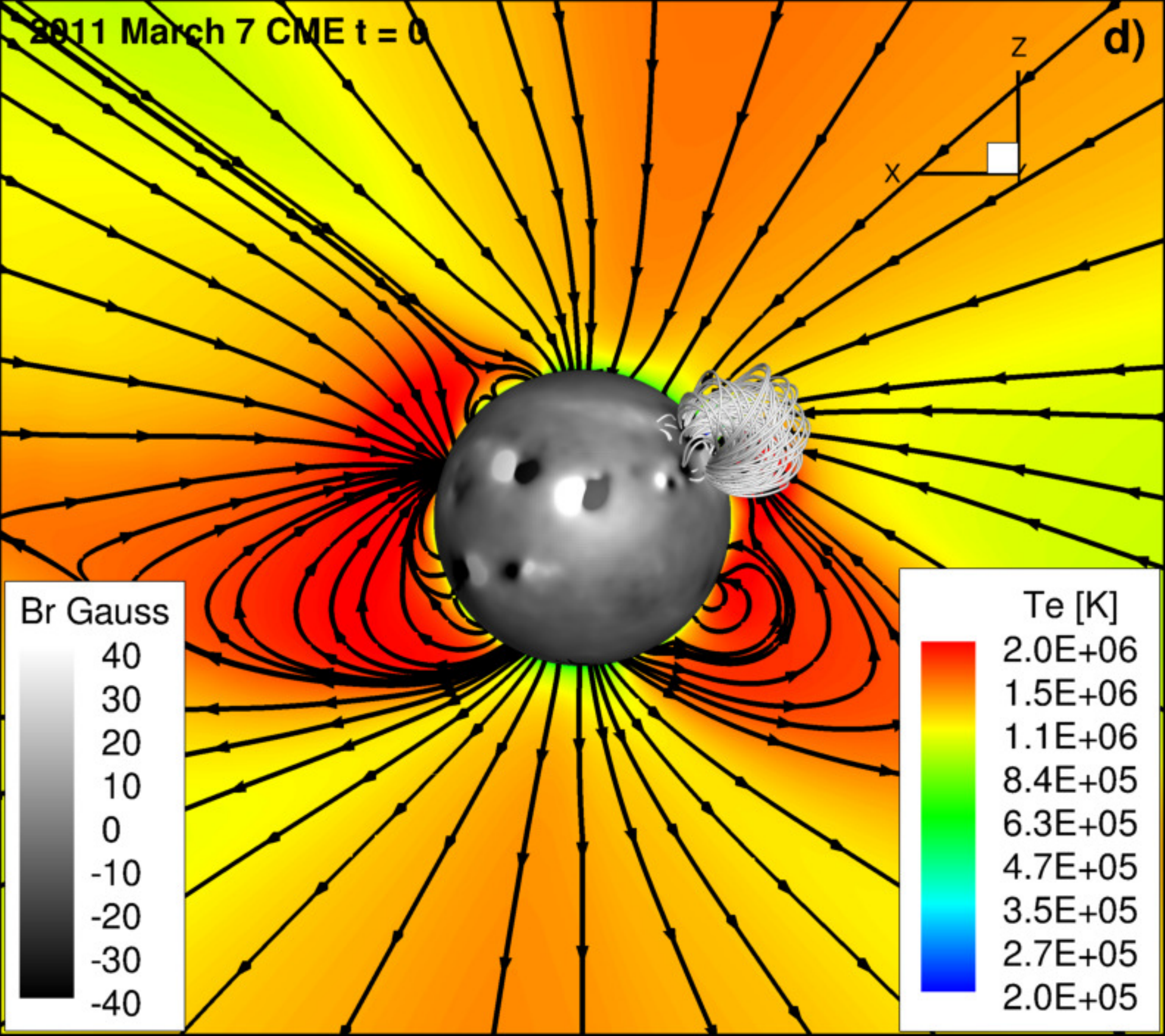} 
\end{array}$
\end{center}
\caption{\label{fig:t0}Meridional slice of the SC showing the (a) radial velocity, (b) plasma density, (c) proton temperature, and (d) electron temperature at t = 0 after GL flux rope implement. The radial magnetic field is shown at r = 1.03 R$_{\odot}$ with gray scale. The white boxes in the velocity map show the grid information for the steady state simulation. The black lines show the projected magnetic field lines on the meridional slice.}
\end{figure}

In Figure \ref{fig:t0}, we overlap the background solar wind solution on a 2D meridional slice with the GL flux rope shown as a bundle of field lines drawn in 3D. The grid information is also shown before the refinement for the CME-driven shock (Figure \ref{fig:t0}a). The flux rope eruption is very close to the north-polar coronal hole and the open-close field boundary. The coronal hole region can be easily identified from the proton temperature figure (Figure \ref{fig:t0}c), in which the temperature of the coronal hole is lower than that of the closed field region.

\subsection{CME Thermodynamic Evolution}
\begin{figure}[h]
\begin{center}$
\begin{array}{cc}
\includegraphics[scale=0.28]{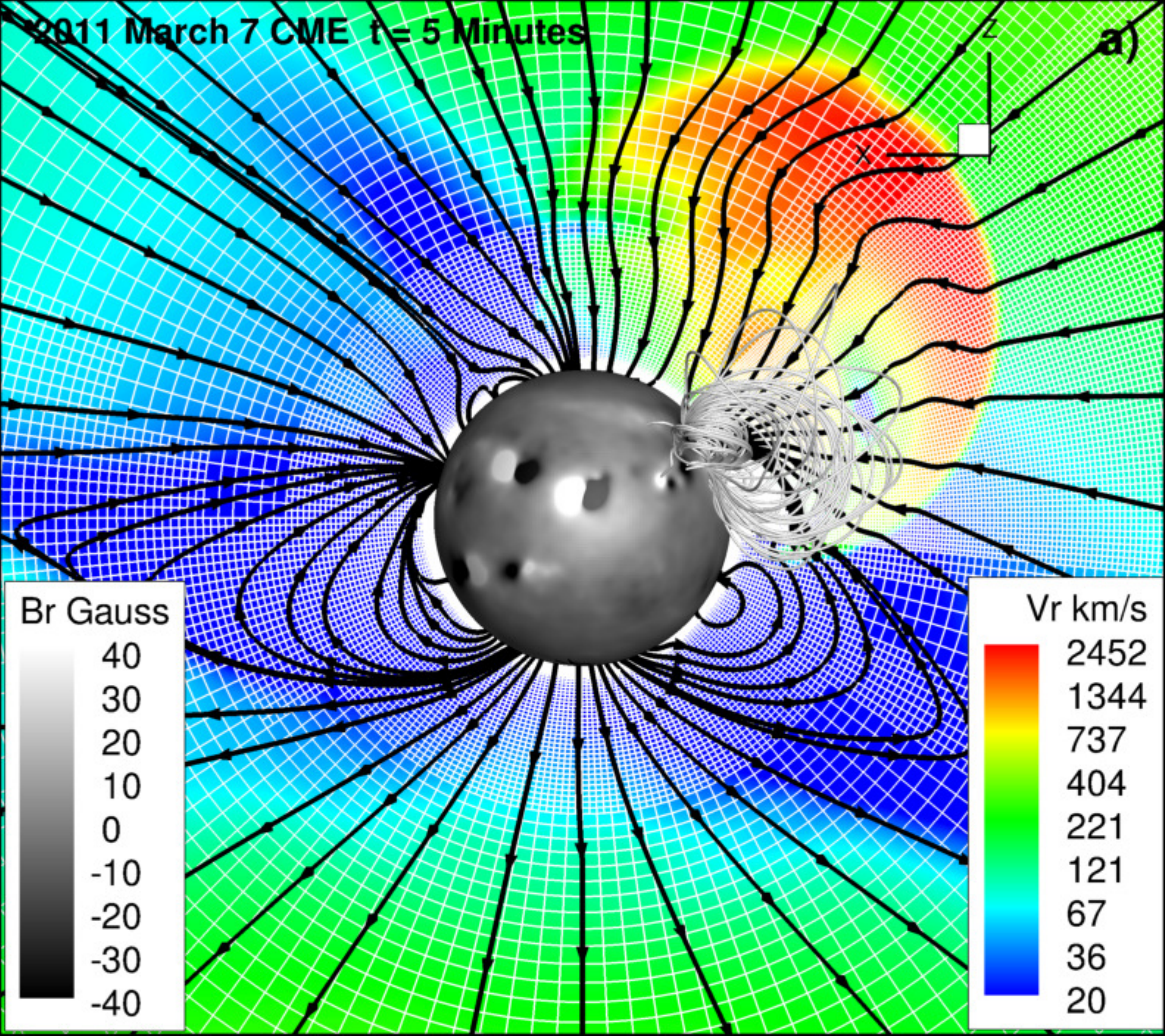} &
\includegraphics[scale=0.28]{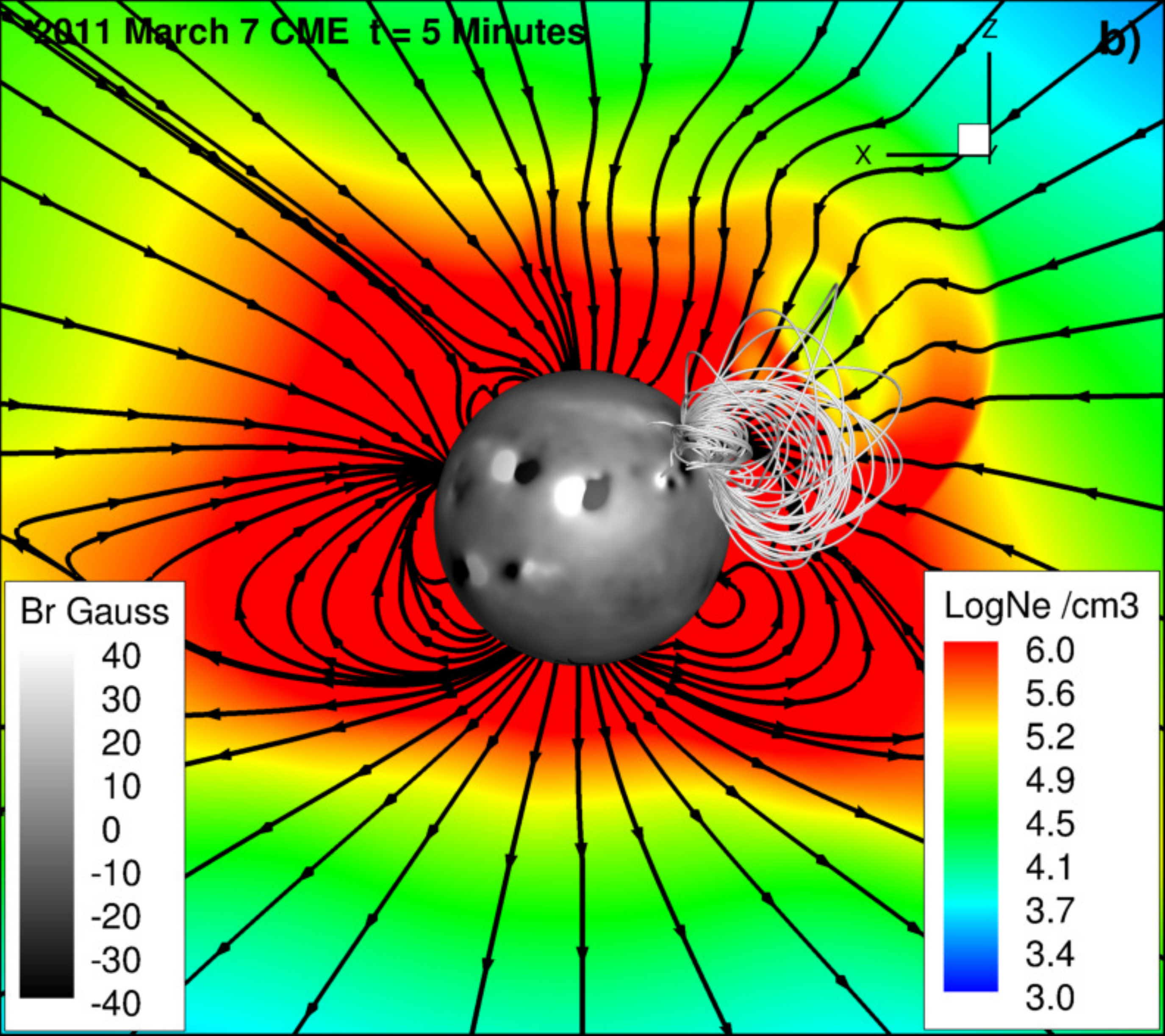} \\
\includegraphics[scale=0.28]{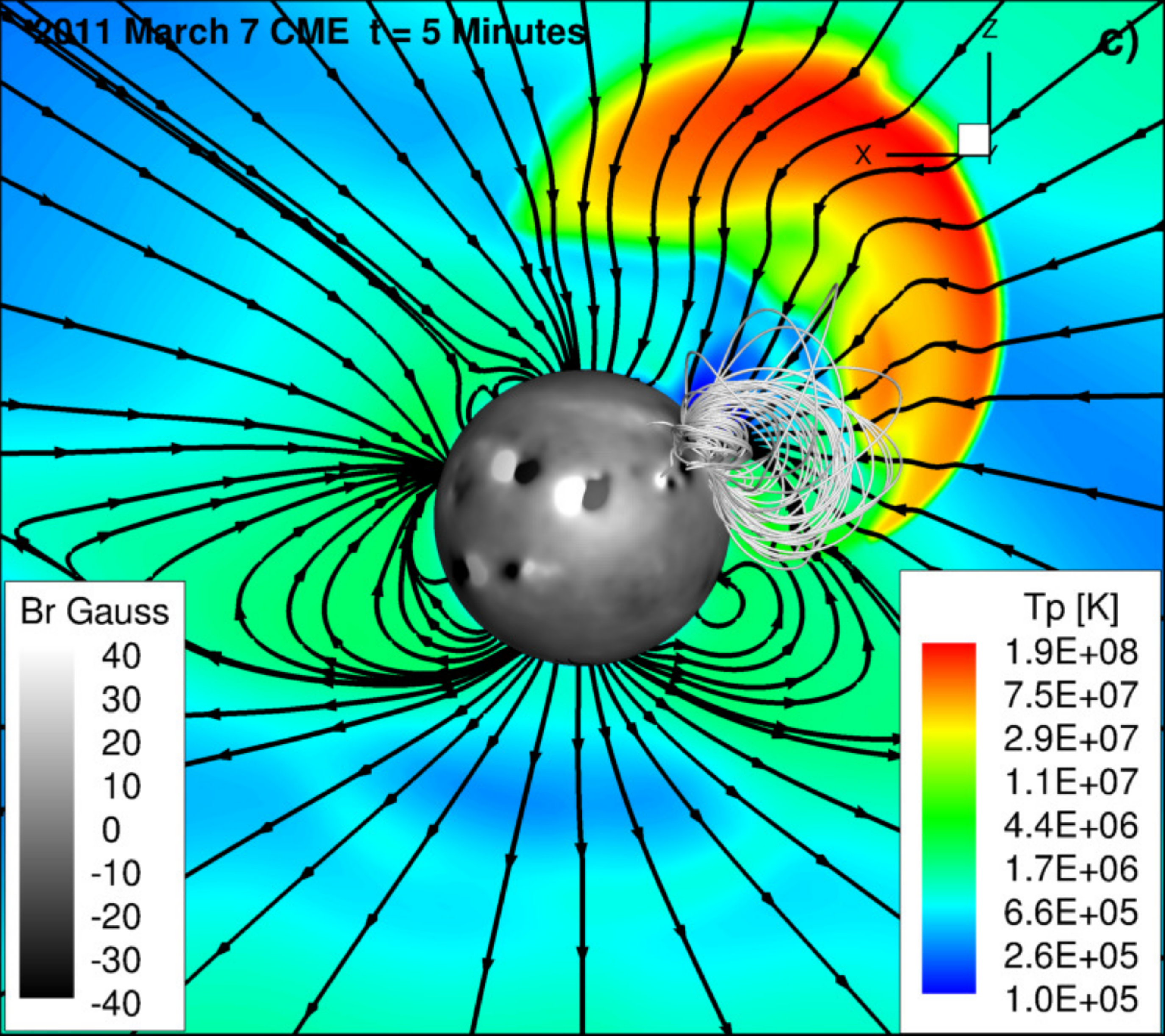} &
\includegraphics[scale=0.28]{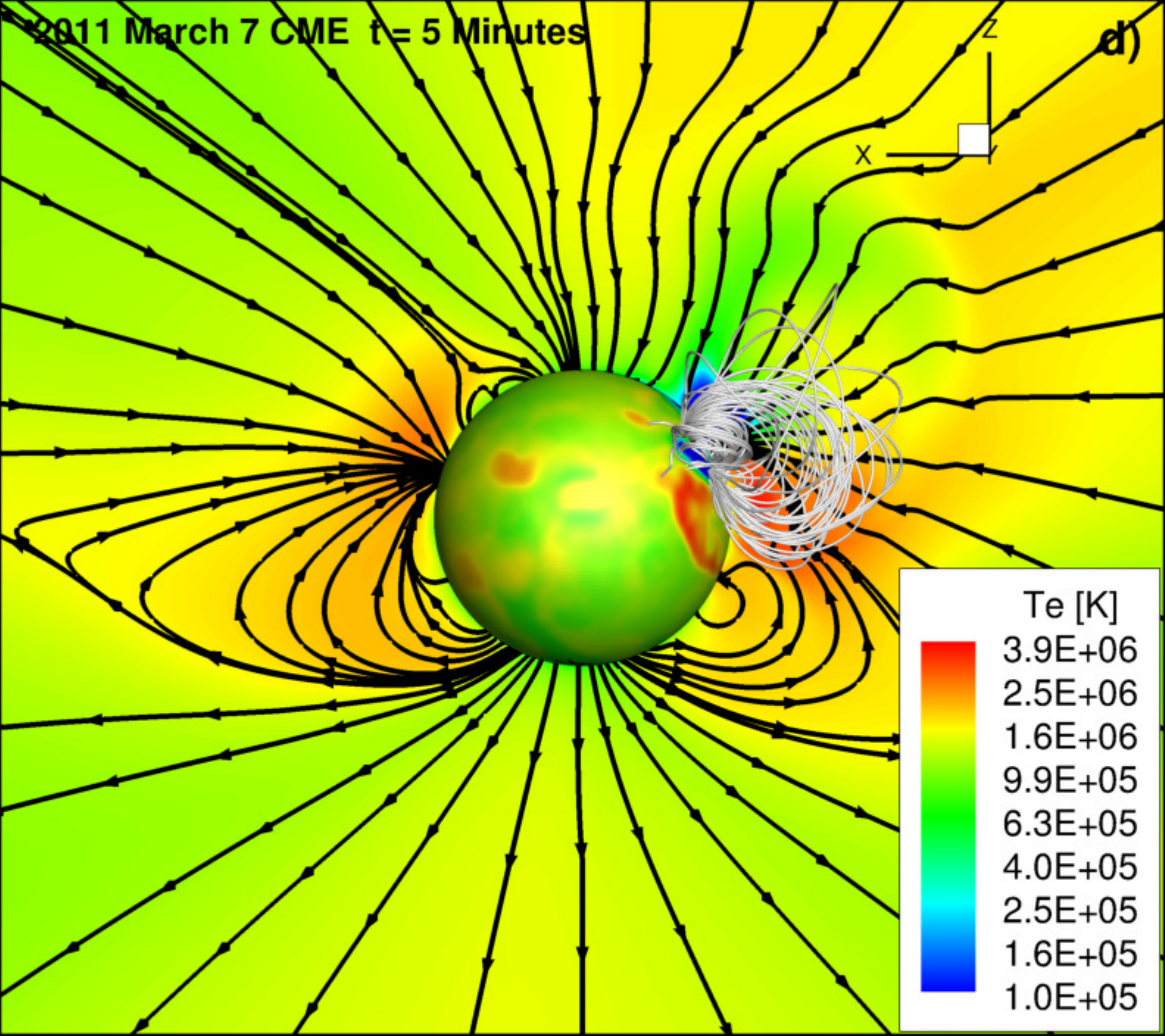} 
\end{array}$
\end{center}
\caption{\label{fig:t5}Meridional slice of the SC showing the (a) radial velocity, (b) plasma density, (c) proton temperature, and (d) electron temperature at t = 5 minutes after GL flux rope implement. The radial magnetic field ((a)-(c))/electron temperature ((d)) is shown at r = 1.03 R$_{\odot}$. The white boxes in the velocity map show the grid information used in the CME simulation. The black lines show the projected magnetic field lines on the meridional slice.}
\end{figure}

In Figure \ref{fig:t5}, we show the CME-driven shock at t = 5 minutes. We can see that the radial velocity of the CME reaches $\sim$2500 km s$^{-1}$, which far exceeds the proton thermal speed of $\sim$100 km s$^{-1}$ in the corona. Therefore, the protons are shock-heated to a temperature of 200 MK after 5 minutes. Due to the close distance to the polar coronal hole, part of the CME-driven shock propagates into the fast wind and obtains a higher velocity and proton temperature. The refined grid information for the CME-driven shock is shown in the radial velocity figure (Figure \ref{fig:t5}a).

\begin{figure}[h]
\begin{center}$
\begin{array}{c}
\includegraphics[scale=0.65]{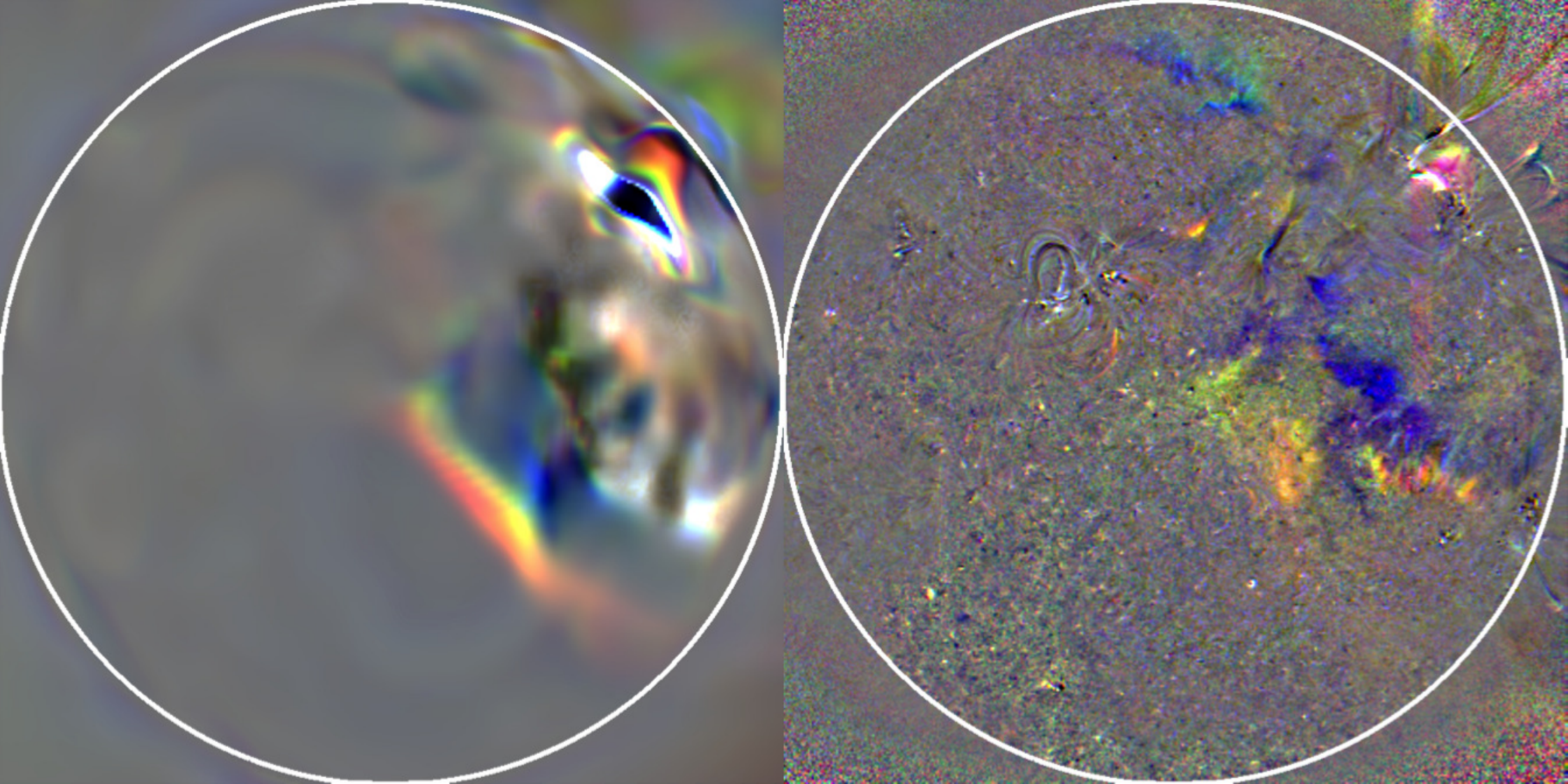}
\end{array}$
\end{center}
\caption{\label{fig:euvwaves}The EUV waves in the simulation (left) and in the SDO/AIA observation. Both the simulation and observation images are produced by tri-ratio running difference method. The tri-color channels are AIA 211 \AA (red), AIA 193 \AA (green), and AIA 171 \AA (blue). The ratio in each channel is identically scaled to 1$\pm$0.2 for both observation and simulation.}
\end{figure}

One of the most intriguing phenomena associated with CMEs is EUV waves, which was first discovered by \citet{moses97} and \citet{thompson98, thompson99} using the data from SOHO/EIT \citep{del95}. The EUV waves are bright fronts that propagate over the solar disk during CME and flare events. There were extensive studies of EUV waves in the past (See reviews by \citealt{chen05, pat12, liu14}). In Figure \ref{fig:euvwaves}, the EUV waves in our simulation and in the observation are shown. Both the simulated and observed images are produced by tri-ratio running difference method. The tricolor channels are AIA 211 \AA (red), AIA 193 \AA (green), and AIA 171 \AA (blue). For both the observation and simulation, the ratio in each channel is identically scaled to 1$\pm$0.2. The white circles show the limb of the Sun. It is clear that our model reproduces many features of the EUV waves in this event: first, the position of the EUV wave front matches the observation. Especially, we notice that part of the wave front is missing in both the simulation and the observation (east of the CME source region AR 11164), which is due to an active region (AR 11167 in AIA 211 \AA~observation of Figure \ref{fig:euv}). In Figure 5d, we also show that this EUV wave front is associated with electron temperature elevation due to the compression by the fast-mode wave. Our simulation result is consistent with previous MHD modeling results (e.g., \citealt{wu01, cohen09, downs11, downs12}) in that the bright EUV waves are driven by the expanding CME and also have a fast-mode wave nature.

\subsection{CME Propagation: Mass \& Velocity}
\begin{figure}[h]
\begin{center}$
\begin{array}{cc}
\includegraphics[scale=0.25]{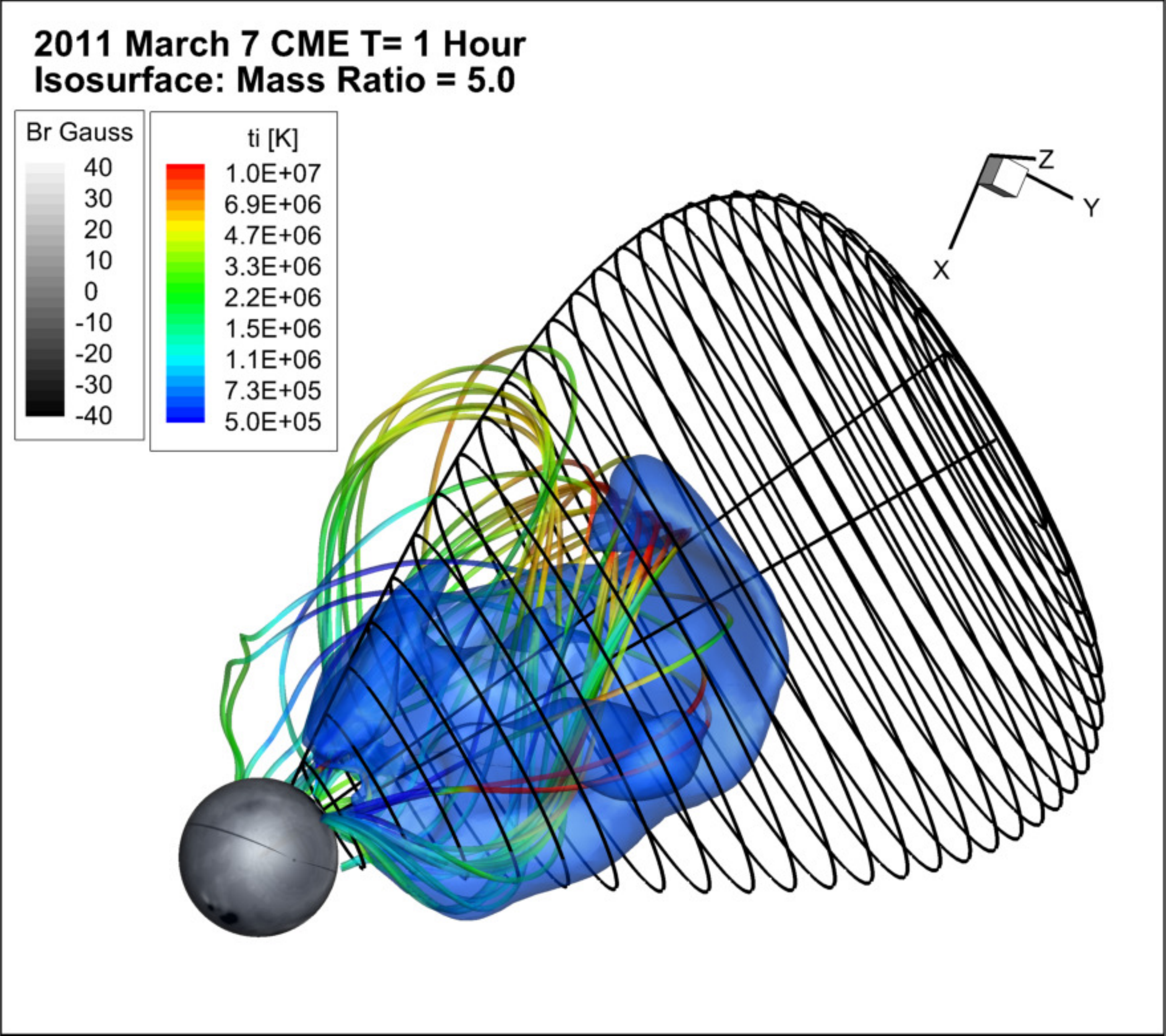} &
\includegraphics[scale=0.25]{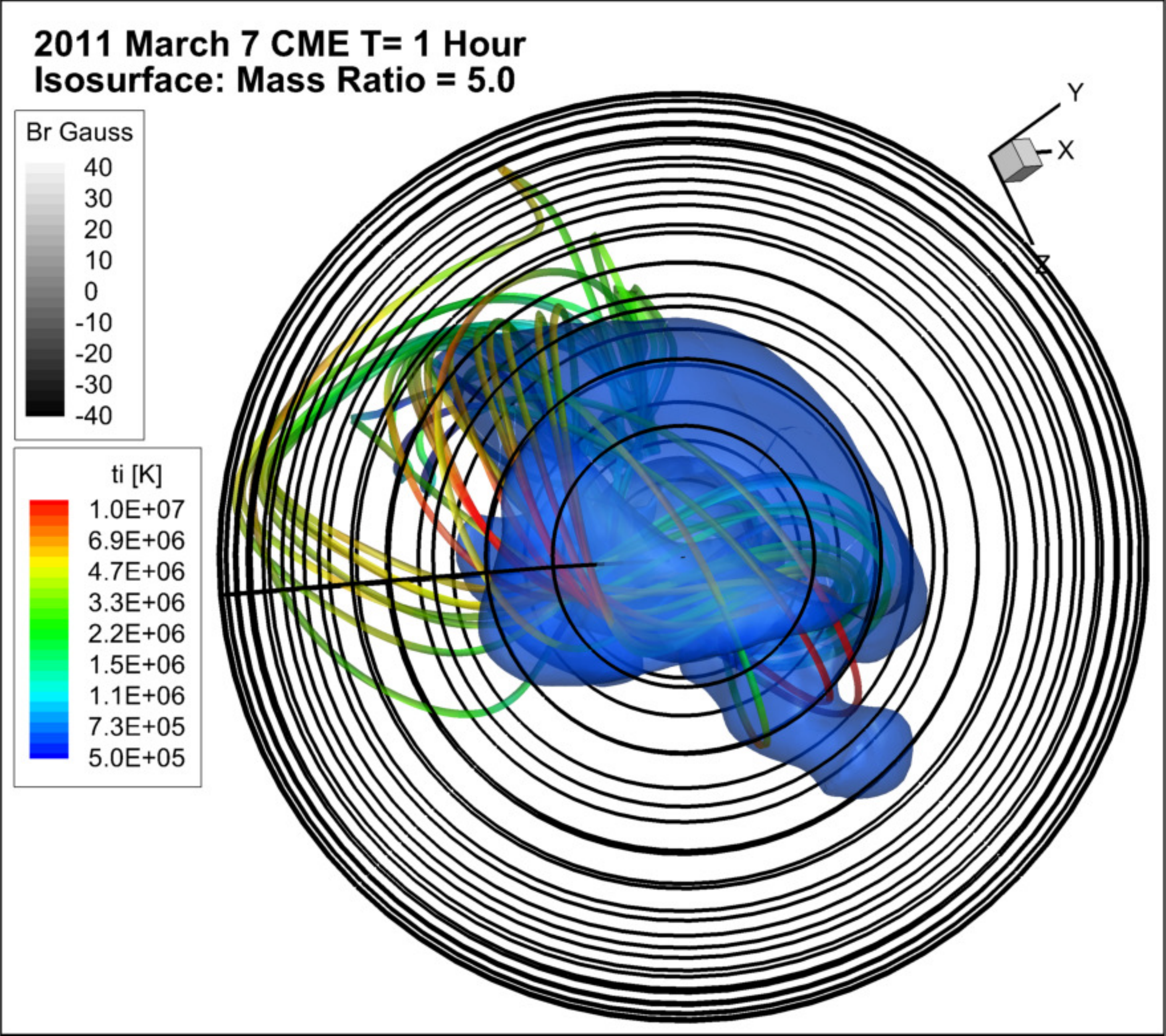}
\end{array}$
\end{center}
\caption{\label{fig:cone}The comparison between the simulated CME and the 3D CME reconstruction of the event from two different viewing angles. The blue isosurface represents the density ratio of 5. The color scale on the selected field lines shows the proton temperature.}
\end{figure}

The direction of CME propagation can be affected by the interaction between the CME and the background solar corona/solar wind structures. In order to validate our model's propagation direction, we compare the simulated CME with the CME model reconstructed from STEREO COR2 observations \citep{koning09, koning11, millward13}. In Figure \ref{fig:cone}, the dense CME material in the model is represented by the density ratio iso-surface of 5.0. The black lines show the model reconstruction of the CME based on a deformed lemniscate. Two viewpoints are shown so we can see that the model CME propagates in the same direction as the model reconstruction. Also, we show several selected field lines in the model. The color scale on the field lines shows the proton temperature. Due to the shock heating, the top of the field lines have the highest temperature $\sim$10 MK.  

\begin{figure}[h]
\begin{center}$
\begin{array}{ccc}
\includegraphics[scale=0.19]{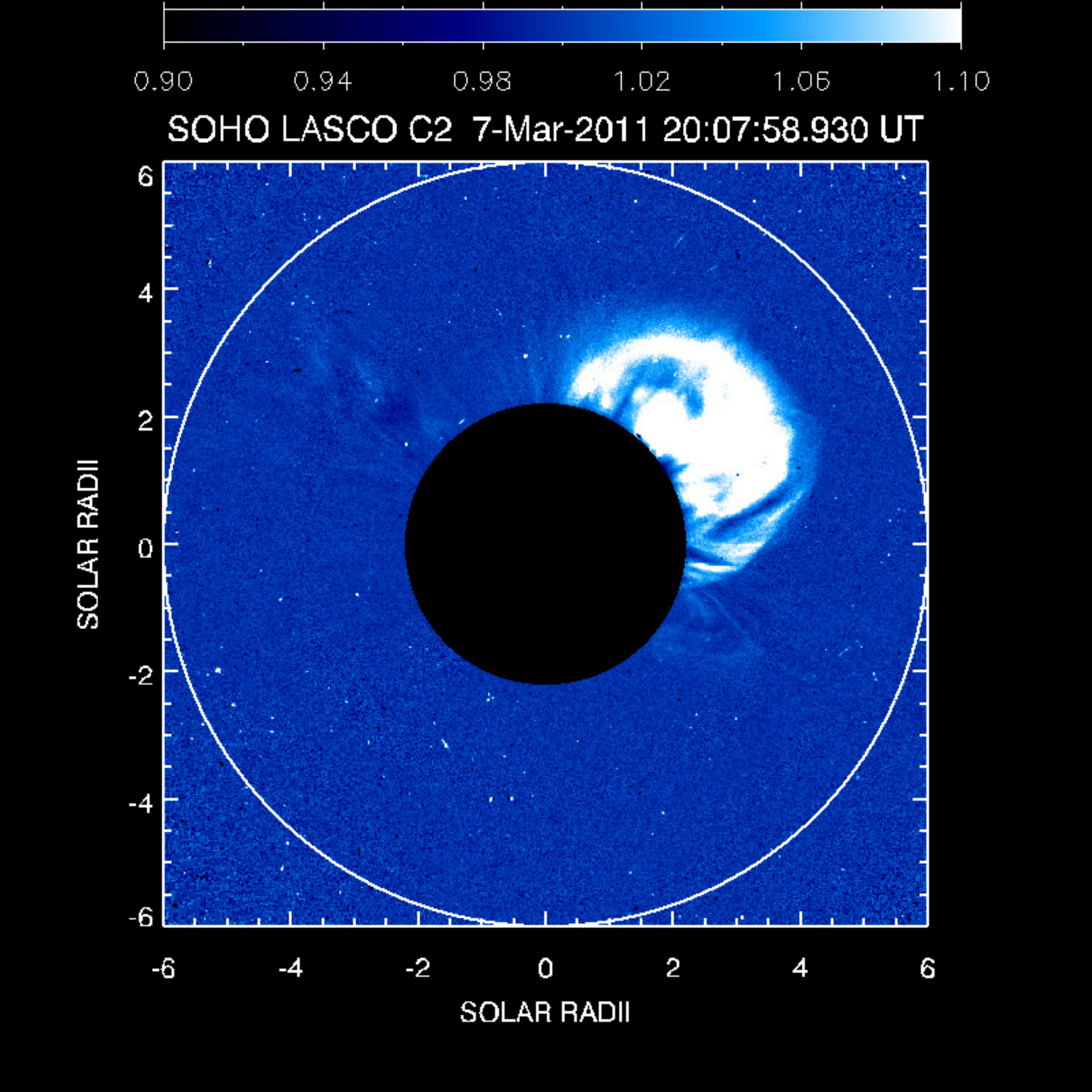} &
\includegraphics[scale=0.19]{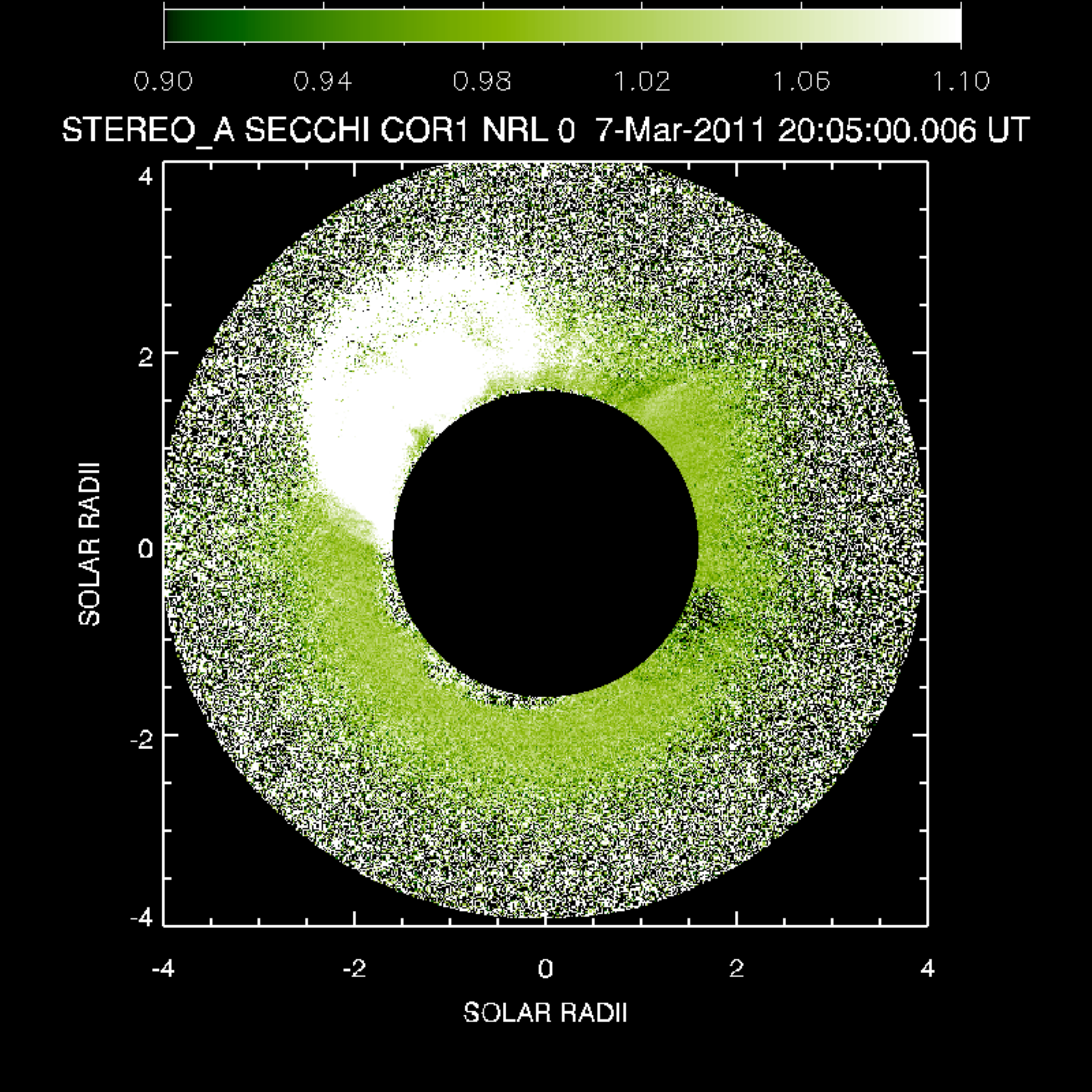} &
\includegraphics[scale=0.19]{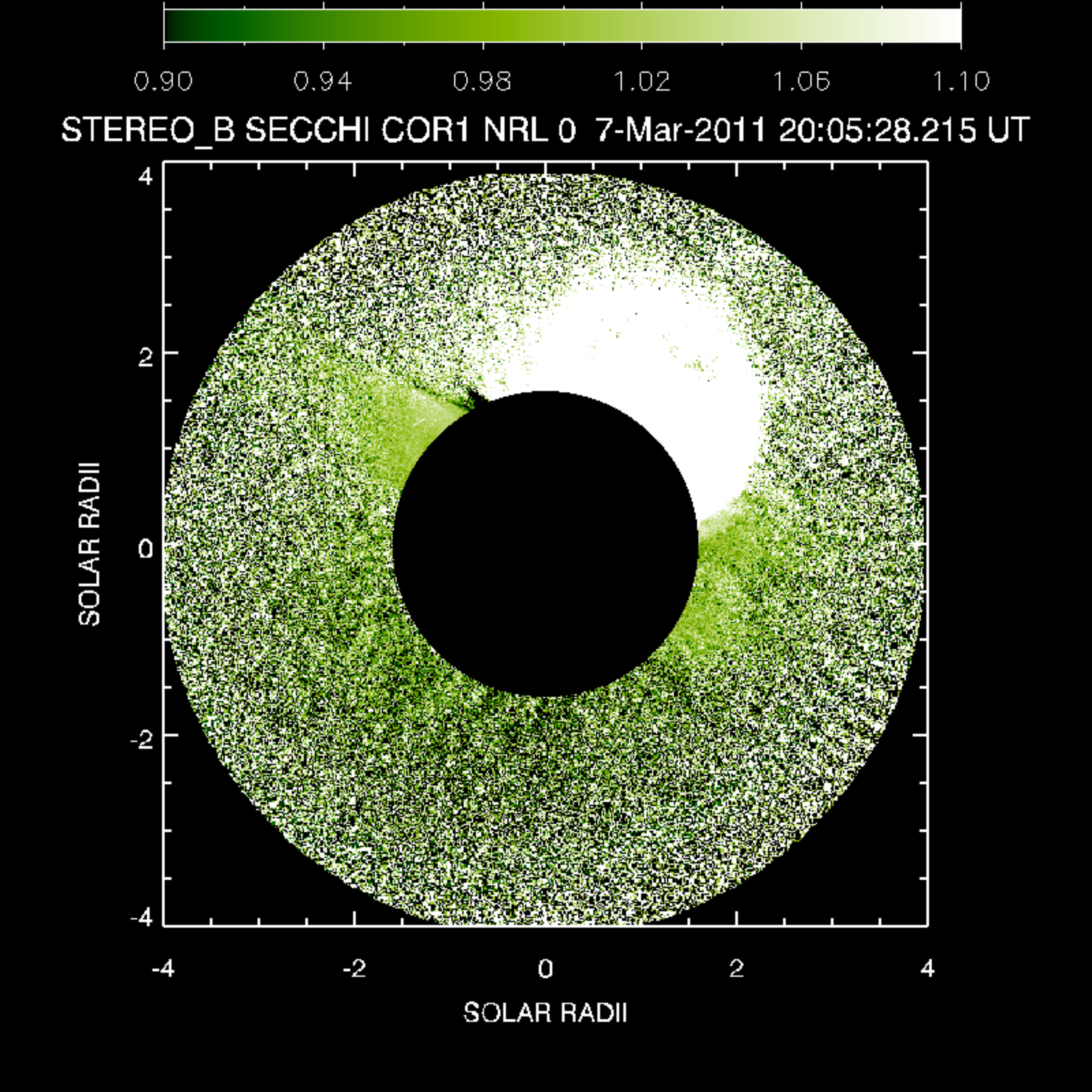}\\
\includegraphics[scale=0.19]{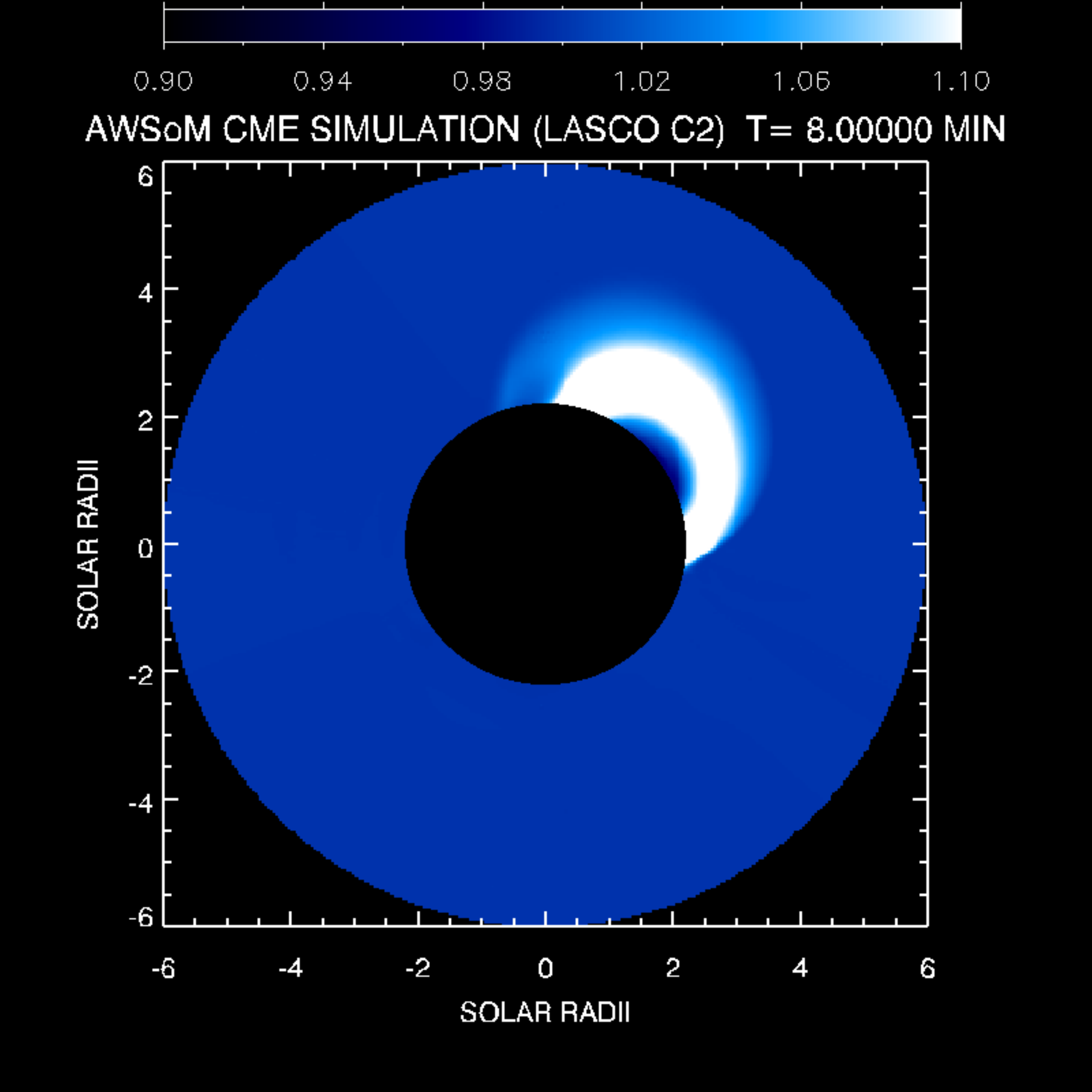} &
\includegraphics[scale=0.19]{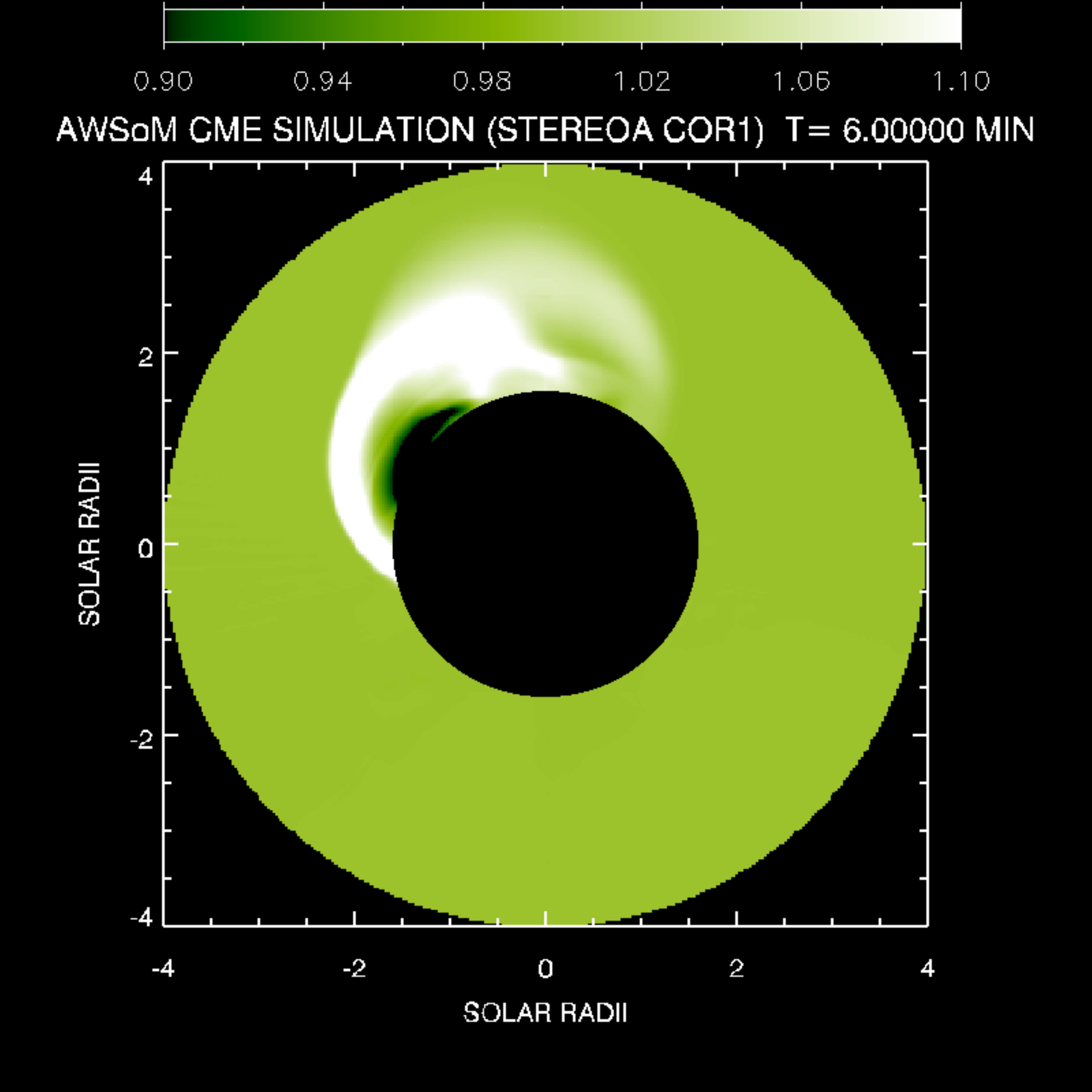} &
\includegraphics[scale=0.19]{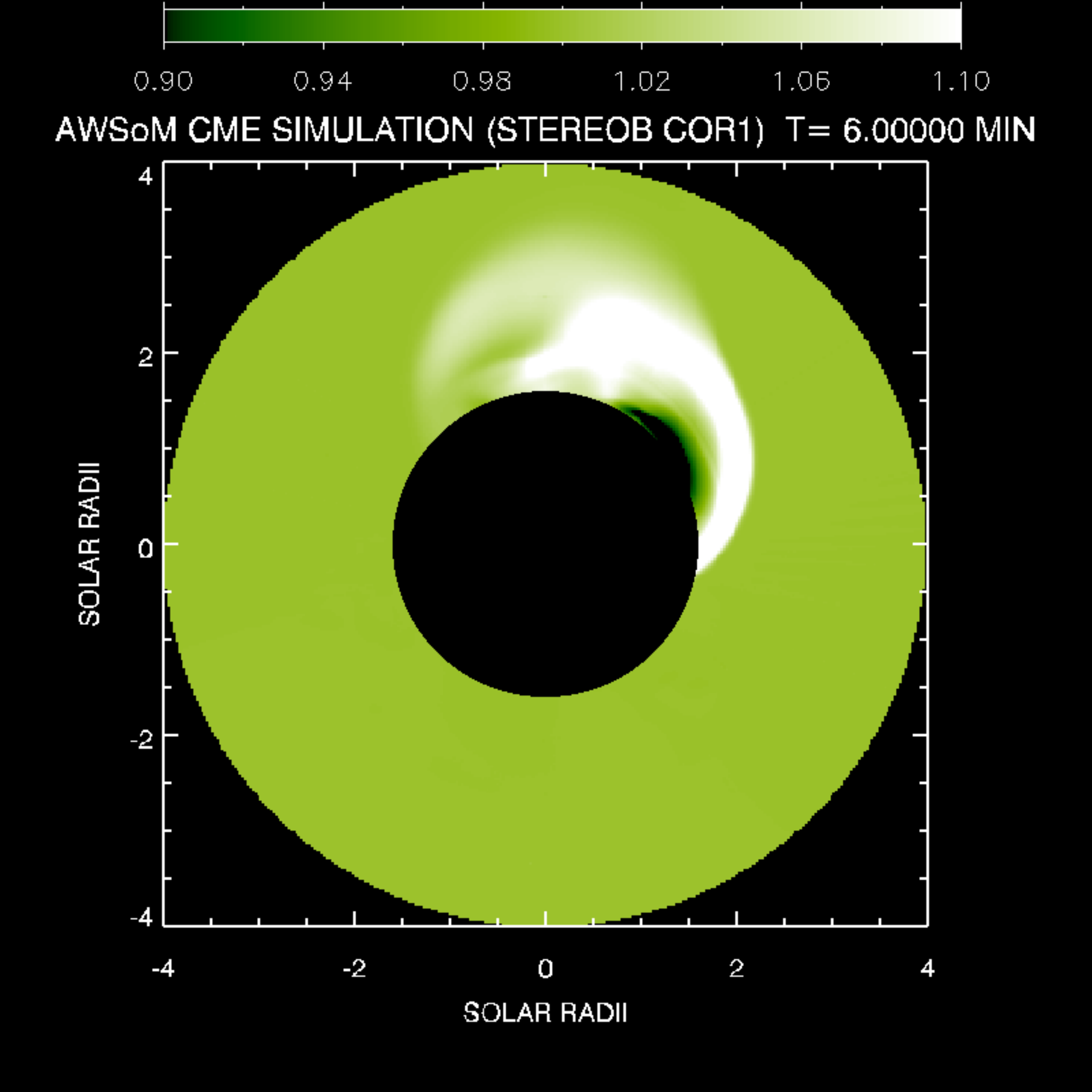}
\end{array}$
\end{center}
\caption{\label{fig:wl1}Comparison between the LASCO C2, COR1A, and COR1B white light images with the model synthesized images for the 2011 March 7 CME event. The color scale shows the white light total brightness divided by that of the pre-event background solar wind.}
\end{figure}

\begin{figure}[h]
\begin{center}$
\begin{array}{ccc}
\includegraphics[scale=0.19]{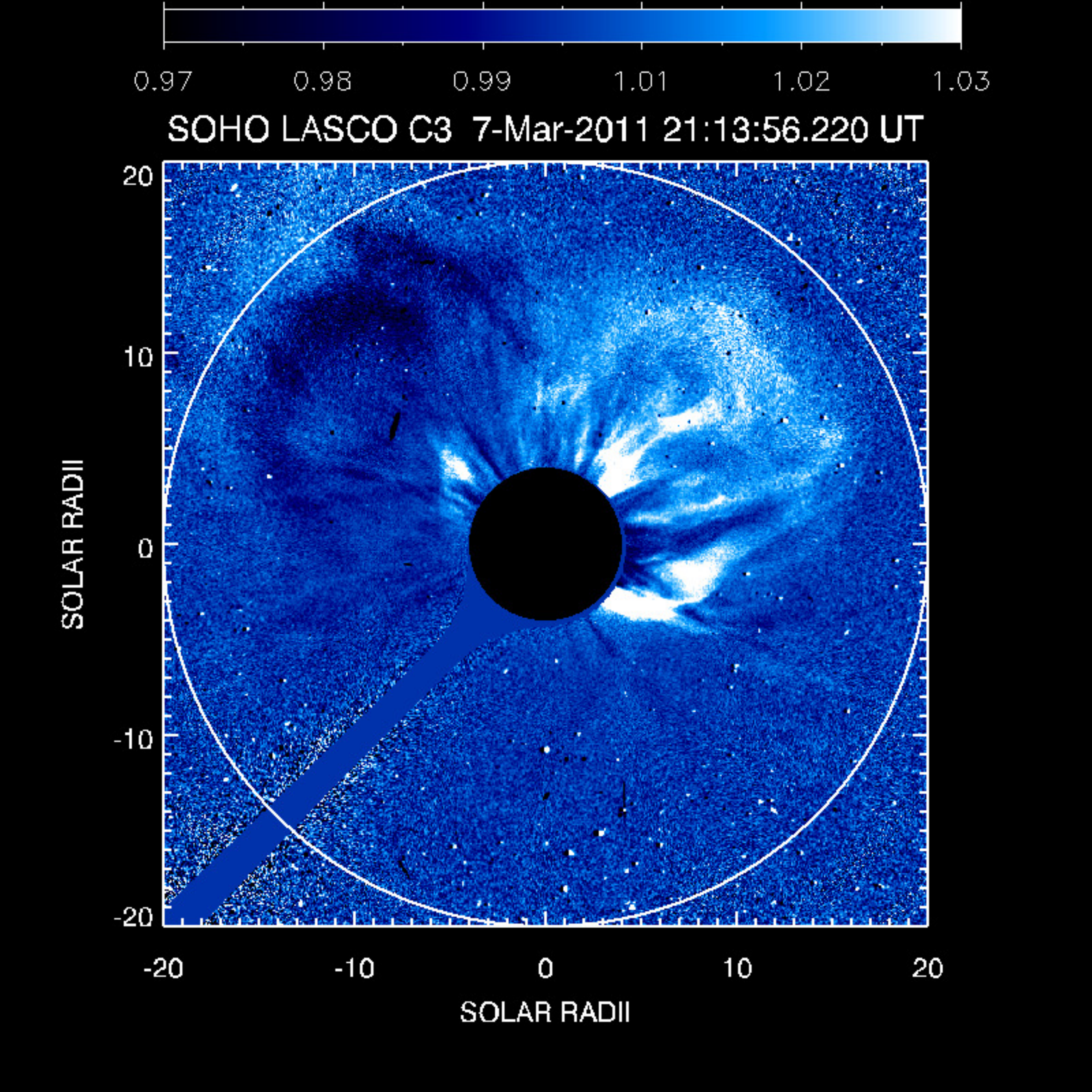} &
\includegraphics[scale=0.19]{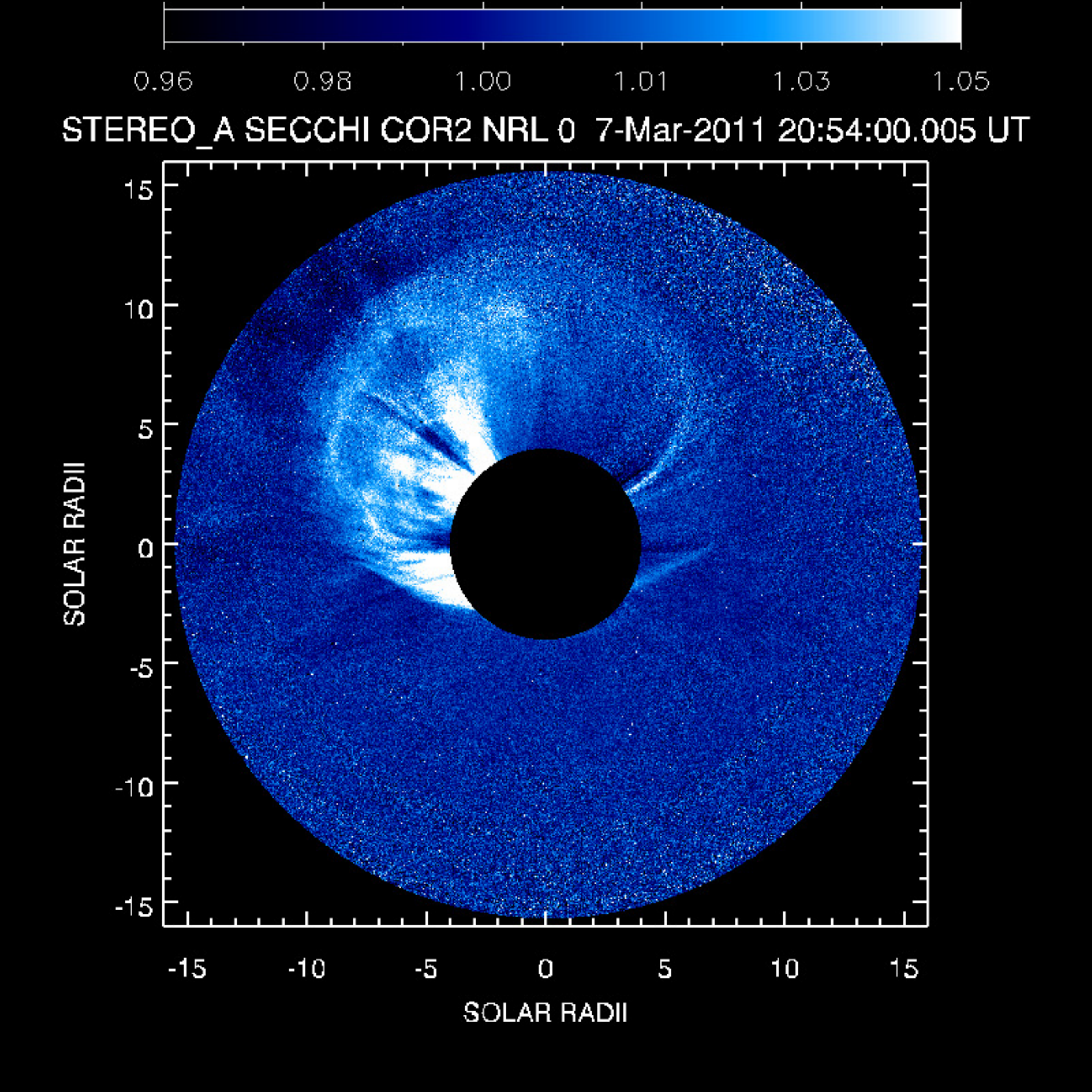} &
\includegraphics[scale=0.19]{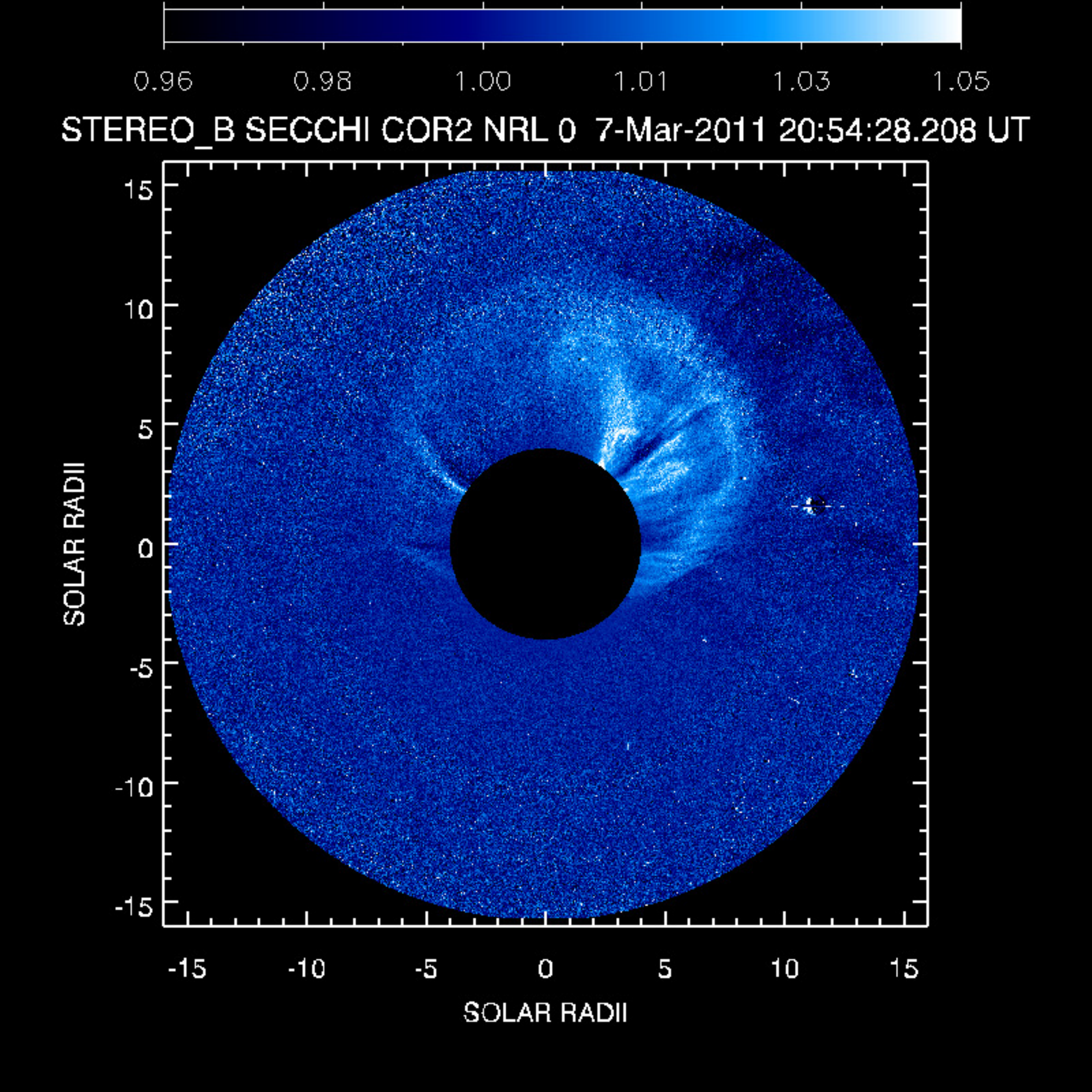}\\
\includegraphics[scale=0.19]{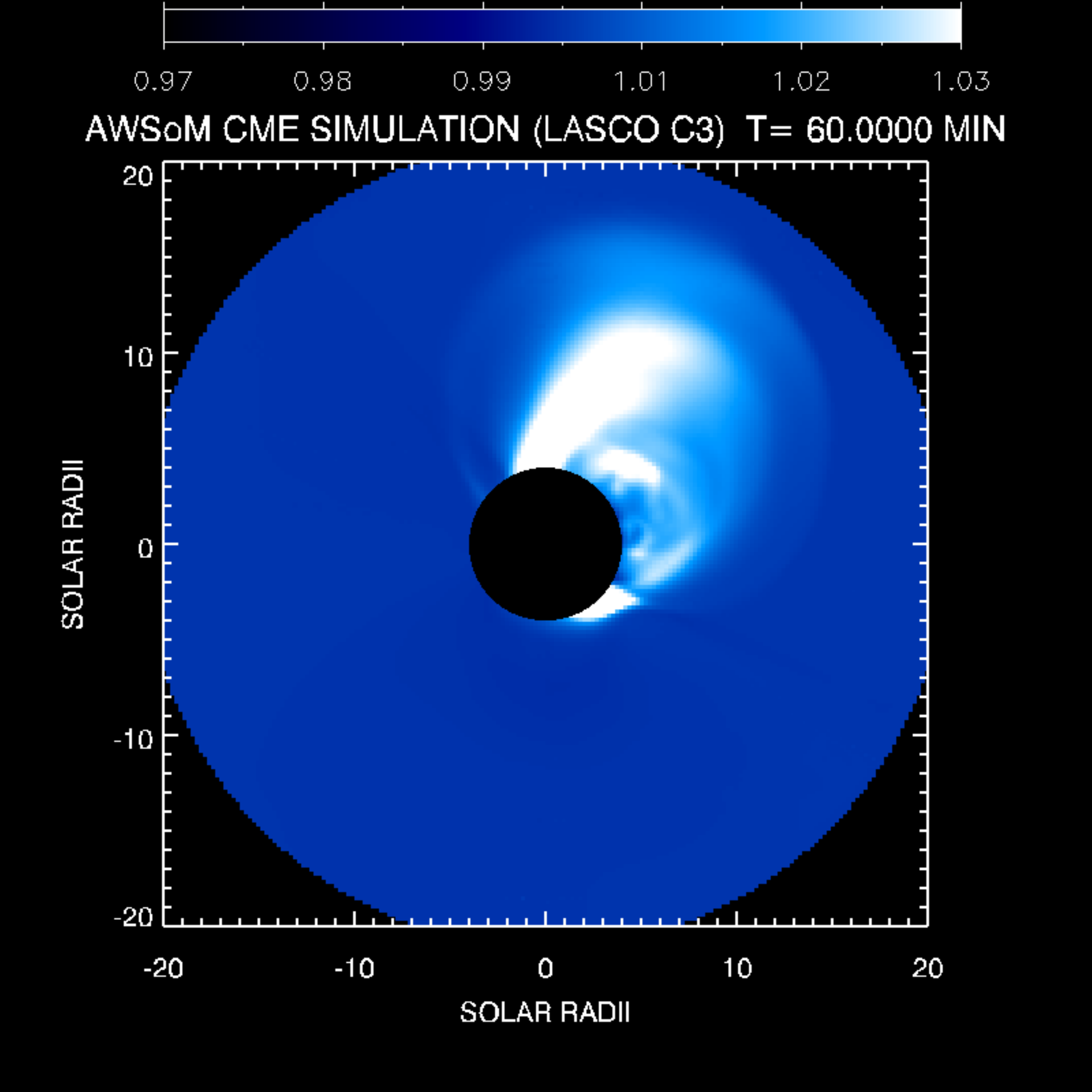} &
\includegraphics[scale=0.19]{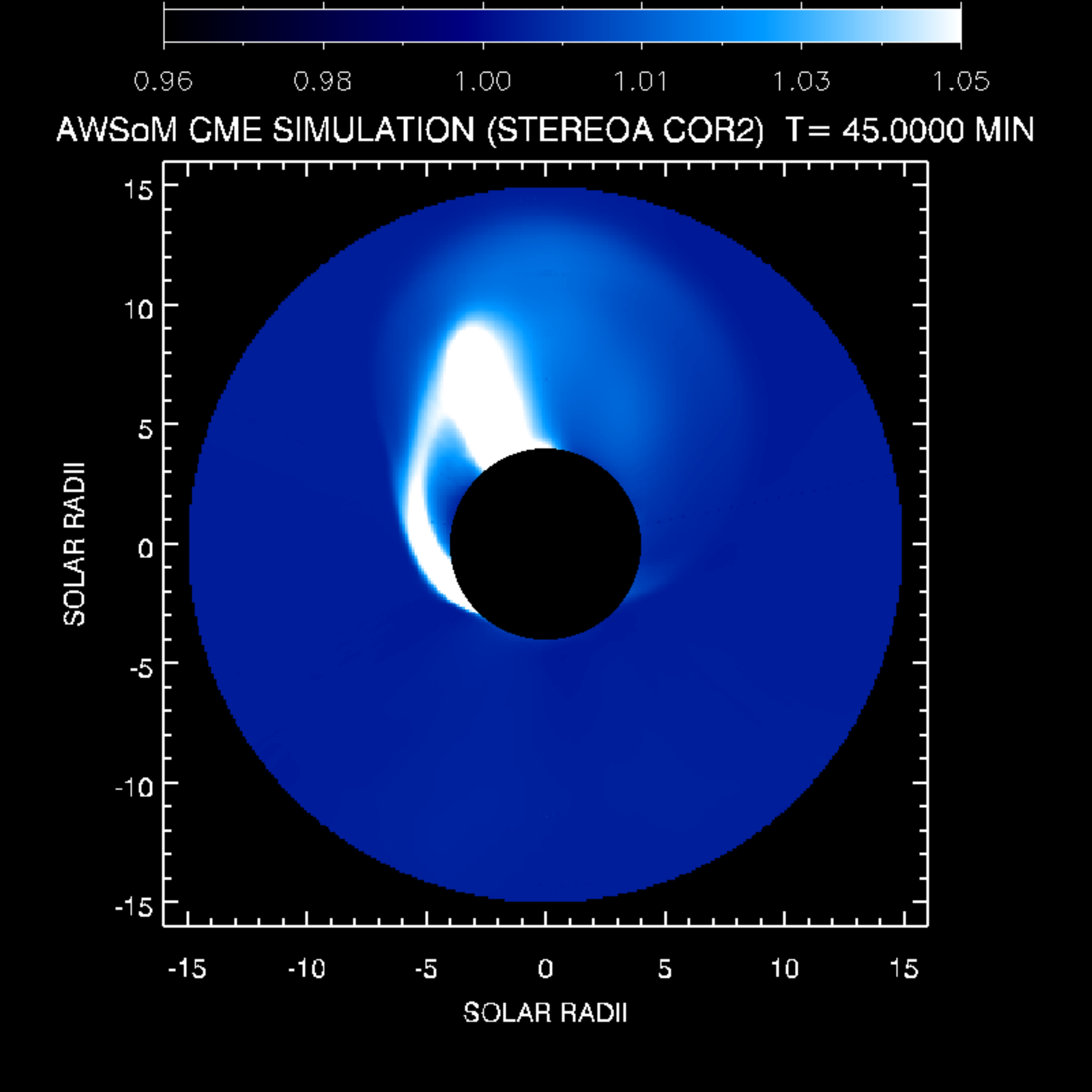} &
\includegraphics[scale=0.19]{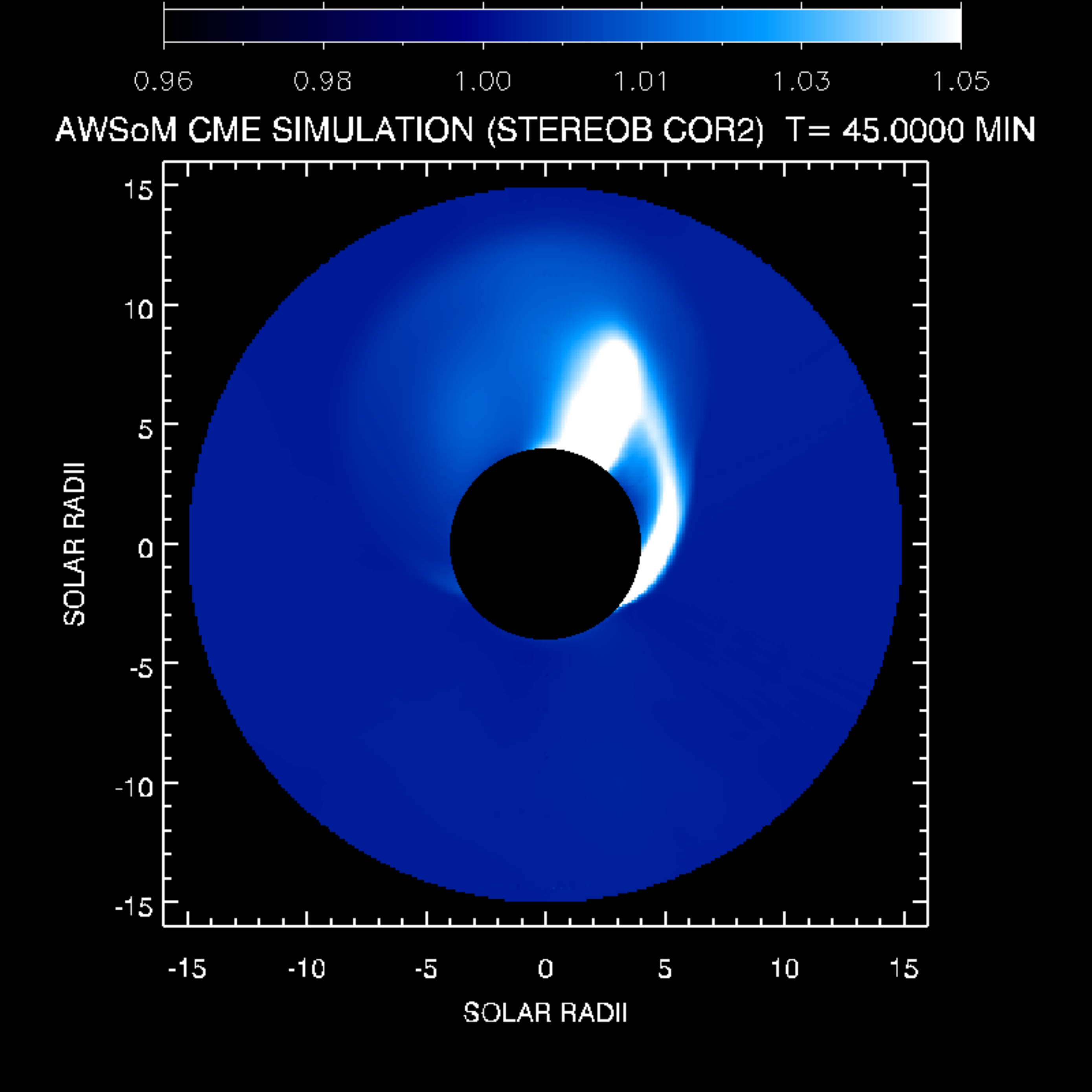}
\end{array}$
\end{center}
\caption{\label{fig:wl2}Comparison between the LASCO C3, COR2A, and COR2B white light images with the model synthesized images for the 2011 March 7 CME event. The color scale shows the white light total brightness divided by that of the pre-event background solar wind.}
\end{figure}

The CME propagation near the Sun and in the heliosphere is mainly observed by white light coronagraphs. For this event, there are six white light observations available from SOHO/LASCO C2/C3, STA COR1/COR2 and STB COR1/COR2. C2 has a field of view (FOV) from 2 R$_{\odot}$ to 6 R$_{\odot}$ and C3 has a FOV from 3 R$_{\odot}$ to 30 R$_{\odot}$. The FOVs of COR1 and COR2 are from 1.5 R$_{\odot}$ to 4 R$_{\odot}$ and 3 R$_{\odot}$ to 15 R$_{\odot}$, respectively. In Figure \ref{fig:wl1} and Figure \ref{fig:wl2}, we show a comparison between the observed white light images and the model synthesized images for the 2011 March 7 event. Both the color scales show the white light total brightness divided by that of the pre-event background solar wind. Note that we take into account the effect of F corona far from the Sun when calculating the synthesized images \citep{chip08}. The higher noise level in the COR1 observation is due to the design on the COR1 coronagraph. With an exposed front lens, it leads to higher instrumental background that needs to be removed to reveal the coronal signal \citep{thompson10}. This decreases the signal-to-noise ratio of the final processed images and results in high level of ``salt-and-pepper" noise in the images. In the observation, we can see clearly that the CME has a typical 3-part structure: the bright core that represents the filament material; the dark cavity that corresponds to the flux rope; the bright front that is due to the mass pile-up in front of the flux rope \citep{llling85}. In the synthesized images, this 3-part structure is also evident. Moreover, both the observation and model show the second faint front that is the outermost part of the increased intensity region. The ``double-front" morphology is consistent with CME-driven shocks \citep{vour03, vour09}, which has been verified with numerical simulations \citep{chip08}. The white light comparison from three points of view confirms that the simulated CME propagates in the correct direction as observed. 

\begin{figure}[h]
\begin{center}$
\begin{array}{c}
\includegraphics[scale=0.45]{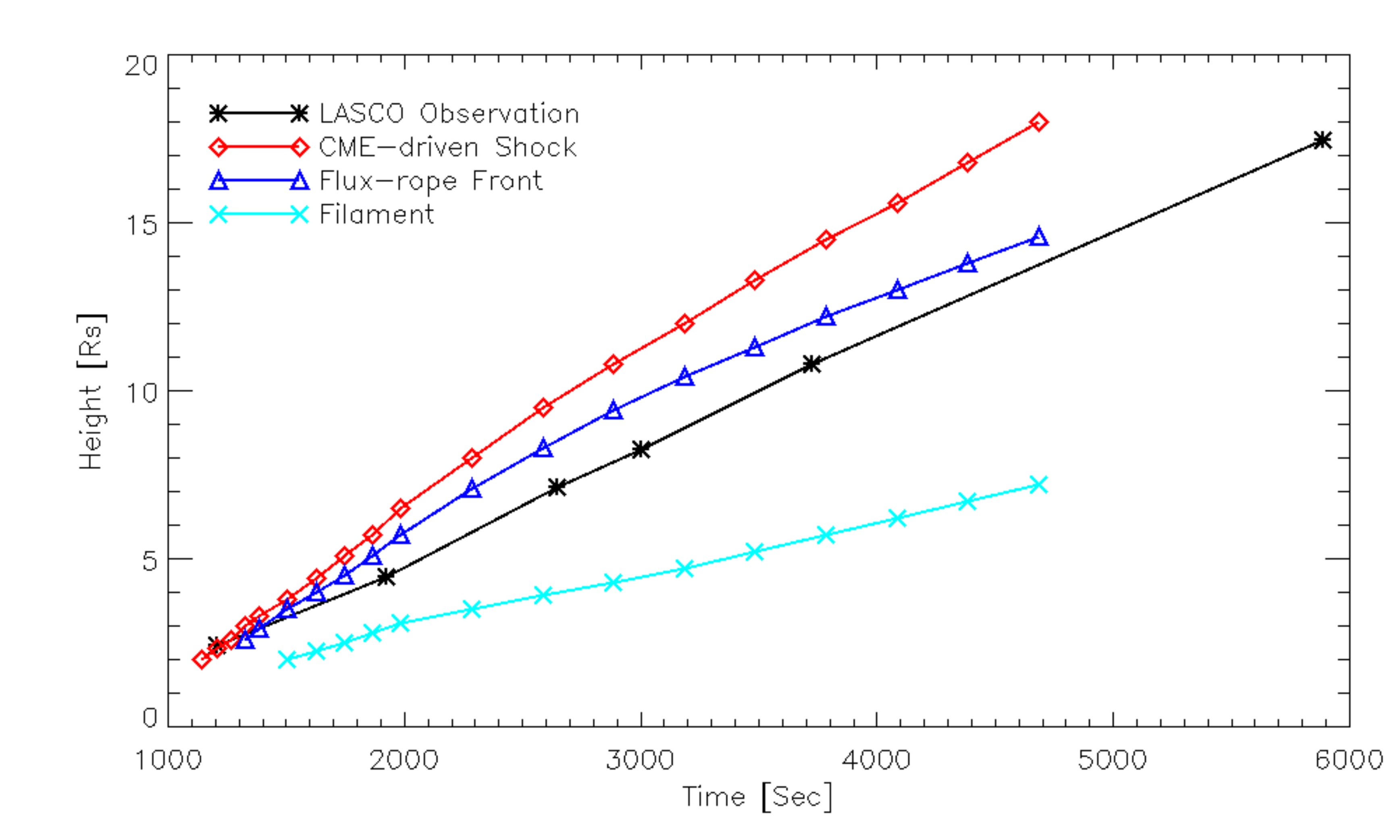}
\end{array}$
\end{center}
\caption{\label{fig:speed}CME speed comparison between the simulation and LASCO observation.}
\end{figure}

The speed of the CME is another important factor for precise space weather forecasts. From synthesized white light images, the height-time (HT) evolution of different structures (CME-driven shock, flux rope front, and filament) is obtained. Due to the complexity of the observation, only the outermost part of white light observation is used to obtain the HT map. The results are shown in Figure \ref{fig:speed}. In the simulation, the faint front related to the CME-driven shock has the largest speed $\sim$2878 km s$^{-1}$. The bright front related to the flux rope pile-up has the second largest speed $\sim$2158 km s$^{-1}$. The filament has the slowest speed $\sim$1089 km s$^{-1}$. The observed CME-driven shock speed (outermost front) is $\sim$2275 km s$^{-1}$, which is close to the speed of the bright front in the simulation, while $\sim$600 km s$^{-1}$ less than the speed of the outermost front in the simulation. All the speeds are derived by linear fitting of the data points in Figure \ref{fig:speed}. Due to the force-imbalance nature of the initial state of the flux rope, all the structures in the simulation experience a deceleration process in the early stage of propagation, which is not obvious in the observation. As a result, the model CME speed needs to be higher at onset to match subsequently velocities far from the Sun. 

\begin{figure}[h]
\begin{center}$
\begin{array}{cc}
\includegraphics[scale=0.25]{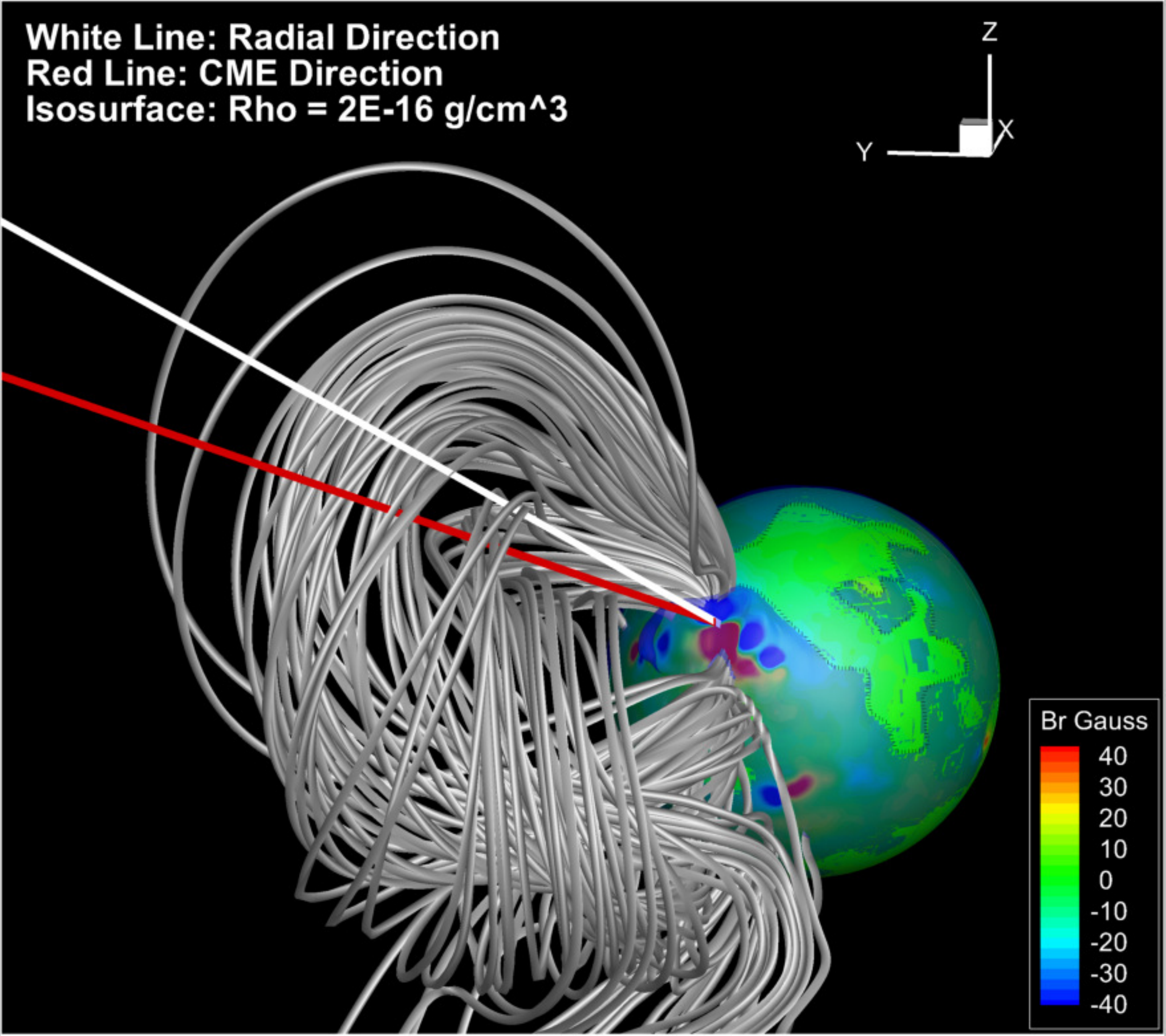} &
\includegraphics[scale=0.25]{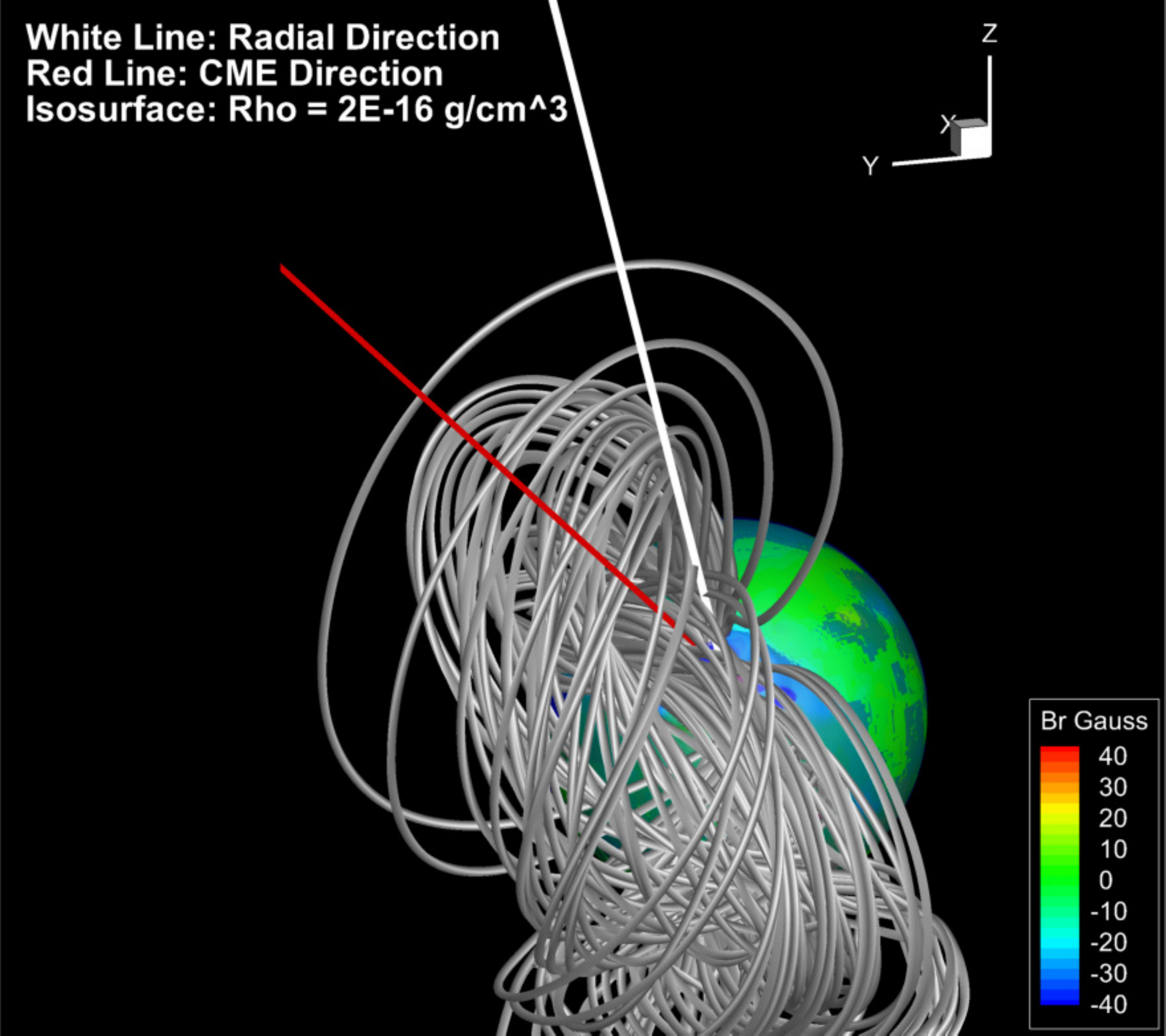}
\end{array}$
\end{center}
\caption{\label{fig:deflection}The CME flux rope deflection by the nearby coronal hole at t = 30 Minutes in the simulation from two different viewing angles. The color scale on the Sun shows the radial magnetic field strength. The isosurface on the Sun shows the density of $2\times10^{-16}$ g cm$^{-3}$. The white line shows the radial direction from the CME source region. The red line shows the direction of flux rope expansion in the simulation.}
\end{figure}

Another observational feature of this CME event is that the CME-driven shock passes STA without the flux rope/magnetic driver structure behind the shock. The interplanetary shocks without magnetic drivers are studied in detail by \cite{gopalswamy09}. They found that 15\% of the driverless shocks occur within 15$^{\circ}$ of the solar central meridian, for which every CME source region is accompanied by a nearby corona hole. Therefore, the authors suggest that the coronal hole may play an important role in deflecting the CME flux rope away from the Sun-Satellite line so that only the shock arrives at the satellite. Our study confirms this point in the 2011 March 7 event simulation. In Figure \ref{fig:deflection}, the CME flux rope deflection in our simulation is shown at t = 30 mintues. We can see that the CME source region is located just to the east of a coronal hole (see also the STA observation in Figure \ref{fig:euv}). In our simulation, the flux rope is deflected by $\sim$8$^{\circ}$ away from the nearby coronal hole 30 minutes after the eruption. The CME deflection phenomenon has also been modeled by previous studies (e.g., \citealt{lugaz11b, kay13}). The global MHD simulations achieved by \citet{lugaz11b} show that the effect of the Lorentz force can deflect the CME $\sim$10$^{\circ}$ after 35 minutes. \citet{kay13} developed a CME deflection model including the effects of magnetic pressure gradient and magnetic tension to predict the observed CME deflection. Due to the complicated solar wind condition in our simulation, it is hard to separate the effects that may play a role in defecting the CME (e.g., varying background solar wind, magnetic reconnection).

\subsection{CME Evolution in the Heliosphere}
\begin{figure}[h]
\begin{center}$
\begin{array}{cc}
\includegraphics[scale=0.28]{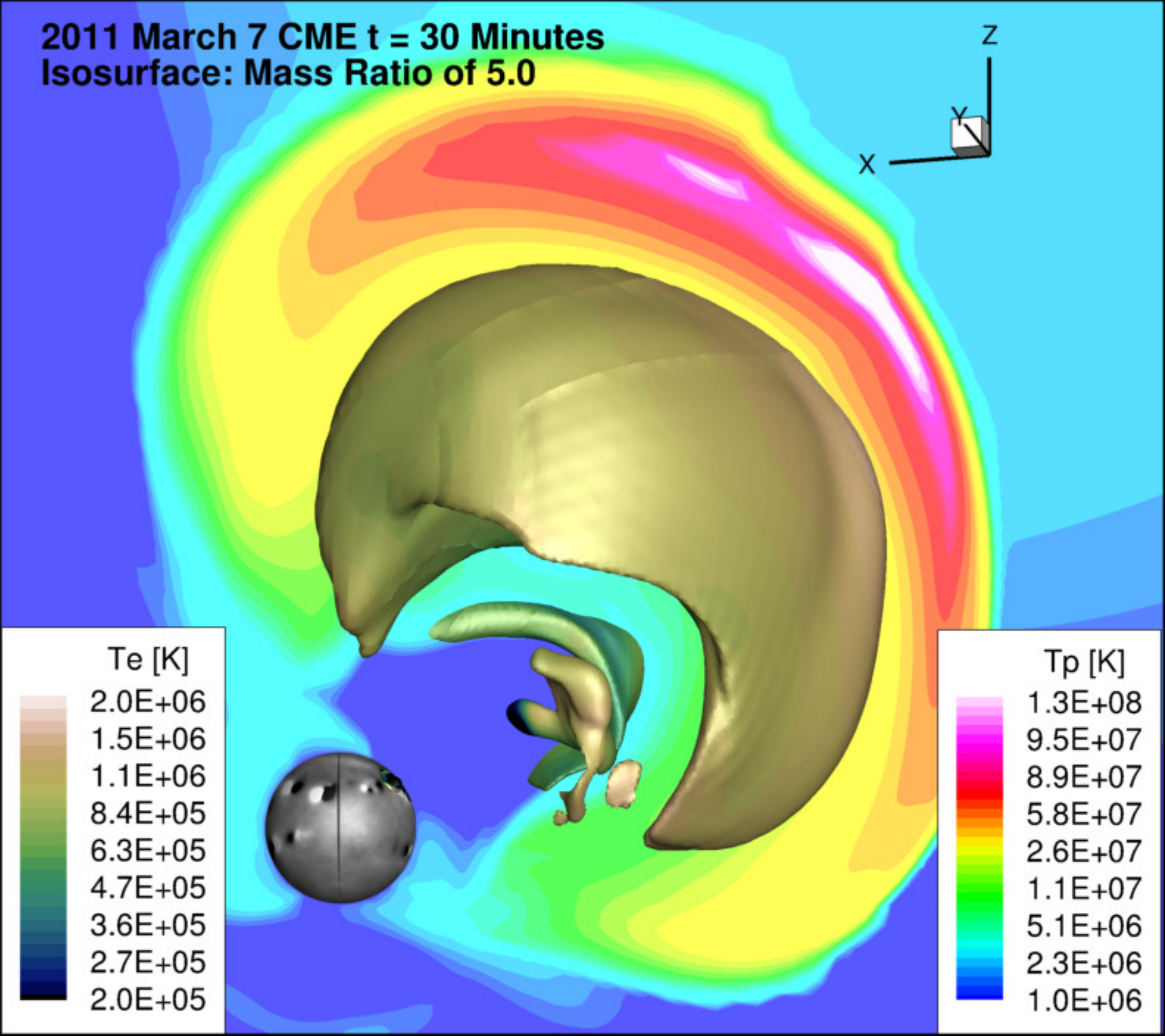} &
\includegraphics[scale=0.365]{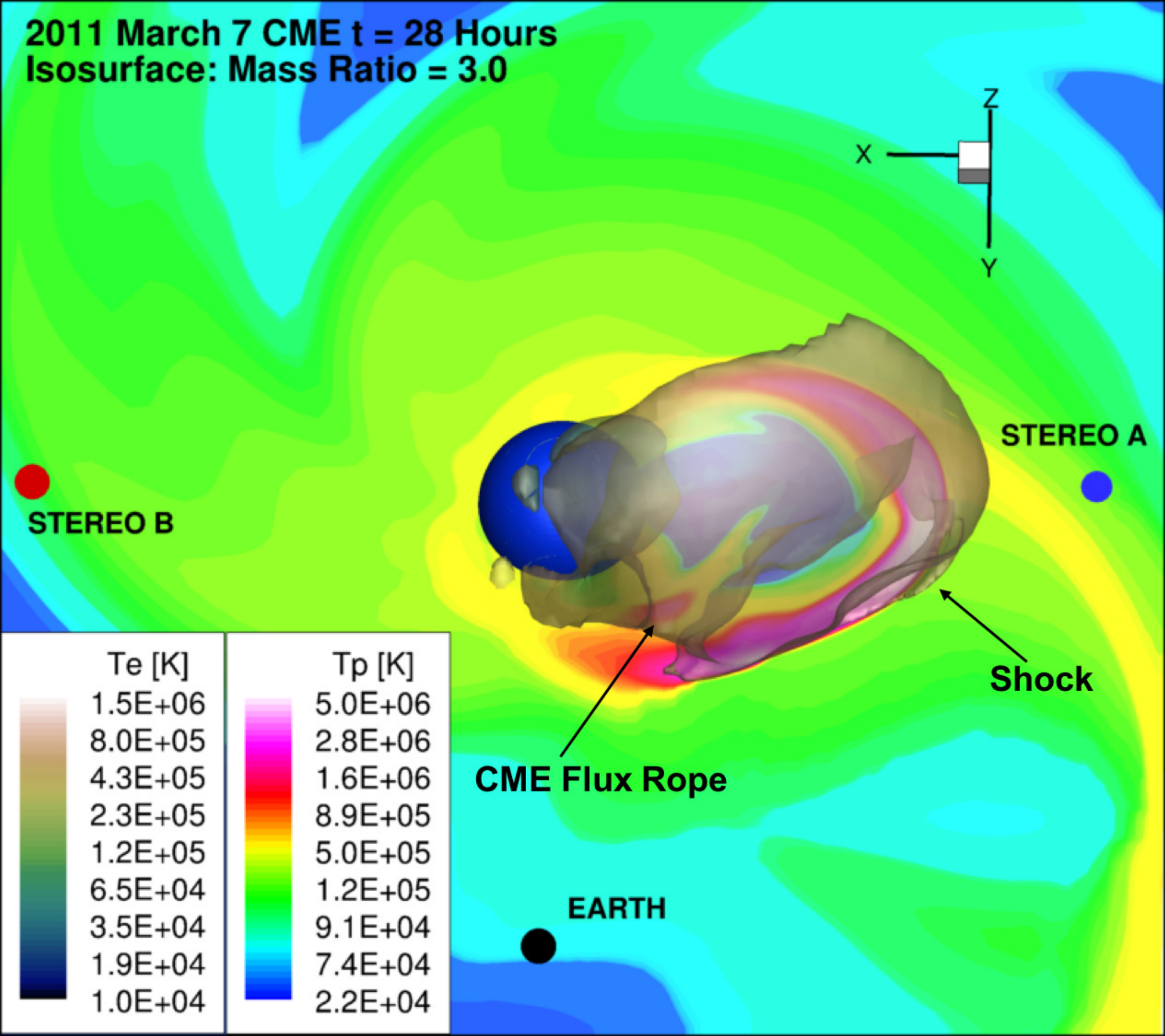}
\end{array}$
\end{center}
\caption{\label{fig:shock}Left panel: CME-driven shock structure in SC at t = 30 Minutes. Right panel: CME-driven shock structure in IH at t = 28 hours. The isosurface in SC shows the density ratio of 5. The isosurface in IH shows the density ratio of 3. The background shows the proton temperature and the color scale on the isosurface shows the electron temperature. The Earth, STEREO A and STEREO B positions are shown in IH with different color spots.}
\end{figure}

In Figure \ref{fig:shock}, we show the CME-driven shock structure both near the Sun and in the heliosphere. In the left panel, the slice shows the proton temperature at t = 30 minutes, while the isosurface (mass density ratio of 5.0 relative to the background) shows the electron temperature. Due to the decoupling between the electrons and protons, their temperatures are an order of magnitude different at the same location \citep{kos91, chip12, jin13}. The CME-driven shock heats the protons to $\sim$130 MK, while the electrons are only heated by adiabatic compression at the shock. In the right panel, the slice shows the proton temperature at t = 28 hours, while the isosurface shows the mass density ratio of 3.0. Again, we see the difference between the electron and proton temperatures. 

Note that the Te/Tp ratio is obtained under single fluid assumption. In our model, we assume the electrons and protons are thermally coupled only through collisions. In reality, there are other mechanisms that can couple the two populations and more rapidly thermalize the electrons (e.g., \citealt{wu84}). The electron heating at collisionless shocks is found to be inversely proportional to Mach number \citep{schwartz88, gha01}. With a low Mach number, half of the heat goes to electrons. With a high Mach number, only less than 10\% of the heat goes to electrons. Therefore, our results can be considered as a limiting case with minimum thermal coupling between electrons and protons, which is most appropriate for strong/parallel shocks. In order to have a more precise Te/Tp ratio, kinetic treatment is needed.

\begin{figure}[h]
\begin{center}$
\begin{array}{cc}
\includegraphics[scale=0.25]{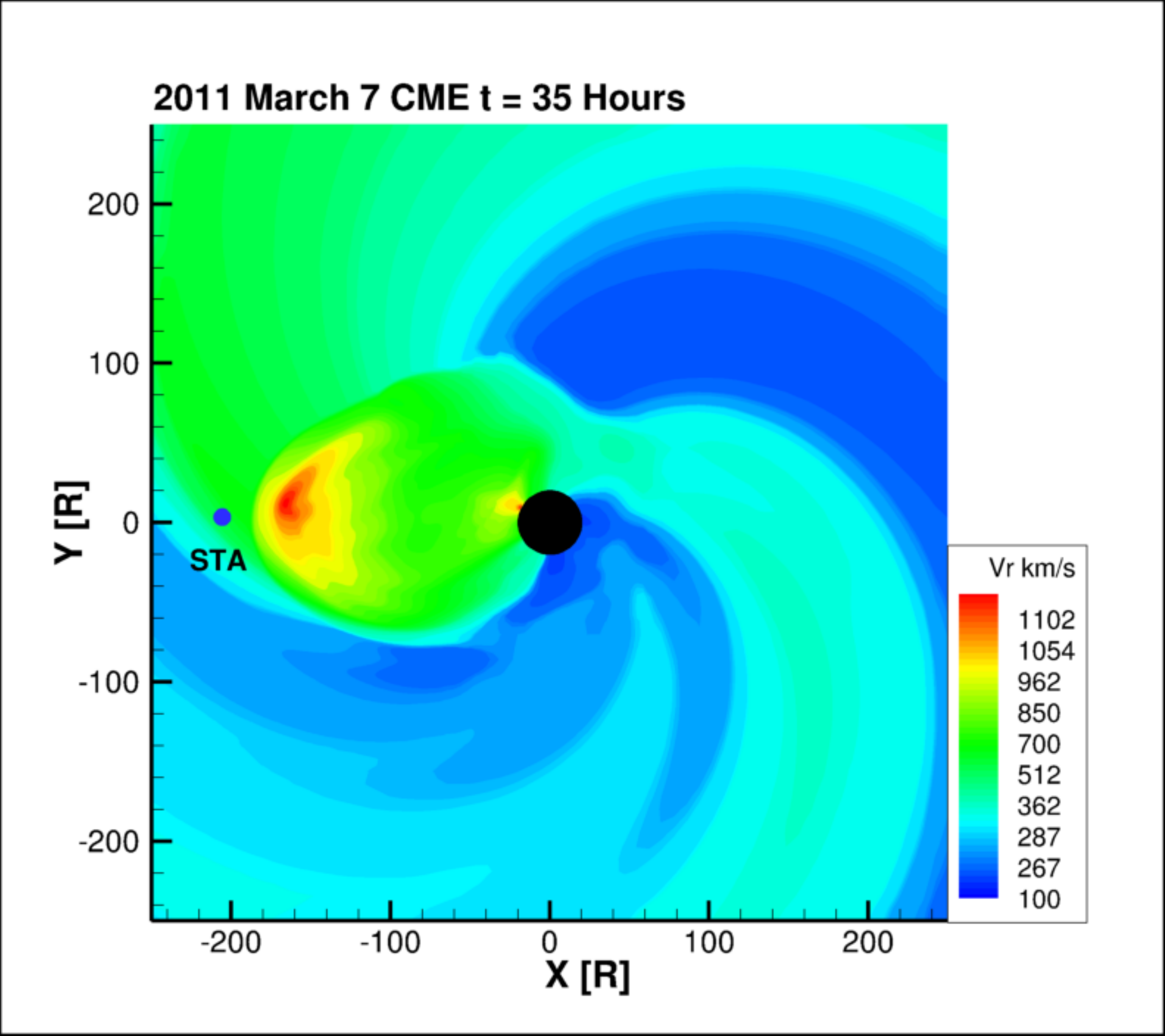} &
\includegraphics[scale=0.25]{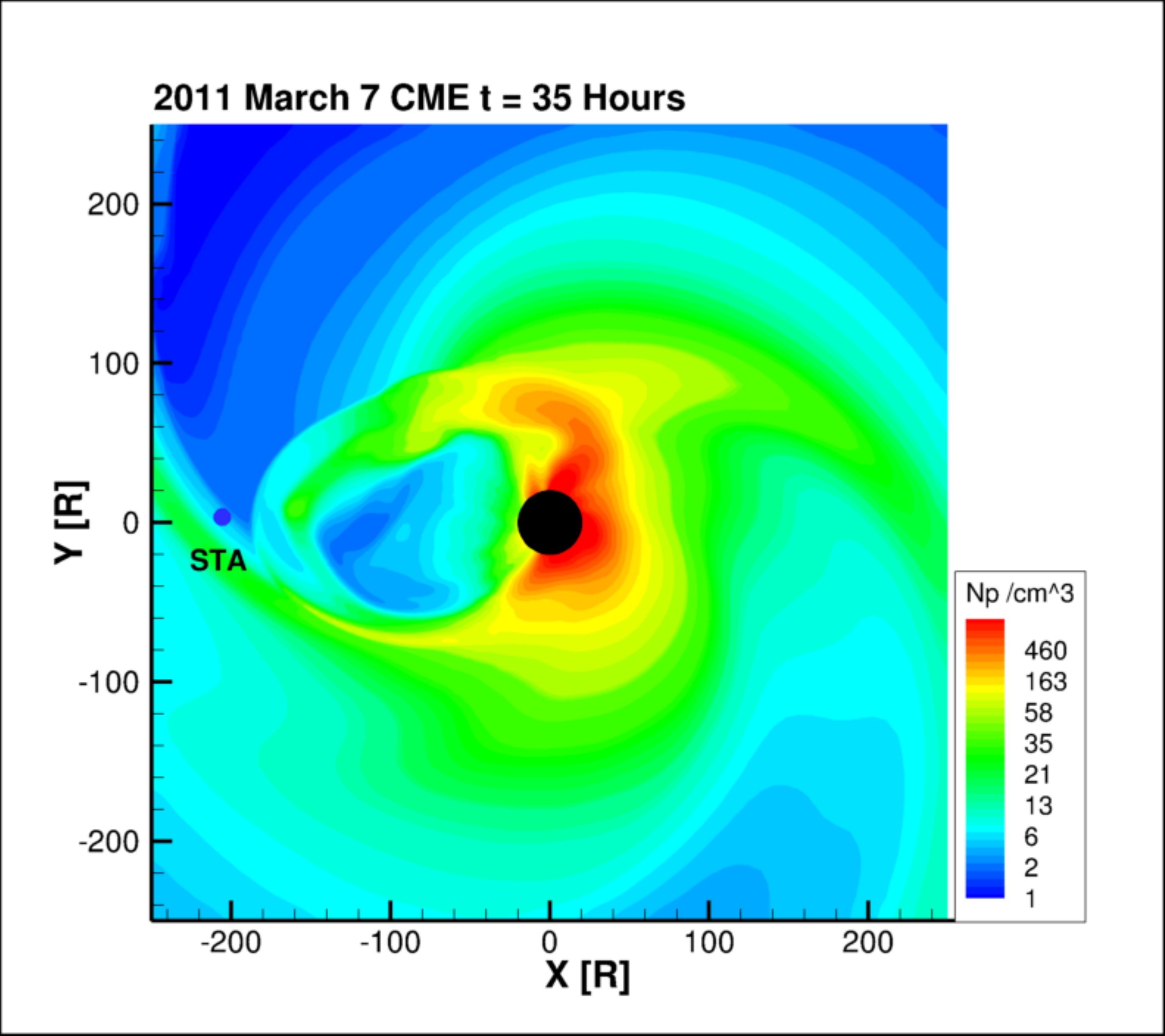} \\
\includegraphics[scale=0.3258]{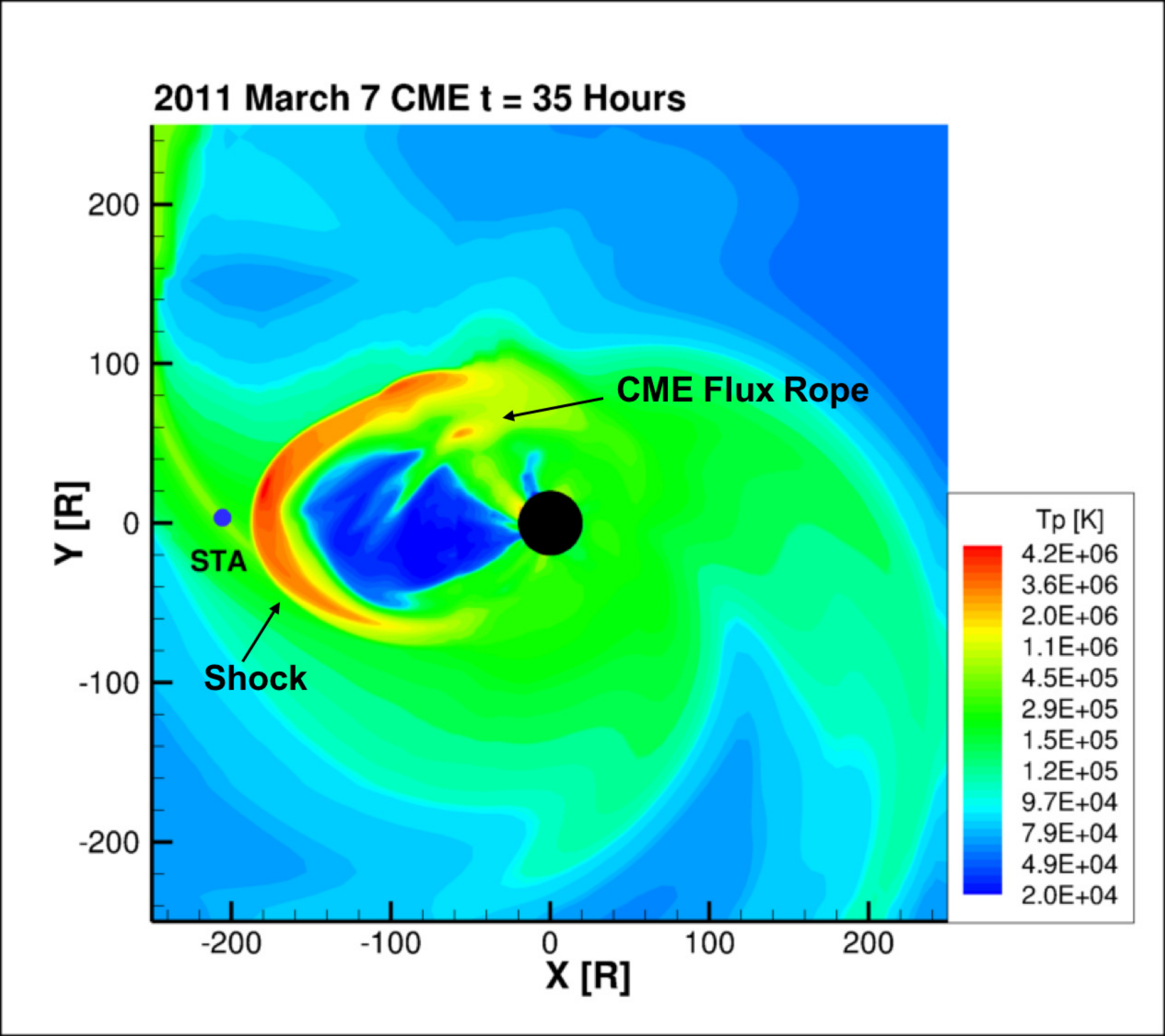} &
\includegraphics[scale=0.25]{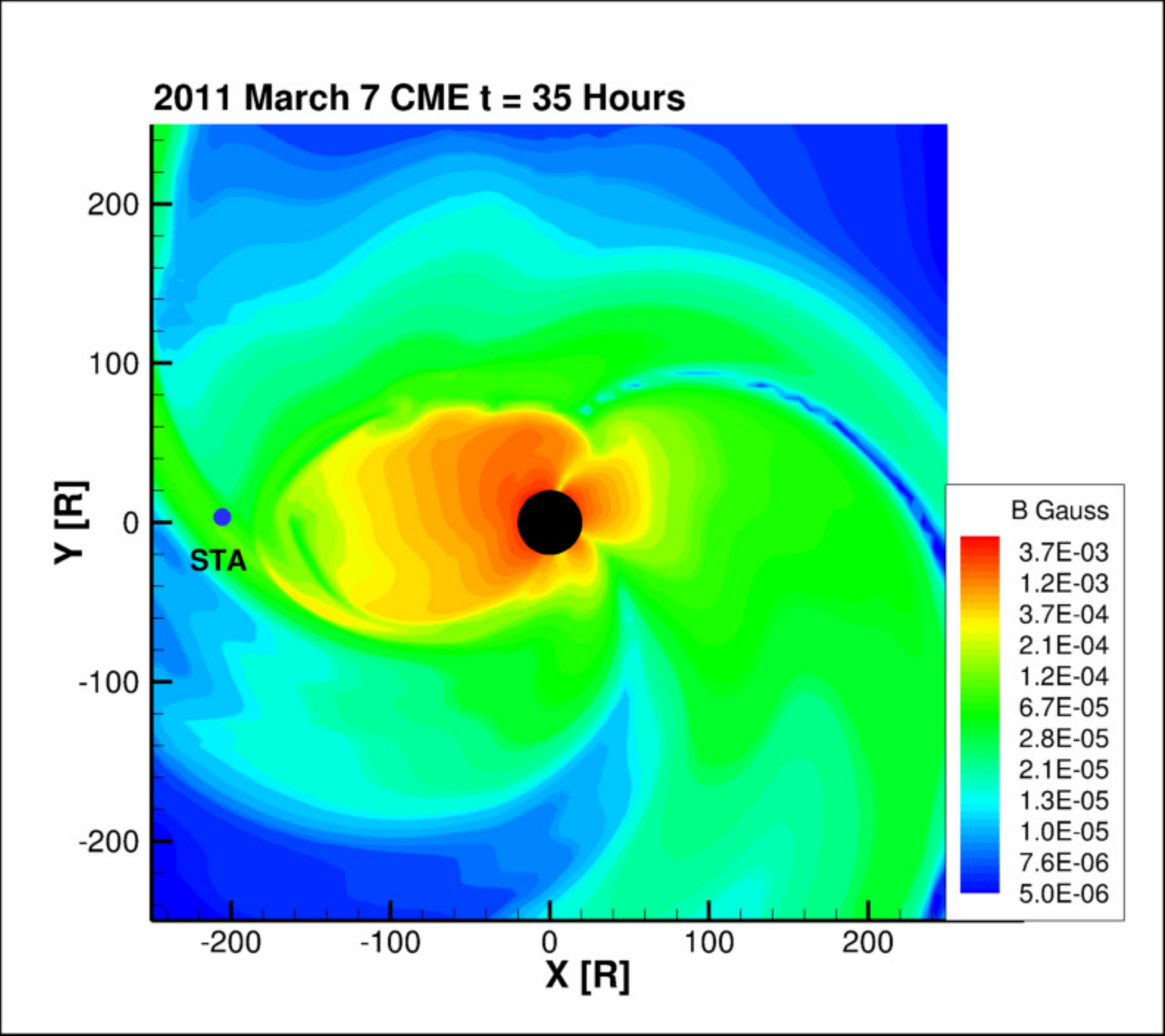} 
\end{array}$
\end{center}
\caption{\label{fig:ih}The radial velocity, proton density, proton temperature, and total magnetic field of the simulated CME at t = 35 hours.}
\end{figure}

The CME evolution in the heliosphere is shown in Figure \ref{fig:shock} and \ref{fig:ih}. In the right panel of Figure \ref{fig:shock}, the Earth, STA, and STB positions are shown, which provide the multi-viewpoints of the CME event. Also, we marked the interplanetary shock and CME ejecta locations in our simulation. We can see in the model that the interplanetary shock mainly propagates toward STA, and that the slower flank of the shock with the CME flux rope in behind propagates toward the Earth. The very different shock speeds toward STA and Earth are caused by the different background solar wind speeds shown in Figure \ref{fig:ih}. The shock toward STA is propagating into a fast velocity stream with speed $>700$ km s$^{-1}$. This stream can be traced back to the on-disk corona hole (CH in STA observation in Figure \ref{fig:euv}). Since the CME happens just east of the corona hole (AR 11164 in STA observation in Figure \ref{fig:euv}), the CME-driven shock expands into the fast stream and propagates toward STA, while the CME-driven shock in the slow velocity stream propagates toward the Earth. This fact is in good agreement with the shock/ejecta reconstruction from the observations (see CME2 in Figure 2 from \citealt{wood12}). The standoff distance is inversely related to the shock Mach number. Therefore, the lower shock Mach number in the fast stream leads to a larger standoff distance of the shock. Since the standoff distance is so large in this event, it is reasonable to say that the shock toward STA has detached from the CME driver. While not a true blast wave (the shock is initially driven by a CME), the detached shock has features similar to a blast wave as suggested by \citet{howard16}. The density and temperature decrease behind the detached shock is a result from the divergent flow (expansion) after the shock passing. Without a driver behind the shock, there is nothing to maintain the compression in the sheath.

Another interesting feature of this CME event is the interplanetary shock-CIR interaction as can be seen in Figures \ref{fig:shock} and \ref{fig:ih}. The shock-CIR interaction phenomenon has been observed in many cases (e.g., \citealt{gomez11}) and is believed to be related to the enhanced local ion acceleration in the hundred-keV energy range \citep{giacalone02}. The shocks that interact with CIRs can be difficult to identify in observations due to their distorted structure after interaction (e.g., \citealt{richardson04, riley06}). The shock-CIR interaction acts as shock-shock collisions (e.g., CME-CME interaction; \citealt{lugaz08}) and will amplify the magnetic fields, plasma temperature, and density of the CIR. We can see the effect of shock-CIR interaction in Figure \ref{fig:ih}. This phenomenon is also found in the interaction of high Mach-number shocks in laser-produced plasma \citep{morita13}. Note that, although the shock-CIR interaction is evident in the simulation, the shock passing through the STA location is not interacting with the leading edge of the CIR but the fast flow behind the CIR front.

\subsection{Shock Structure at 1 AU}
\begin{figure}[h]
\begin{center}$
\begin{array}{c}
\includegraphics[scale=0.55]{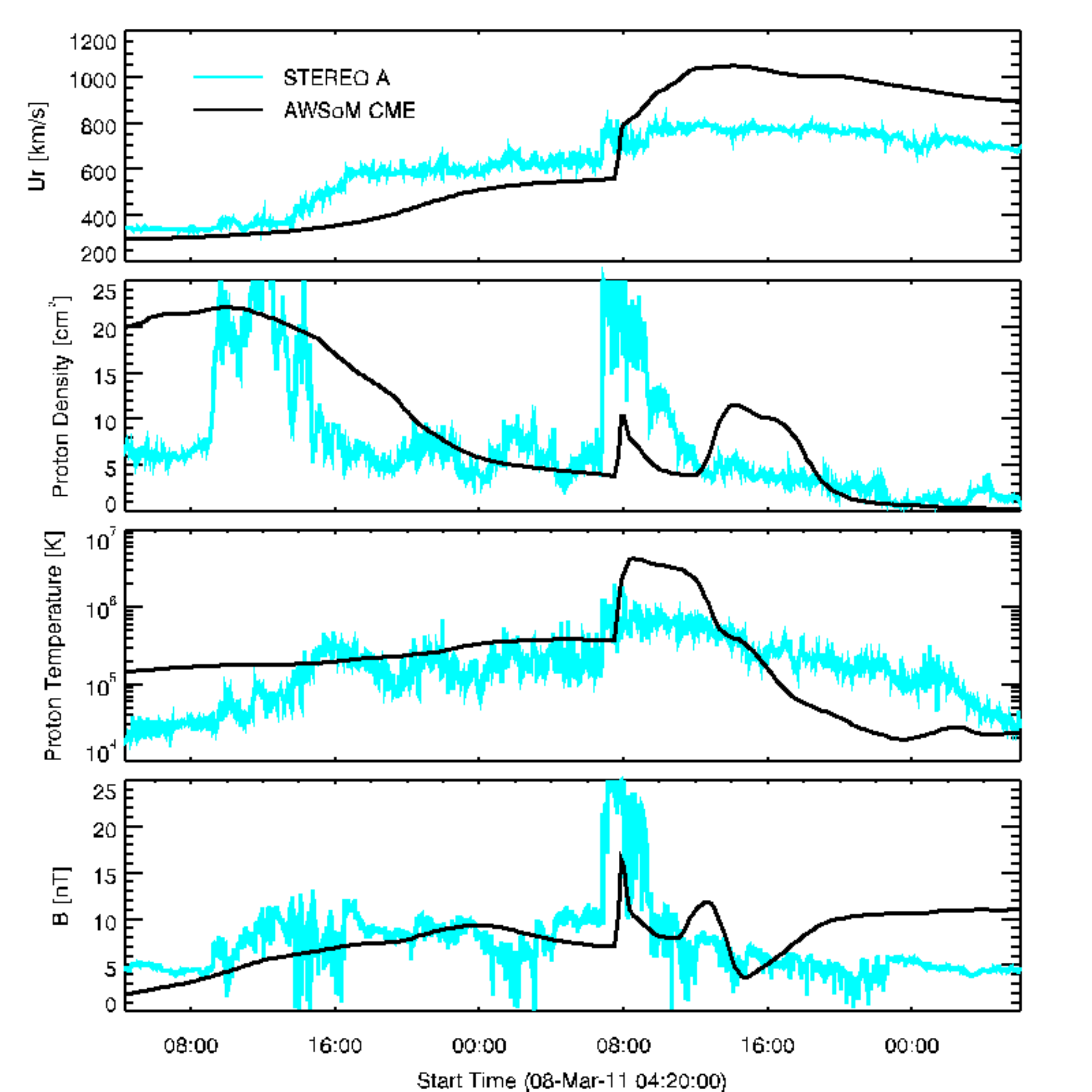}
\end{array}$
\end{center}
\caption{\label{fig:1aucme1}Comparison of the CME \textit{in situ} observation with the simulation for radial velocity, proton density, proton temperature, and total magnetic field.}
\end{figure}

In Figure \ref{fig:1aucme1}, we show the comparison of the CME \textit{in situ} observations with the simulation for radial velocity, proton density, proton temperature, and total magnetic field. The detached shock hits STA at $\sim$6:50 UT on 2011 March 9 in the observation. For the simulation, the shock arrives within $\sim$1 hour later. Considering the discrepancy between the shock speed in the simulation and in the observation near the Sun, it indicates that the simulated shock suffers more deceleration in the heliosphere than the observed shock. This effect could be caused by the relatively higher background solar wind density in the simulation along the shock propagation path. In the radial velocity comparison, we can see that the simulation reproduces the velocity jump of $\sim$200 km s$^{-1}$ at the shock as well as the gradual decrease in velocity after the shock passing. The most significant difference is that in the simulation there is another more gradual velocity increase by $\sim$200 km s$^{-1}$ after the shock, where the velocity is higher in the simulation than in the observation after shock. This difference is due to the numerical reconnection behind the shock. Numerical reconnection is also responsible for the density peak at 14:00 UT in the simulation. These ``features" are formed near the Sun due to the post-eruption reconnection between the flux rope and background fields \citep{jin13}. Although magnetic reconnection does exist and is observed during the CME propagation (e.g., Webb et al. 2003, Ko et al. 2003, Gosling et al. 2005), the reconnection prescription in our model may not be sufficient to address the physics behind this phenomenon. Note that the energy released by numerical reconnection in our model only heats the protons. Since the heat condition is not applied to protons, the dissipated energy cannot transfer back to the Sun therefore leads to an elevated proton temperature as well as velocity and density increases. To improve the current prescription, a finer grid should be used to reduce numerical reconnection. Also, explicit resistivity for the Joule heating of the electrons in the reconnection region should be included in the future. In the observed density plot, we can see two peaks in the data. The first one (around 2011 March 8 10:00 UT) is related to the CIR structure, and the second one is related to the CME-driven shock. The density jump at the shock is a factor of $\sim$4 in the observation, while it is $\sim$2 in the simulation. The proton temperature at the shock jumps from $\sim$0.1 MK to $\sim$1 MK in the observation, while in the simulation it jumps from $\sim$0.3 MK to $\sim$3 MK. The magnetic field in both the simulation and the observation has a jump of factor $\sim$2.5 at the shock with the magnitude slightly smaller in the simulation.

In order to explain the discrepancy between the compression ratios in the simulation and the observation, we show the evolution of the compression ratio and the shock Mach number (acoustic) at the shock front toward the location of STA in the simulation. At t = 9 hours, the shock compression ratio is $\sim$4 and the shock Mach number is $\sim$6. Both the compression ratio and the shock Mach number gradually decrease. At t = 36 hours (near the STA impact), the compression ratio is $\sim$2.6 with the shock Mach number of $\sim$3.5. The shock changes from a strong shock to a moderate shock before hitting STA in the simulation. This is caused by the elevated proton temperature in the CIR region, which dramatically increases the local acoustic speed and decreases the shock Mach number. However, the proton temperature in the observation is lower by a factor of $\sim$5-10 in the CIR region than in the simulation. The lower acoustic speed in reality leads to a higher shock Mach number, and therefore, a higher compression ratio when the shock hits STA. In order to capture the correct shock compression ratio, the background proton temperature in the CIR region needs to be improved in the future.

Another phenomenon that we need to understand in the simulation is the inconsistency between the shock Mach number and the proton temperature jump. Based on the Mach number of the simulated shock, the temperature jump should be $\sim$4 (for Mach number = 3.5). However, as we can see in Figure \ref{fig:1aucme1}, the actual temperature jump in the simulation is $\sim$10. The higher temperature jump is caused by the local Alfv\'{e}n wave dissipation formulation we used in this study (see Eq. 1). The energy of the reflected Alfv\'{e}n wave behind the shock is immediately dissipated, which leads to extensive proton heating and therefore the elevated proton temperature. Note that the local dissipation formulation assumes strongly imbalanced and balanced turbulence. Our results suggest that this assumption cannot be applied to the CME-driven shock region, where the turbulence could be moderately imbalanced. Therefore, a more physically complete dissipation formulation described by \citet{bart14} should be used for future CME simulations.
 
In Figure \ref{fig:1aucme2}, we further compare the three components of the magnetic field between the simulation and the observed event. As we can see, the simulation successfully captures the overall variation of the magnetic field at the shock passing. The magnitudes of all three components increase at the shock. The Bx component has a positive direction, while By and Bz have negative directions. The negative Bz does not last long in this event at STA. In Figure \ref{fig:1aucme3}, we show the comparison of velocity components between the simulation and observation. Again, our simulation shows consistency with the observations for all three velocity components. The velocity information is critically important to determine the shock normal. Based on the comparison, our simulation catches the shock normal correctly at STA for this event. 

\section{Summary \& Conclusion}
In this study, the 2011 March 7 CME event is simulated from the chromosphere to 1AU where we capture shocks at the flank of the CME. We do not assess model performance in the vicinity of the CME driver due to lack of relevant observations. Comparing the model with previous work \citep{jin13}, the new AWSoM model incorporates physically consistent wave reflection/dissipation and spatially dependent heat partitioning based on linear wave theory and stochastic heating. Moreover, collisionless electron heat conduction is taken into account and combined with the collisional Spitzer heat conduction. Our simulation results are compared using multi-spacecraft observations from SOHO, SDO, STEREO A/B, and OMNI. The new model shows capability to reproduce many observed features of this CME near the Sun and of its disturbance at 1 AU in this event. We summarize the major conclusions as follows:

1. Near the Sun, the synthesized EUV images of the model can reproduce most of the active regions and on-disk/polar coronal holes. Also, the intensity of the active region is comparable with the observation thanks to the enhanced wave reflection around the active regions.

2. The 3D CME reconstruction and white-light comparison from three different viewpoints show that the simulated CME propagates in the same direction as the observed event to a very high degree. The GL flux rope shows the capacity to reproduce the observed white-light features of the CME (e.g., double-front morphology, dark cavity, dense core), which was also shown in the previous work by \citet{lugaz05}. Within 20 R$_{\odot}$, the simulated CME-driven shock front is $\sim$600 km$^{-1}$ faster than the observed CME shock front, but the speed is comparable with the second front in the simulation.

3. A comprehensive 1 AU \textit{in situ} comparison shows that our simulation captures all the shock features of this event with varying degrees of accuracy. The deflection of the CME away from the coronal hole is evident both in the observation and in the simulation. The CME-driven shock expands into the coronal hole's fast outflow and travels far from the ejecta where it is observed by STA.

4. Although initially driven by the CME flux rope close to the Sun, the shock toward STA becomes detached from the driver in the heliosphere and has features similar to a blast wave.

Based on these promising results, our future work will focus on the following directions: First, since the CME flux rope is not well observed in this event due to its propagation direction, we will conduct more benchmark case studies as suggested by \citet{mostl12} to validate both the CME-driven shocks and flux rope structures from multiple \textit{in situ} observations \citep{liu13}. Second, the gradual SEP events are believed to be accelerated by CME-driven shocks through the diffusive shock acceleration mechanism \citep{reames99}. \citet{mewaldt08} pointed out that the total energy budget of the energetic particles can be 10\% or more of the CME kinetic energy. Therefore, coupling the CME model with the SEP model \citep{sokolov04, sokolov09} needs to be pursued. By separating the electron and proton temperatures in our CME model, the CME-driven shock is well reproduced both near the Sun and in the heliosphere, which could lead to an effective acceleration by DSA. However, the shock jump condition was not well reproduced at Earth. Work remains on getting the correct CIR conditions in the model. Third, AWSoM shows a new capacity for investigating turbulence phenomenon related to the CME-driven shocks. With higher temporal and spatial resolution in the simulation as well as with pressure anisotropy \citep{bart14, meng15}, the CME-turbulence interactions will be investigated and compared with observations (e.g., \citealt{liu06}). Finally, we should note that the flux rope parameters are chosen in an \textit{ad-hoc} way to best reproduce the observed travel time at STA.  Our companion paper details a methodology to automatically set these parameters based on empirical relationships between the flux rope parameters and solar observations (Jin et al. 2016b). 

At last, we should note that it is still unclear if all CMEs can be represented by a flux rope model as used in this study. Although, the bright twisted loops, which are believed to be flux rope structures, are frequently observed at the CME onset (e.g., \citealt{zhang12, cheng13, nindos15}), we still do not know whether the magnetic structure is consistent with our flux rope model. Therefore, there is still a long way to go before claiming the predictive value of the simulation model. Meanwhile, it will keep playing an important role for understanding the physical processes of CME propagation that may be included in future operational space weather models.

\begin{acknowledgements}
We are very grateful to the referee for invaluable comments that helped improve the paper. M.Jin is supported by NASA/UCAR Jack Eddy Postdoctoral Fellowship and NASA's SDO/AIA contract (NNG04EA00C) to LMSAL. The work performed at the University of Michigan was partially supported by National Science Foundation grants cAGS-1322543, AGS1408789, AGS1459862 and PHY-1513379, NASA grant NNX13AG25G, the European Union's Horizon 2020 research and innovation program under grant agreement No 637302 PROGRESS. C.A. de Koning was supported by NASA grant LWS NNX09AJ84G. We would also like to acknowledge high-performance computing support from: (1) Yellowstone (ark:/85065/d7wd3xhc) provided by NCAR's Computational and Information Systems Laboratory, sponsored by the National Science Foundation, and (2) Pleiades operated by NASA's Advanced Supercomputing Division. 

This work utilizes data obtained by the Global Oscillation Network Group (GONG) Program, managed by the National Solar Observatory, which is operated by AURA, Inc. under a cooperative agreement with the National Science Foundation. SDO is the first mission to be launched for NASA's Living With a Star (LWS) Program. LASCO was built by a consortium of the Naval Research Laboratory, USA, the Laboratoire d'Astrophysique de Marseille (formerly Laboratoire d'Astronomie Spatiale), France, the Max-Planck-Institut f\"ur Sonnensystemforschung (formerly Max Planck Institute f\"ur Aeronomie), Germany, and the School of Physics and Astronomy, University of Birmingham, UK. SoHO is a project of joint collaboration by ESA and NASA. STEREO (Solar TErrestrial RElations Observatory) is the third mission in NASA's Solar Terrestrial Probes program (STP). The STEREO/SECCHI data are produced by a consortium of NRL (U.S.), LMSAL (U.S.), NASA/GSFC (U.S.), RAL (UK), UBHAM (UK), MPS (Germany), CSL (Belgium), IOTA (France), and IAS (France). This work has also made use of data provided by the STEREO PLASTIC and IMPACT teams, supported by NASA contracts NAS5-00132 and NAS5-00133. The OMNI data access is provided by the NASA Goddard Space Flight Center Space Physics Data Facility (SPDF).

\end{acknowledgements}

\newpage
%\bibliographystyle{apj}
%\bibliography{ref}

\begin{thebibliography}{92}
\expandafter\ifx\csname natexlab\endcsname\relax\def\natexlab#1{#1}\fi

\bibitem[{{Antiochos} {et~al.}(1999){Antiochos}, {DeVore}, \&
  {Klimchuk}}]{anti99}
{Antiochos}, S.~K., {DeVore}, C.~R., \& {Klimchuk}, J.~A. 1999, \apj, 510, 485

\bibitem[{{Burlaga}(1971)}]{burlaga71}
{Burlaga}, L.~F. 1971, \ssr, 12, 600

\bibitem[{{Chandran} {et~al.}(2011){Chandran}, {Dennis}, {Quataert}, \&
  {Bale}}]{chandran11}
{Chandran}, B.~D.~G., {Dennis}, T.~J., {Quataert}, E., \& {Bale}, S.~D. 2011,
  \apj, 743, 197

\bibitem[{{Chen} {et~al.}(2005){Chen}, {Fang}, \& {Shibata}}]{chen05}
{Chen}, P.~F., {Fang}, C., \& {Shibata}, K. 2005, \apj, 622, 1202

\bibitem[Chen et al.(2016)]{chen16} Chen, Y., T{\'o}th, G., \& Gombosi, T.~I.\ 2016, Journal of Computational Physics, 305, 604 

\bibitem[Cheng et al.(2013)]{cheng13} Cheng, X., Zhang, J., Ding, M.~D., Liu, Y., \& Poomvises, W.\ 2013, \apj, 763, 43

\bibitem[{{Cohen} {et~al.}(2009){Cohen}, {Attrill}, {Manchester}, \&
  {Wills-Davey}}]{cohen09}
{Cohen}, O., {Attrill}, G.~D.~R., {Manchester}, IV, W.~B., \& {Wills-Davey},
  M.~J. 2009, \apj, 705, 587

\bibitem[{{Cohen} {et~al.}(2007){Cohen}, {Sokolov}, {Roussev}, {Arge},
  {Manchester}, {Gombosi}, {Frazin}, {Park}, {Butala}, {Kamalabadi}, \&
  {Velli}}]{cohen07}
{Cohen}, O., {et~al.} 2007, \apjl, 654, L163

\bibitem[de Koning et al.(2009)]{koning09} de Koning, C.~A., 
Pizzo, V.~J., \& Biesecker, D.~A.\ 2009, \solphys, 256, 167 

\bibitem[de Koning \& Pizzo (2011)]{koning11} de Koning, C. A., \& Pizzo, V. J. 2011, SpWea, 9, S03001

\bibitem[{{Delaboudini{\`e}re} {et~al.}(1995){Delaboudini{\`e}re}, {Artzner},
  {Brunaud}, {Gabriel}, {Hochedez}, {Millier}, {Song}, {Au}, {Dere}, {Howard},
  {Kreplin}, {Michels}, {Moses}, {Defise}, {Jamar}, {Rochus}, {Chauvineau},
  {Marioge}, {Catura}, {Lemen}, {Shing}, {Stern}, {Gurman}, {Neupert},
  {Maucherat}, {Clette}, {Cugnon}, \& {van Dessel}}]{del95}
{Delaboudini{\`e}re}, J.-P., {et~al.} 1995, \solphys, 162, 291

\bibitem[{{Downs} {et~al.}(2012){Downs}, {Roussev}, {van der Holst}, {Lugaz},
  \& {Sokolov}}]{downs12}
{Downs}, C., {Roussev}, I.~I., {van der Holst}, B., {Lugaz}, N., \& {Sokolov},
  I.~V. 2012, \apj, 750, 134

\bibitem[{{Downs} {et~al.}(2011){Downs}, {Roussev}, {van der Holst}, {Lugaz},
  {Sokolov}, \& {Gombosi}}]{downs11}
{Downs}, C., {Roussev}, I.~I., {van der Holst}, B., {Lugaz}, N., {Sokolov},
  I.~V., \& {Gombosi}, T.~I. 2011, \apj, 728, 2

\bibitem[{{Dryer} {et~al.}(2004){Dryer}, {Smith}, {Fry}, {Sun}, {Deehr}, \&
  {Akasofu}}]{dryer04}
{Dryer}, M., {Smith}, Z., {Fry}, C.~D., {Sun}, W., {Deehr}, C.~S., \&
  {Akasofu}, S.-I. 2004, Space Weather, 2, 9001

\bibitem[{{Evans} {et~al.}(2012){Evans}, {Opher}, {Oran}, {van der Holst},
  {Sokolov}, {Frazin}, {Gombosi}, \& {V{\'a}squez}}]{evans12}
{Evans}, R.~M., {Opher}, M., {Oran}, R., {van der Holst}, B., {Sokolov}, I.~V.,
  {Frazin}, R., {Gombosi}, T.~I., \& {V{\'a}squez}, A. 2012, \apj, 756, 155

\bibitem[{{Feng} {et~al.}(2012){Feng}, {Jiang}, {Xiang}, {Zhao}, \&
  {Wu}}]{feng12}
{Feng}, X., {Jiang}, C., {Xiang}, C., {Zhao}, X., \& {Wu}, S.~T. 2012, \apj,
  758, 62

\bibitem[{{Feng} {et~al.}(2011){Feng}, {Zhang}, {Xiang}, {Yang}, {Jiang}, \&
  {Wu}}]{feng11}
{Feng}, X., {Zhang}, S., {Xiang}, C., {Yang}, L., {Jiang}, C., \& {Wu}, S.~T.
  2011, \apj, 734, 50

\bibitem[{{Fry} {et~al.}(2001){Fry}, {Sun}, {Deehr}, {Dryer}, {Smith},
  {Akasofu}, {Tokumaru}, \& {Kojima}}]{fry01}
{Fry}, C.~D., {Sun}, W., {Deehr}, C.~S., {Dryer}, M., {Smith}, Z., {Akasofu},
  S.-I., {Tokumaru}, M., \& {Kojima}, M. 2001, \jgr, 106, 20985

\bibitem[{{Gallagher} {et~al.}(2002){Gallagher}, {Moon}, \& {Wang}}]{gal02}
{Gallagher}, P.~T., {Moon}, Y.-J., \& {Wang}, H. 2002, \solphys, 209, 171

\bibitem[{{Galvin} {et~al.}(2008){Galvin}, {Kistler}, {Popecki}, {Farrugia},
  {Simunac}, {Ellis}, {M{\"o}bius}, {Lee}, {Boehm}, {Carroll}, {Crawshaw},
  {Conti}, {Demaine}, {Ellis}, {Gaidos}, {Googins}, {Granoff}, {Gustafson},
  {Heirtzler}, {King}, {Knauss}, {Levasseur}, {Longworth}, {Singer}, {Turco},
  {Vachon}, {Vosbury}, {Widholm}, {Blush}, {Karrer}, {Bochsler}, {Daoudi},
  {Etter}, {Fischer}, {Jost}, {Opitz}, {Sigrist}, {Wurz}, {Klecker}, {Ertl},
  {Seidenschwang}, {Wimmer-Schweingruber}, {Koeten}, {Thompson}, \&
  {Steinfeld}}]{galvin08}
{Galvin}, A.~B., {et~al.} 2008, \ssr, 136, 437

\bibitem[Ghavamian et al.(2001)]{gha01} Ghavamian, P., 
Raymond, J., Smith, R.~C., \& Hartigan, P.\ 2001, \apj, 547, 995 

\bibitem[{{Giacalone} {et~al.}(2002){Giacalone}, {Jokipii}, \&
  {K{\'o}ta}}]{giacalone02}
{Giacalone}, J., {Jokipii}, J.~R., \& {K{\'o}ta}, J. 2002, \apj, 573, 845

\bibitem[{{Gibson} \& {Low}(1998)}]{gibson98}
{Gibson}, S.~E., \& {Low}, B.~C. 1998, \apj, 493, 460

\bibitem[{{G{\'o}mez-Herrero} {et~al.}(2011){G{\'o}mez-Herrero}, {Malandraki},
  {Dresing}, {Kilpua}, {Heber}, {Klassen}, {M{\"u}ller-Mellin}, \&
  {Wimmer-Schweingruber}}]{gomez11}
{G{\'o}mez-Herrero}, R., {Malandraki}, O., {Dresing}, N., {Kilpua}, E.,
  {Heber}, B., {Klassen}, A., {M{\"u}ller-Mellin}, R., \&
  {Wimmer-Schweingruber}, R.~F. 2011, Journal of Atmospheric and
  Solar-Terrestrial Physics, 73, 551

\bibitem[{{Gopalswamy} {et~al.}(2001){Gopalswamy}, {Lara}, {Yashiro}, {Kaiser},
  \& {Howard}}]{gopalswamy01}
{Gopalswamy}, N., {Lara}, A., {Yashiro}, S., {Kaiser}, M.~L., \& {Howard},
  R.~A. 2001, \jgr, 106, 29207

\bibitem[{{Gopalswamy} {et~al.}(2009){Gopalswamy}, {M{\"a}kel{\"a}}, {Xie},
  {Akiyama}, \& {Yashiro}}]{gopalswamy09}
{Gopalswamy}, N., {M{\"a}kel{\"a}}, P., {Xie}, H., {Akiyama}, S., \& {Yashiro},
  S. 2009, Journal of Geophysical Research (Space Physics), 114, 0

\bibitem[Gosling(1993)]{gosling93} Gosling, J.~T.\ 1993, \jgr, 98, 18937 

\bibitem[{{Groth} {et~al.}(2000){Groth}, {De Zeeuw}, {Gombosi}, \&
  {Powell}}]{groth00}
{Groth}, C.~P.~T., {De Zeeuw}, D.~L., {Gombosi}, T.~I., \& {Powell}, K.~G.
  2000, \jgr, 105, 25053

\bibitem[{{Hakamada} \& {Akasofu}(1982)}]{hakamada82}
{Hakamada}, K., \& {Akasofu}, S.-I. 1982, \ssr, 31, 3

\bibitem[{{Han} {et~al.}(1988){Han}, {Wu}, \& {Dryer}}]{han88}
{Han}, S.~M., {Wu}, S.~T., \& {Dryer}, M. 1988, Computers and Fluids, 16, 81

\bibitem[{{Hayashi} {et~al.}(2006){Hayashi}, {Zhao}, \& {Liu}}]{hayashi06}
{Hayashi}, K., {Zhao}, X.~P., \& {Liu}, Y. 2006, \grl, 33, 20103

\bibitem[{{Howard} {et~al.}(2008){Howard}, {Moses}, {Vourlidas}, {Newmark},
  {Socker}, {Plunkett}, {Korendyke}, {Cook}, {Hurley}, {Davila}, {Thompson},
  {St Cyr}, {Mentzell}, {Mehalick}, {Lemen}, {Wuelser}, {Duncan}, {Tarbell},
  {Wolfson}, {Moore}, {Harrison}, {Waltham}, {Lang}, {Davis}, {Eyles},
  {Mapson-Menard}, {Simnett}, {Halain}, {Defise}, {Mazy}, {Rochus}, {Mercier},
  {Ravet}, {Delmotte}, {Auchere}, {Delaboudiniere}, {Bothmer}, {Deutsch},
  {Wang}, {Rich}, {Cooper}, {Stephens}, {Maahs}, {Baugh}, {McMullin}, \&
  {Carter}}]{howard08}
{Howard}, R.~A., {et~al.} 2008, \ssr, 136, 67

\bibitem[Howard \& Pizzo(2016)]{howard16} Howard, T.~A., \& Pizzo, V.~J.\ 2016, \apj, 824, 92 

\bibitem[{{Illing} \& {Hundhausen}(1985)}]{llling85}
{Illing}, R.~M.~E., \& {Hundhausen}, A.~J. 1985, \jgr, 90, 275

\bibitem[{Jian {et~al.}(2015)Jian, MacNeice, Taktakishvili, Odstrcil, Jackson,
  Yu, Riley, Sokolov, \& Evans}]{jian2015}
Jian, L.~K., {et~al.} 2015, Space Weather, 2015SW001174

\bibitem[{{Jin} {et~al.}(2012){Jin}, {Manchester}, {van der Holst},
  {Gruesbeck}, {Frazin}, {Landi}, {Vasquez}, {Lamy}, {Llebaria}, {Fedorov},
  {Toth}, \& {Gombosi}}]{jin12}
{Jin}, M., {et~al.} 2012, \apj, 745, 64

\bibitem[{{Jin} {et~al.}(2013){Jin}, {Manchester}, {van der Holst}, {Oran},
  {Sokolov}, {Toth}, {Liu}, {Sun}, \& {Gombosi}}]{jin13}
{Jin}, M., Manchester, IV, W. B., van der Holst, B., {et~al.} 2013, \apj, 773, 50

\bibitem[Jin et al.(2016a)]{jin16} Jin, M., Schrijver, C.~J., Cheung, M.~C.~M., DeRosa, M.~L., Nitta, N.~V., \& Title, A.~M. 2016a, \apj, 820, 16

\bibitem[Jin et al.(2016b)]{jin16b} Jin, M., Manchester, W.~B., van der Holst, B., {et~al.} 2016b, \apj, \textit{in press}

\bibitem[{{Karpen} {et~al.}(2012){Karpen}, {Antiochos}, \& {DeVore}}]{karpen12}
{Karpen}, J.~T., {Antiochos}, S.~K., \& {DeVore}, C.~R. 2012, \apj, 760, 81

\bibitem[Kay et al.(2013)]{kay13} Kay, C., Opher, M., 
\& Evans, R.~M.\ 2013, \apj, 775, 5 

\bibitem[{{Kosovichev} \& {Stepanova}(1991)}]{kos91}
{Kosovichev}, A.~G., \& {Stepanova}, T.~V. 1991, \sovast, 35, 646

\bibitem[{{Kozarev} {et~al.}(2013){Kozarev}, {Evans}, {Schwadron}, {Dayeh},
  {Opher}, {Korreck}, \& {van der Holst}}]{kozarev13}
{Kozarev}, K.~A., {Evans}, R.~M., {Schwadron}, N.~A., {Dayeh}, M.~A., {Opher},
  M., {Korreck}, K.~E., \& {van der Holst}, B. 2013, \apj, 778, 43

\bibitem[{{Lemen} {et~al.}(2012){Lemen}, {Title}, {Akin}, {Boerner}, {Chou},
  {Drake}, {Duncan}, {Edwards}, {Friedlaender}, {Heyman}, {Hurlburt}, {Katz},
  {Kushner}, {Levay}, {Lindgren}, {Mathur}, {McFeaters}, {Mitchell}, {Rehse},
  {Schrijver}, {Springer}, {Stern}, {Tarbell}, {Wuelser}, {Wolfson}, {Yanari},
  {Bookbinder}, {Cheimets}, {Caldwell}, {Deluca}, {Gates}, {Golub}, {Park},
  {Podgorski}, {Bush}, {Scherrer}, {Gummin}, {Smith}, {Auker}, {Jerram},
  {Pool}, {Soufli}, {Windt}, {Beardsley}, {Clapp}, {Lang}, \&
  {Waltham}}]{lemen12}
{Lemen}, J.~R., {et~al.} 2012, \solphys, 275, 17

\bibitem[{{Lionello} {et~al.}(2009){Lionello}, {Linker}, \&
  {Miki{\'c}}}]{lio09}
{Lionello}, R., {Linker}, J.~A., \& {Miki{\'c}}, Z. 2009, \apj, 690, 902

\bibitem[Liu 
\& Ofman(2014)]{liu14} Liu, W., \& Ofman, L.\ 2014, \solphys, 289, 3233 

\bibitem[{{Liu} {et~al.}(2006){Liu}, {Richardson}, {Belcher}, {Kasper}, \&
  {Skoug}}]{liu06}
{Liu}, Y., {Richardson}, J.~D., {Belcher}, J.~W., {Kasper}, J.~C., \& {Skoug},
  R.~M. 2006, Journal of Geophysical Research (Space Physics), 111, 9108

\bibitem[{{Liu} {et~al.}(2013){Liu}, {Luhmann}, {Lugaz}, {M{\"o}stl}, {Davies},
  {Bale}, \& {Lin}}]{liu13}
{Liu}, Y.~D., {Luhmann}, J.~G., {Lugaz}, N., {M{\"o}stl}, C., {Davies}, J.~A.,
  {Bale}, S.~D., \& {Lin}, R.~P. 2013, \apj, 769, 45
  
\bibitem[{{Lugaz} {et~al.}(2013){Lugaz}, {Farrugia}, {Manchester}, \&
  {Schwadron}}]{lugaz13}
{Lugaz}, N., {Farrugia}, C.~J., {Manchester}, IV, W.~B., \& {Schwadron}, N.
  2013, \apj, 778, 20

\bibitem[{{Lugaz} {et~al.}(2005){Lugaz}, {Manchester}, \& {Gombosi}}]{lugaz05}
{Lugaz}, N., {Manchester}, IV, W.~B., \& {Gombosi}, T.~I. 2005, \apj, 634, 651

\bibitem[{{Lugaz} {et~al.}(2008){Lugaz}, {Manchester}, {Roussev}, \&
  {Gombosi}}]{lugaz08}
{Lugaz}, N., {Manchester}, IV, W.~B., {Roussev}, I.~I., \& {Gombosi}, T.~I.
  2008, Journal of Atmospheric and Solar-Terrestrial Physics, 70, 598

\bibitem[{{Lugaz} {et~al.}(2007){Lugaz}, {Manchester}, {Roussev}, {T{\'o}th},
  \& {Gombosi}}]{lugaz07}
{Lugaz}, N., {Manchester}, IV, W.~B., {Roussev}, I.~I., {T{\'o}th}, G., \&
  {Gombosi}, T.~I. 2007, \apj, 659, 788

\bibitem[{{Lugaz} \& {Roussev}(2011)}]{lugaz11a}
{Lugaz}, N., \& {Roussev}, I.~I. 2011, Journal of Atmospheric and
  Solar-Terrestrial Physics, 73, 1187
  
 \bibitem[Lugaz et al.(2011)]{lugaz11b} Lugaz, N., Downs, C., 
Shibata, K., et al.\ 2011, \apj, 738, 127 

\bibitem[{{Luhmann} {et~al.}(2008){Luhmann}, {Curtis}, {Schroeder}, {McCauley},
  {Lin}, {Larson}, {Bale}, {Sauvaud}, {Aoustin}, {Mewaldt}, {Cummings},
  {Stone}, {Davis}, {Cook}, {Kecman}, {Wiedenbeck}, {von Rosenvinge}, {Acuna},
  {Reichenthal}, {Shuman}, {Wortman}, {Reames}, {Mueller-Mellin}, {Kunow},
  {Mason}, {Walpole}, {Korth}, {Sanderson}, {Russell}, \&
  {Gosling}}]{luhmann08}
{Luhmann}, J.~G., {et~al.} 2008, \ssr, 136, 117

\bibitem[Luhmann et al.(2010)]{luhmann10} Luhmann, J.~G., 
Ledvina, S.~A., Odstrcil, D., et al.\ 2010, Advances in Space Research, 46, 
1 

\bibitem[{{MacNeice} {et~al.}(2004){MacNeice}, {Antiochos}, {Phillips},
  {Spicer}, {DeVore}, \& {Olson}}]{mac04}
{MacNeice}, P., {Antiochos}, S.~K., {Phillips}, A., {Spicer}, D.~S., {DeVore},
  C.~R., \& {Olson}, K. 2004, \apj, 614, 1028

\bibitem[{{Manchester} {et~al.}(2004{\natexlab{a}}){Manchester}, {Gombosi},
  {DeZeeuw}, \& {Fan}}]{chip04c}
{Manchester}, IV, W., {Gombosi}, T., {DeZeeuw}, D., \& {Fan}, Y.
  2004{\natexlab{a}}, \apj, 610, 588

\bibitem[{{Manchester} {et~al.}(2004{\natexlab{b}}){Manchester}, {Gombosi},
  {Roussev}, {de Zeeuw}, {Sokolov}, {Powell}, {T{\'o}th}, \& {Opher}}]{chip04a}
{Manchester}, W.~B., {Gombosi}, T.~I., {Roussev}, I., {de Zeeuw}, D.~L.,
  {Sokolov}, I.~V., {Powell}, K.~G., {T{\'o}th}, G., \& {Opher}, M.
  2004{\natexlab{b}}, Journal of Geophysical Research (Space Physics), 109,
  1102

\bibitem[{{Manchester} {et~al.}(2004{\natexlab{c}}){Manchester}, {Gombosi},
  {Roussev}, {Ridley}, {de Zeeuw}, {Sokolov}, {Powell}, \&
  {T{\'o}th}}]{chip04b}
{Manchester}, W.~B., {Gombosi}, T.~I., {Roussev}, I., {Ridley}, A., {de Zeeuw},
  D.~L., {Sokolov}, I.~V., {Powell}, K.~G., \& {T{\'o}th}, G.
  2004{\natexlab{c}}, Journal of Geophysical Research (Space Physics), 109,
  2107

\bibitem[{{Manchester} {et~al.}(2012){Manchester}, {van der Holst}, {T{\'o}th},
  \& {Gombosi}}]{chip12}
{Manchester}, IV, W.~B., {van der Holst}, B., {T{\'o}th}, G., \& {Gombosi},
  T.~I. 2012, \apj, 756, 81

\bibitem[{{Manchester} {et~al.}(2005){Manchester}, {Gombosi}, {De Zeeuw},
  {Sokolov}, {Roussev}, {Powell}, {K{\'o}ta}, {T{\'o}th}, \&
  {Zurbuchen}}]{chip05}
{Manchester}, IV, W.~B., {et~al.} 2005, \apj, 622, 1225

\bibitem[Manchester et al.(2008)]{chip08} Manchester, W.~B., IV, Vourlidas, A., T{\'o}th, G., et al.\ 2008, \apj, 684, 1448-1460

\bibitem[Manchester et al.(2014)]{chip14} Manchester, IV, W.~B., van der Holst, B., \& Lavraud, B.\ 2014, Plasma Physics and Controlled Fusion, 56, 064006 

\bibitem[{{Martin}(1998)}]{martin98}
{Martin}, S.~F. 1998, in Astronomical Society of the Pacific Conference Series,
  Vol. 150, IAU Colloq. 167: New Perspectives on Solar Prominences, ed. D.~F.
  {Webb}, B.~{Schmieder}, \& D.~M. {Rust}, 419

\bibitem[{{Mewaldt} {et~al.}(2008){Mewaldt}, {Cohen}, {Giacalone}, {Mason},
  {Chollet}, {Desai}, {Haggerty}, {Looper}, {Selesnick}, \&
  {Vourlidas}}]{mewaldt08}
{Mewaldt}, R.~A., {et~al.} 2008, in American Institute of Physics Conference
  Series, Vol. 1039, American Institute of Physics Conference Series, ed.
  G.~{Li}, Q.~{Hu}, O.~{Verkhoglyadova}, G.~P. {Zank}, R.~P. {Lin}, \&
  J.~{Luhmann}, 111--117

\bibitem[Meng et al.(2015)]{meng15} Meng, X., van der Holst, B., T{\'o}th, G., \& Gombosi, T.~I.\ 2015, \mnras, 454, 3697 

\bibitem[Michalek et al.(2007)]{michalek07} Michalek, G., 
Gopalswamy, N., \& Yashiro, S.\ 2007, \solphys, 246, 399 

\bibitem[Miki{\'c} et al.(1999)]{mikic99} Miki{\'c}, Z., 
Linker, J.~A., Schnack, D.~D., Lionello, R., 
\& Tarditi, A.\ 1999, Physics of Plasmas, 6, 2217

\bibitem[Millward et al.(2013)]{millward13} Millward, G., 
Biesecker, D., Pizzo, V., \& Koning, C.~A.\ 2013, Space Weather, 11, 57 

\bibitem[{{Morita} {et~al.}(2013){Morita}, {Sakawa}, {Kuramitsu}, {Dono},
  {Tanji}, {Aoki}, {Ide}, {Nishio}, {Gregory}, {Waugh}, {Woolsey},
  {Dizi{\'e}re}, {Koenig}, {Ide}, {Tsubouchi}, \& {Takabe}}]{morita13}
{Morita}, T., {et~al.} 2013, High Energy Density Physics, 9, 187

\bibitem[{{Moses} {et~al.}(1997){Moses}, {Clette}, {Delaboudini{\`e}re},
  {Artzner}, {Bougnet}, {Brunaud}, {Carabetian}, {Gabriel}, {Hochedez},
  {Millier}, {Song}, {Au}, {Dere}, {Howard}, {Kreplin}, {Michels}, {Defise},
  {Jamar}, {Rochus}, {Chauvineau}, {Marioge}, {Catura}, {Lemen}, {Shing},
  {Stern}, {Gurman}, {Neupert}, {Newmark}, {Thompson}, {Maucherat},
  {Portier-Fozzani}, {Berghmans}, {Cugnon}, {van Dessel}, \&
  {Gabryl}}]{moses97}
{Moses}, D., {et~al.} 1997, \solphys, 175, 571

\bibitem[{{M{\"o}stl} {et~al.}(2012){M{\"o}stl}, {Farrugia}, {Kilpua}, {Jian},
  {Liu}, {Eastwood}, {Harrison}, {Webb}, {Temmer}, {Odstrcil}, {Davies},
  {Rollett}, {Luhmann}, {Nitta}, {Mulligan}, {Jensen}, {Forsyth}, {Lavraud},
  {de Koning}, {Veronig}, {Galvin}, {Zhang}, \& {Anderson}}]{mostl12}
{M{\"o}stl}, C., {et~al.} 2012, \apj, 758, 10

\bibitem[Nindos et al.(2015)]{nindos15} Nindos, A., Patsourakos, S., Vourlidas, A., \& Tagikas, C.\ 2015, \apj, 808, 117 

\bibitem[{{Odstrcil} {et~al.}(2005){Odstrcil}, {Pizzo}, \& {Arge}}]{ods05}
{Odstrcil}, D., {Pizzo}, V.~J., \& {Arge}, C.~N. 2005, Journal of Geophysical
  Research (Space Physics), 110, 2106

\bibitem[{{Oran} {et~al.}(2013){Oran}, {van der Holst}, {Landi}, {Jin},
  {Sokolov}, \& {Gombosi}}]{oran13}
{Oran}, R., {van der Holst}, B., {Landi}, E., {Jin}, M., {Sokolov}, I.~V., \&
  {Gombosi}, T.~I. 2013, \apj, 778, 176

\bibitem[{{Patsourakos} \& {Vourlidas}(2012)}]{pat12}
{Patsourakos}, S., \& {Vourlidas}, A. 2012, \solphys, 281, 187

\bibitem[Pizzo et al.(2011)]{pizzo11} Pizzo, V., Millward, G., 
Parsons, A., et al.\ 2011, Space Weather, 9, 3004 

\bibitem[{{Powell} {et~al.}(1999){Powell}, {Roe}, {Linde}, {Gombosi}, \& {de
  Zeeuw}}]{powell99}
{Powell}, K.~G., {Roe}, P.~L., {Linde}, T.~J., {Gombosi}, T.~I., \& {de Zeeuw},
  D.~L. 1999, Journal of Computational Physics, 154, 284

\bibitem[Qiu et al.(2007)]{qiu07} Qiu, J., Hu, Q., Howard, 
T.~A., \& Yurchyshyn, V.~B.\ 2007, \apj, 659, 758 

\bibitem[{{Reames}(1999)}]{reames99}
{Reames}, D.~V. 1999, \ssr, 90, 413

\bibitem[{{Richardson} \& {Cane}(2004)}]{richardson04}
{Richardson}, I.~G., \& {Cane}, H.~V. 2004, Journal of Geophysical Research
  (Space Physics), 109, 9104

\bibitem[{{Riley} {et~al.}(2006){Riley}, {Schatzman}, {Cane}, {Richardson}, \&
  {Gopalswamy}}]{riley06}
{Riley}, P., {Schatzman}, C., {Cane}, H.~V., {Richardson}, I.~G., \&
  {Gopalswamy}, N. 2006, \apj, 647, 648

\bibitem[{{Roussev} {et~al.}(2004){Roussev}, {Sokolov}, {Forbes}, {Gombosi},
  {Lee}, \& {Sakai}}]{rou04}
{Roussev}, I.~I., {Sokolov}, I.~V., {Forbes}, T.~G., {Gombosi}, T.~I., {Lee},
  M.~A., \& {Sakai}, J.~I. 2004, \apjl, 605, L73

\bibitem[{{Roussev} {et~al.}(2003){Roussev}, {Gombosi}, {Sokolov}, {Velli},
  {Manchester}, {DeZeeuw}, {Liewer}, {T{\'o}th}, \& {Luhmann}}]{rou03}
{Roussev}, I.~I., {et~al.} 2003, \apjl, 595, L57 

\bibitem[Schwartz et al.(1988)]{schwartz88} Schwartz, S.~J., 
Thomsen, M.~F., Bame, S.~J., \& Stansberry, J.\ 1988, \jgr, 93, 12923 

\bibitem[{{Sime} \& {Hundhausen}(1987)}]{sime87}
{Sime}, D.~G., \& {Hundhausen}, A.~J. 1987, \jgr, 92, 1049

\bibitem[{{Sokolov} {et~al.}(2004){Sokolov}, {Roussev}, {Gombosi}, {Lee},
  {K{\'o}ta}, {Forbes}, {Manchester}, \& {Sakai}}]{sokolov04}
{Sokolov}, I.~V., {Roussev}, I.~I., {Gombosi}, T.~I., {Lee}, M.~A., {K{\'o}ta},
  J., {Forbes}, T.~G., {Manchester}, W.~B., \& {Sakai}, J.~I. 2004, \apjl, 616,
  L171

\bibitem[{{Sokolov} {et~al.}(2009){Sokolov}, {Roussev}, {Skender}, {Gombosi},
  \& {Usmanov}}]{sokolov09}
{Sokolov}, I.~V., {Roussev}, I.~I., {Skender}, M., {Gombosi}, T.~I., \&
  {Usmanov}, A.~V. 2009, \apj, 696, 261

\bibitem[{{Sokolov} {et~al.}(2013){Sokolov}, {van der Holst}, {Oran}, {Downs},
  {Roussev}, {Jin}, {Manchester}, {Evans}, \& {Gombosi}}]{sokolov13}
{Sokolov}, I.~V., {et~al.} 2013, \apj, 764, 23

\bibitem[{{Suresh} \& {Huynh}(1997)}]{suresh97}
{Suresh}, A., \& {Huynh}, H.~T. 1997, Journal of Computational Physics, 136, 83

\bibitem[{{Thompson} {et~al.}(1998){Thompson}, {Plunkett}, {Gurman}, {Newmark},
  {St.~Cyr}, \& {Michels}}]{thompson98}
{Thompson}, B.~J., {Plunkett}, S.~P., {Gurman}, J.~B., {Newmark}, J.~S.,
  {St.~Cyr}, O.~C., \& {Michels}, D.~J. 1998, \grl, 25, 2465

\bibitem[{{Thompson} {et~al.}(1999){Thompson}, {Gurman}, {Neupert}, {Newmark},
  {Delaboudini{\`e}re}, {St.~Cyr}, {Stezelberger}, {Dere}, {Howard}, \&
  {Michels}}]{thompson99}
{Thompson}, B.~J., {et~al.} 1999, \apjl, 517, L151

\bibitem[Thompson et al.(2010)]{thompson10} Thompson, W.~T., Wei, 
K., Burkepile, J.~T., Davila, J.~M., 
\& St.~Cyr, O.~C.\ 2010, \solphys, 262, 213 

\bibitem[{{Titov} \& {D{\'e}moulin}(1999)}]{titov99}
{Titov}, V.~S., \& {D{\'e}moulin}, P. 1999, \aap, 351, 707

\bibitem[Titov et al.(2014)]{titov14} Titov, V.~S., T{\"o}r{\"o}k, T., Mikic, Z., \& Linker, J.~A.\ 2014, \apj, 790, 163 

\bibitem[{{T{\'o}th} {et~al.}(2007){T{\'o}th}, {de Zeeuw}, {Gombosi},
  {Manchester}, {Ridley}, {Sokolov}, \& {Roussev}}]{toth07}
{T{\'o}th}, G., {de Zeeuw}, D.~L., {Gombosi}, T.~I., {Manchester}, W.~B.,
  {Ridley}, A.~J., {Sokolov}, I.~V., \& {Roussev}, I.~I. 2007, Space Weather,
  5, 6003

\bibitem[{{T{\'o}th} {et~al.}(2011){T{\'o}th}, {van der Holst}, \&
  {Huang}}]{toth11b}
{T{\'o}th}, G., {van der Holst}, B., \& {Huang}, Z. 2011, \apj, 732, 102

\bibitem[{{T{\'o}th} {et~al.}(2005){T{\'o}th}, {Sokolov}, {Gombosi}, {Chesney},
  {Clauer}, {De Zeeuw}, {Hansen}, {Kane}, {Manchester}, {Oehmke}, {Powell},
  {Ridley}, {Roussev}, {Stout}, {Volberg}, {Wolf}, {Sazykin}, {Chan}, {Yu}, \&
  {K{\'o}ta}}]{toth05}
{T{\'o}th}, G., {et~al.} 2005, Journal of Geophysical Research (Space Physics),
  110, 12226

\bibitem[T{\'o}th et al.(2012)]{toth12} T{\'o}th, G., van der Holst, B., Sokolov, I.~V., et al.\ 2012, Journal of Computational Physics, 231, 870 

\bibitem[{{van der Holst} {et~al.}(2009){van der Holst}, {Manchester},
  {Sokolov}, {T{\'o}th}, {Gombosi}, {DeZeeuw}, \& {Cohen}}]{bart09}
{van der Holst}, B., {Manchester}, IV, W., {Sokolov}, I.~V., {T{\'o}th}, G.,
  {Gombosi}, T.~I., {DeZeeuw}, D., \& {Cohen}, O. 2009, \apj, 693, 1178

\bibitem[{{van der Holst} {et~al.}(2010){van der Holst}, {Manchester},
  {Frazin}, {V{\'a}squez}, {T{\'o}th}, \& {Gombosi}}]{bart10}
{van der Holst}, B., {Manchester}, W.~B., {Frazin}, R.~A., {V{\'a}squez},
  A.~M., {T{\'o}th}, G., \& {Gombosi}, T.~I. 2010, \apj, 725, 1373

\bibitem[{{van der Holst} {et~al.}(2014){van der Holst}, {Sokolov}, {Meng},
  {Jin}, {Manchester}, {T{\'o}th}, \& {Gombosi}}]{bart14}
{van der Holst}, B., {Sokolov}, I.~V., {Meng}, X., {Jin}, M., {Manchester}, IV,
  W.~B., {T{\'o}th}, G., \& {Gombosi}, T.~I. 2014, \apj, 782, 81

\bibitem[{{Vourlidas} \& {Ontiveros}(2009)}]{vour09}
{Vourlidas}, A., \& {Ontiveros}, V. 2009, in American Institute of Physics
  Conference Series, Vol. 1183, American Institute of Physics Conference
  Series, ed. X.~{Ao} \& G.~Z.~R. {Burrows}, 139--146

\bibitem[{{Vourlidas} {et~al.}(2003){Vourlidas}, {Wu}, {Wang}, {Subramanian},
  \& {Howard}}]{vour03}
{Vourlidas}, A., {Wu}, S.~T., {Wang}, A.~H., {Subramanian}, P., \& {Howard},
  R.~A. 2003, \apj, 598, 1392

\bibitem[Vr{\v s}nak et al.(2014)]{vrsnak14} Vr{\v s}nak, B., 
Temmer, M., {\v Z}ic, T., et al.\ 2014, \apjs, 213, 21 

\bibitem[{{Wood} {et~al.}(2012){Wood}, {Wu}, {Rouillard}, {Howard}, \&
  {Socker}}]{wood12}
{Wood}, B.~E., {Wu}, C.-C., {Rouillard}, A.~P., {Howard}, R.~A., \& {Socker},
  D.~G. 2012, \apj, 755, 43

\bibitem[{{Wu} {et~al.}(2007{\natexlab{a}}){Wu}, {Fry}, {Dryer}, {Wu},
  {Thompson}, {Liou}, \& {Feng}}]{wu07a}
{Wu}, C.-C., {Fry}, C.~D., {Dryer}, M., {Wu}, S.~T., {Thompson}, B., {Liou},
  K., \& {Feng}, X.~S. 2007{\natexlab{a}}, Advances in Space Research, 40, 1827

\bibitem[{{Wu} {et~al.}(2007{\natexlab{b}}){Wu}, {Fry}, {Wu}, {Dryer}, \&
  {Liou}}]{wu07b}
{Wu}, C.-C., {Fry}, C.~D., {Wu}, S.~T., {Dryer}, M., \& {Liou}, K.
  2007{\natexlab{b}}, Journal of Geophysical Research (Space Physics), 112,
  9104

\bibitem[Wu et al.(1984)]{wu84} Wu, C.~S., Winske, D., 
Tanaka, M., et al.\ 1984, \ssr, 37, 63

\bibitem[Wu et al.(2001)]{wu01} Wu, S.~T., Zheng, H., Wang, 
S., et al.\ 2001, \jgr, 106, 25089

\bibitem[{{Xie} {et~al.}(2004){Xie}, {Ofman}, \& {Lawrence}}]{xie04}
{Xie}, H., {Ofman}, L., \& {Lawrence}, G. 2004, Journal of Geophysical Research
  (Space Physics), 109, 3109

\bibitem[Zhang et al.(2012)]{zhang12} Zhang, J., Cheng, X., \& Ding, M.-D.\ 2012, Nature Communications, 3, 747

\bibitem[{{Zhao} {et~al.}(2002){Zhao}, {Plunkett}, \& {Liu}}]{zhao02}
{Zhao}, X.~P., {Plunkett}, S.~P., \& {Liu}, W. 2002, Journal of Geophysical
  Research (Space Physics), 107, 1223

\end{thebibliography}

\newpage

\newpage

\clearpage
\begin{deluxetable}{cccccccccccc}
\tablecolumns{12}
\tablewidth{0pt}
\tabletypesize{\footnotesize}
\tablecaption{Simulated/Observed Solar Wind Statistics}
\tablehead{
\colhead{}    &  \multicolumn{5}{c}{STEREO A} &   \colhead{}   &
\multicolumn{5}{c}{OMNI} \\
\cline{2-6} \cline{8-12} \\
\colhead{Parameters} & \colhead{MSE\tablenotemark{a}}   & \colhead{Median\tablenotemark{b}}    & \colhead{Mean} &
\colhead{Max}    & \colhead{Min}   & \colhead{}    & \colhead{MSE} & \colhead{Median}    & \colhead{Mean} &
\colhead{Max}    & \colhead{Min}}
\startdata
$U$     & 0.79  & 0.84 & 0.85 & 0.79 & 0.85 & & 2.01 & 0.72 & 0.75 & 0.77 & 0.82 \\
$\rho$ & 36.99 & 2.02 & 1.61 & 1.15 & 1.66 & & 64.27  & 1.93 & 1.49 & 0.68 & 3.22 \\
$T_{p}$ & 133 & 0.19 & 0.33 & 0.46 & 0.30 & & 195 & 0.13 & 0.23 & 0.14 & 0.35 \\
$B_{r}$ & 3.80 & 0.71 & 0.80 & 0.64 & 0.25\tablenotemark{c} & & 7.07 & 0.75 & 0.75 & 0.51 & 1.11\\
\enddata
\tablenotetext{a}{MSE: mean square error. The units for velocity, density, proton temperature, and magnetic field are: 10$^4$ km$^2$ s$^{-2}$, cm$^{-6}$, nT$^2$, and 10$^8$ K$^2$, respectively.}
\tablenotetext{b}{The values (Median/Mean/Max/Min) showed in the table are the ratios between model and observation.}
\tablenotetext{c}{The daily-averaged data is used to obtained the minimum $B_{r}$.}
\end{deluxetable}

\newpage

\newpage

\newpage

\newpage

\newpage

\newpage

\newpage

\newpage

\newpage

\newpage

\newpage

\newpage

\newpage
\begin{figure}[h]
\begin{center}$
\begin{array}{c}
\includegraphics[scale=0.9]{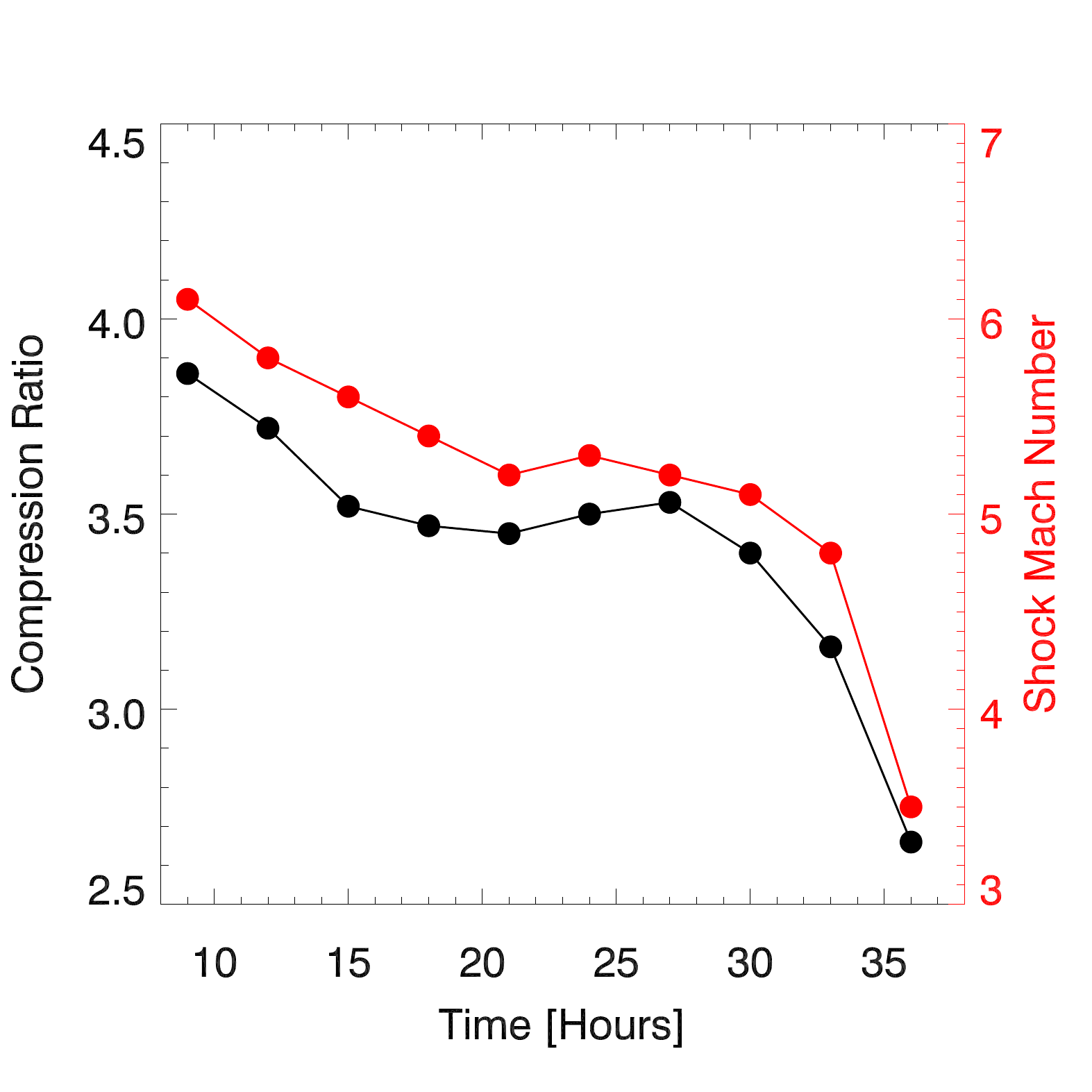}
\end{array}$
\end{center}
\caption{\label{fig:cratio_mach}The compression ratio and shock Mach number evolution at the shock front toward the STA. The black line shows the shock compression ratio and the red line shows the shock Mach number.}
\end{figure}

\newpage
\begin{figure}[h]
\begin{center}$
\begin{array}{c}
\includegraphics[scale=0.55]{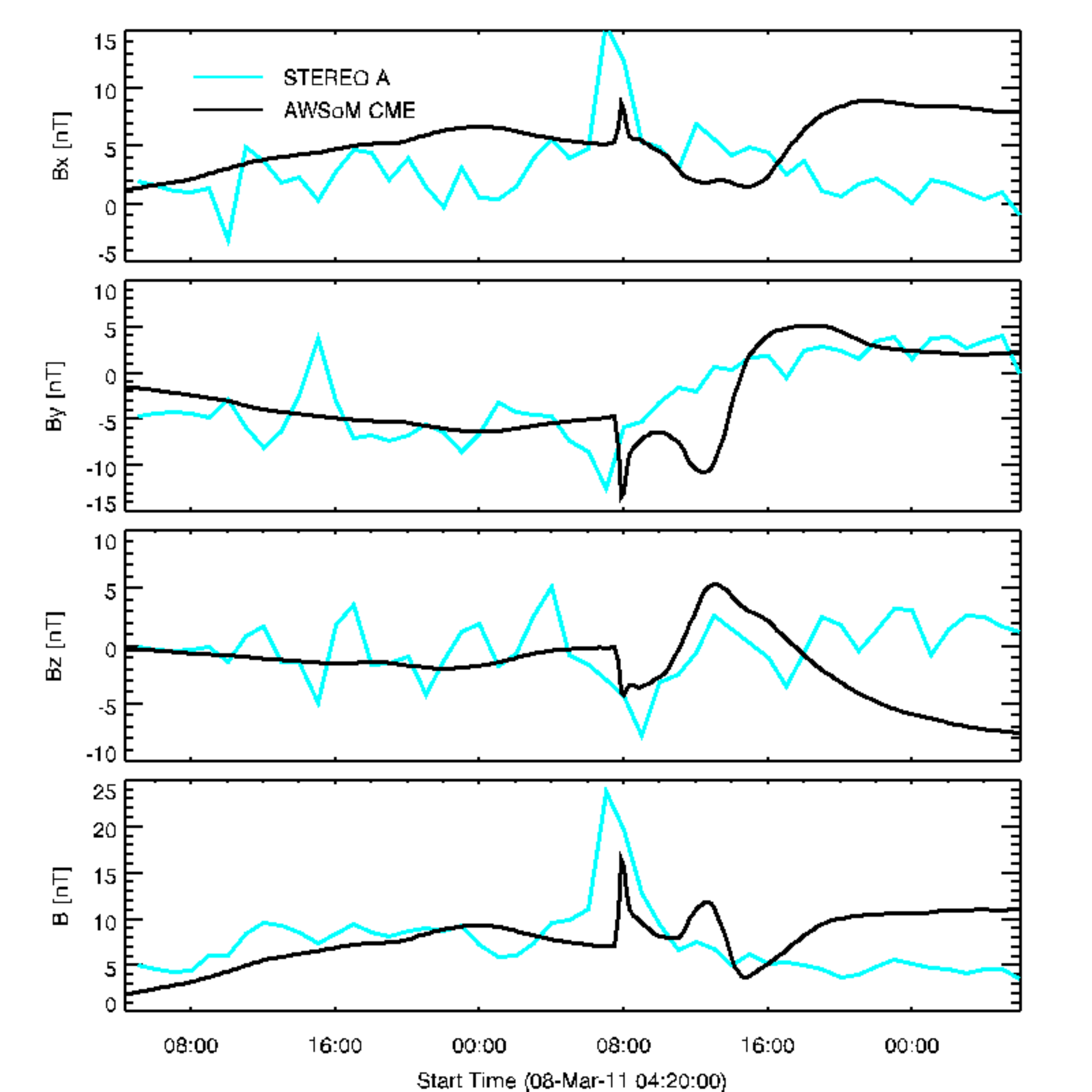}
\end{array}$
\end{center}
\caption{\label{fig:1aucme2}Comparison of the CME \textit{in situ} observation with the simulation for Bx, By, Bz, and total magnetic field.}
\end{figure}

\newpage
\begin{figure}[h]
\begin{center}$
\begin{array}{c}
\includegraphics[scale=0.55]{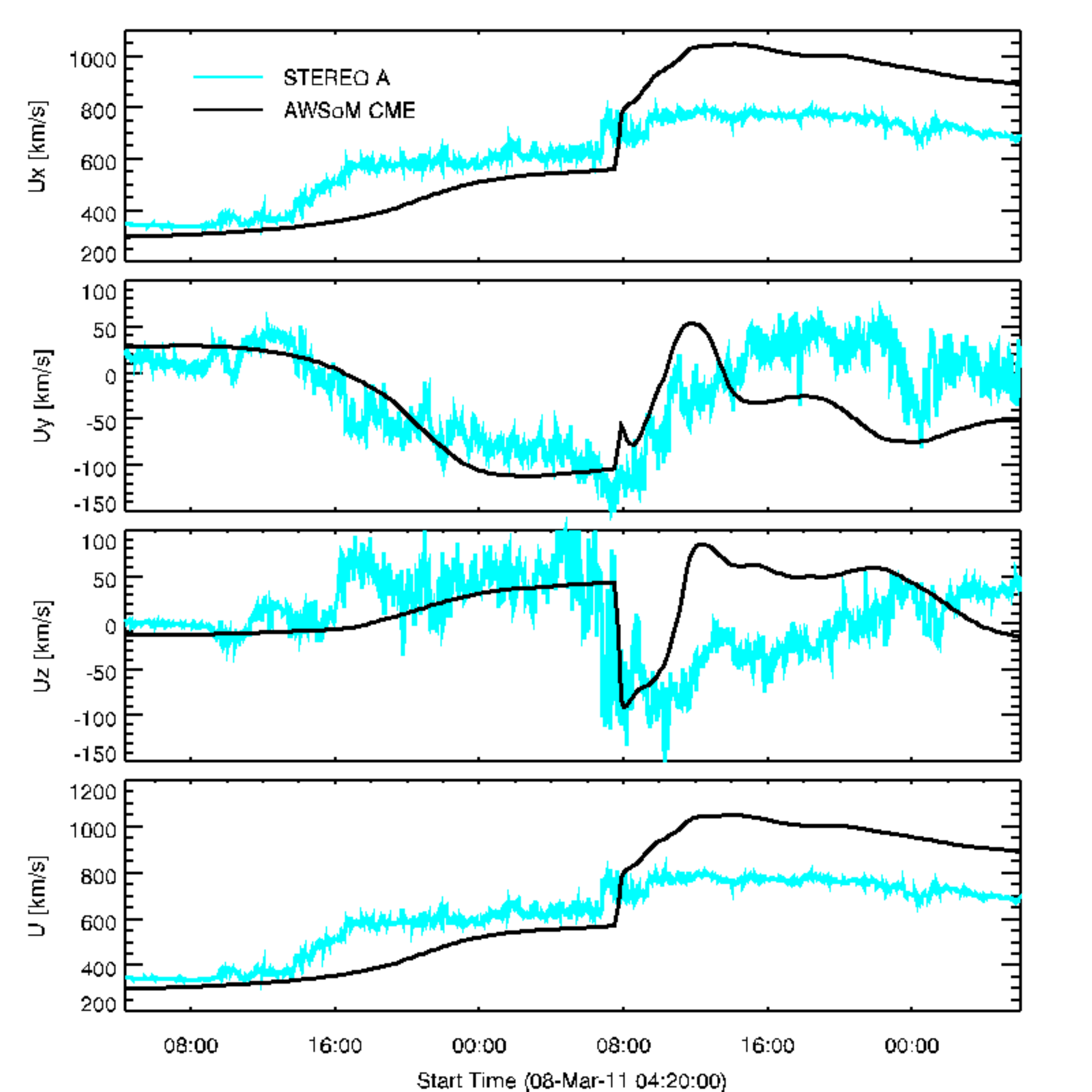}
\end{array}$
\end{center}
\caption{\label{fig:1aucme3}Comparison of the CME \textit{in situ} observation with the simulation for Vx, Vy, Vz, and total total velocity field.}
\end{figure}

\end{document}